\documentclass{aa}
\usepackage[varg]{txfonts}
\usepackage{graphicx}
\usepackage{natbib}
\usepackage{booktabs}
\usepackage{placeins}
\usepackage{float}
\usepackage{morefloats}
\bibpunct{(}{)}{;}{a}{}{,}

\begin{document}

\title{The Type IIb SN 2011dh - Two years of observations and modelling of the lightcurves.}

\author{M.~Ergon\inst{\ref{inst1}} \and A.~Jerkstrand\inst{\ref{inst2}} \and J.~Sollerman\inst{\ref{inst1}} \and N.~Elias-Rosa\inst{\ref{inst3}} \and C.~Fransson\inst{\ref{inst1}} \and M.~Fraser\inst{\ref{inst10},\ref{inst2}} \and A.~Pastorello\inst{\ref{inst3}} \and R.~Kotak\inst{\ref{inst2}} \and S.~Taubenberger\inst{\ref{inst4}} \and L.~Tomasella\inst{\ref{inst3}} \and S.~Valenti\inst{\ref{inst7},\ref{inst8}} \and S.~Benetti\inst{\ref{inst3}}  \and G.~Helou\inst{\ref{inst11}} \and M.M.~Kasliwal\inst{\ref{inst12}} \and J. Maund\inst{\ref{inst2}} \and S.J.~Smartt\inst{\ref{inst2}} \and J.~Spyromilio\inst{\ref{inst9}}}

\institute{The Oskar Klein Centre, Department of Astronomy, AlbaNova, Stockholm University, 106 91 Stockholm, Sweden 
\label{inst1}
\and Astrophysics Research Centre, School of Mathematics and Physics, Queens University Belfast, Belfast, BT7 1NN, UK 
\label{inst2}
\and INAF, Osservatorio Astronomico di Padova, vicolo dell'Osservatorio n. 5, 35122 Padua, Italy 
\label{inst3}
\and Institute of Astronomy, University of Cambridge, Madingley Road, Cambridge, CB3 0HA
\label{inst10}
\and Max-Planck-Institut für Astrophysik, Karl-Schwarzschild-Str. 1, D-85741 Garching, Germany 
\label{inst4}
\and Las Cumbres Observatory Global Telescope Network, 6740 Cortona Dr., Suite 102, Goleta, CA 93117, USA
\label{inst7}
\and Department of Physics, University of California, Santa Barbara, Broida Hall, Mail Code 9530, Santa Barbara, CA 93106-9530, USA
\label{inst8}
\and Infrared Processing and Analysis Center, California Institute of Technology, Pasadena, CA 91125, USA
\label{inst11}
\and Observatories of the Carnegie Institution for Science, 813 Santa Barbara St, Pasadena, CA 91101, USA
\label{inst12}
\and ESO, Karl-Schwarzschild-Strasse 2, 85748 Garching, Germany
\label{inst9}}
\date{Accepted for publication by Astronomy and Astrophysics.}

\abstract{We present optical and near-infrared (NIR) photometry and spectroscopy as well as modelling of the lightcurves of the Type IIb supernova (SN) 2011dh. Our extensive dataset, for which we present the observations obtained after day 100, spans two years, and complemented with Spitzer mid-infrared (MIR) data, we use it to build an optical-to-MIR bolometric lightcurve between days 3 and 732. To model the bolometric lightcurve before day 400 we use a grid of hydrodynamical SN models, which allows us to determine the errors in the derived quantities, and a bolometric correction determined with steady-state NLTE modelling. Using this method we find a helium core mass of 3.1$^{+0.7}_{-0.4}$~M$_\odot$ for SN 2011dh, consistent within error bars with previous results obtained using the bolometric lightcurve before day 80. We present bolometric and broad-band lightcurves between days 100 and 500 for the \citet{Jer14} steady-state NLTE models. The preferred 12 M$_\odot$ (initial mass) model, previously found to agree well with the observed nebular spectra, shows a good overall agreement with the observed lightcurves, although some discrepancies exist. Time-dependent NLTE modelling shows that after day $\sim$600 a steady-state assumption is no longer valid. The radioactive energy deposition in this phase is likely dominated by the positrons emitted in the decay of \element[ ][56]{Co}, but seems insufficient to reproduce the lightcurves, and what energy source is dominating the emitted flux is unclear. We find an excess in the $K$ and the MIR bands developing between days 100 and 250, during which an increase in the optical decline rate is also observed. A local origin of the excess is suggested by the depth of the \ion{He}{i} 20581 \AA~absorption. Steady-state NLTE models with a modest dust opacity in the core ($\tau=0.44$), turned on during this period, reproduce the observed behaviour, but an additional excess in the Spitzer 4.5 $\mu$m band remains. Carbon-monoxide (CO) first-overtone band emission is detected at day 206, and possibly at day 89, and assuming the additional excess to be dominated by CO fundamental band emission, we find fundamental to first-overtone band ratios considerably higher than observed in SN 1987A. The profiles of the [\ion{O}{i}] 6300 \AA~and \ion{Mg}{i}] 4571 \AA~lines show a remarkable similarity, suggesting that these lines originate from a common nuclear burning zone (O/Ne/Mg), and using small scale fluctuations in the line profiles we estimate a filling factor of $\lesssim$0.07 for the emitting material. This paper concludes our extensive observational and modelling work on SN 2011dh. The results from hydrodynamical modelling, steady-state NLTE modelling, and stellar evolutionary progenitor analysis are all consistent, and suggest an initial mass of $\sim$12 M$_\odot$ for the progenitor.}

\keywords{supernovae: general --- supernovae: individual (SN 2011dh)  --- supernovae: individual (SN 1993J)  --- supernovae: individual (SN 2008ax) --- galaxies: individual (M51)}

\titlerunning{The Type IIb SN 2011dh.}
\authorrunning{M. Ergon et al.}
\maketitle

\defcitealias{Erg14a}{E14}
\defcitealias{Erg15}{E15}
\defcitealias{Ber12}{B12}
\defcitealias{Arc11}{A11}
\defcitealias{Mau11}{M11}
\defcitealias{Tsv12}{T12}
\defcitealias{Vin12}{V12}
\defcitealias{Sch98}{S98}
\defcitealias{Sch11}{SF11}
\defcitealias{Mun97}{MZ97}
\defcitealias{Poz12}{P12}
\defcitealias{Val11}{V11}
\defcitealias{Bes12}{BM12}
\defcitealias{Pas09}{P09}
\defcitealias{Cho11}{C11}
\defcitealias{Mar13}{M13}
\defcitealias{Dyk13b}{D13}
\defcitealias{Sah13}{SA13}
\defcitealias{Shi13}{SH13}
\defcitealias{Jer14}{J14}

\section{Introduction}

Type IIb supernovae (SNe) are observationally characterized by a transition from Type II (with hydrogen lines) at early times to Type Ib (without hydrogen lines but with helium lines) at later times. The physical interpretation is that these SNe arise from stars that have lost most of their hydrogen envelope, either through stellar winds or interaction with a binary companion. Which of these production channels dominates is still debated, but for SN 1993J, the prime example of such a SN, a companion star was detected by direct observations \citep{Mau04,Fox14}. Observations of bright, nearby Type IIb SNe such as 1993J, 2008ax, and the recent 2011dh are essential to improve our understanding of this class. Identification of the progenitor star in pre-explosion images, a search for the companion star when the SN has faded, and multi-method modelling of high quality data, all provide important clues to the nature of Type IIb SNe and their progenitor stars.

In this paper we present the late-time part of the extensive optical and near-infrared (NIR) dataset, covering nearly two years, that we have obtained for SN 2011dh. The first 100 days of this dataset were presented in \citet[hereafter \citetalias{Erg14a}]{Erg14a}. Detailed hydrodynamical modelling of the SN using those data were presented in \citet[hereafter \citetalias{Ber12}]{Ber12}, and steady-state NLTE modelling of nebular spectra in \citet[hereafter \citetalias{Jer14}]{Jer14}. Identification and analysis of the plausible progenitor star was presented in \citet[hereafter \citetalias{Mau11}]{Mau11}, and confirmation of the progenitor identification through its disappearance in \citetalias{Erg14a}.

SN 2011dh was discovered on 2011 May 31.893 UT \citep{Gri11,Arc11} in the nearby galaxy M51, and has been extensively monitored from X-ray to radio wavelengths. Most observations cover the period between days 3 and 100 but late-time data have been published by \citet{Tsv12}, \citet{Dyk13b}, \citet{Sah13}, \citet{Shi13} and \citet{Hel13}. The explosion epoch (May 31.5 UT), the distance to M51 (7.8$^{+1.1}_{-0.9}$ Mpc), and the interstellar line-of-sight extinction towards the SN ($E$($B$-$V$)=0.07$^{+0.07}_{-0.04}$ mag) are all adopted from \citetalias{Erg14a}. The phase of the SN is expressed relative to this explosion epoch throughout the paper.

The nature of the progenitor star has been a key issue since the identification of a yellow supergiant in pre-explosion images, coincident with the SN (\citetalias{Mau11}; \citealt{Dyk11}). Recent progress in modelling of the SN (\citetalias{Ber12}; \citealt{Shi13}; \citetalias{Jer14}) and the disappearance of the progenitor candidate (\citealt{Dyk13b}; \citetalias{Erg14a}) strengthen the hypothesis that the progenitor was a yellow supergiant of moderate mass, as was originally proposed in \citetalias{Mau11}. As shown in \citet{Ben12}, a binary interaction scenario that reproduces the observed and modelled properties of the yellow supergiant is possible.

The paper is organized as follows. In Sects.~\ref{s_phot} and \ref{s_spec} we present and analyse our photometric and spectroscopic observations, respectively, and compare these to SNe 1993J and 2008ax. In Sect.~\ref{s_modelling} we model the lightcurves before day 500 and in Sect.~\ref{s_500_days_lightcurves} we discuss the lightcurves after day 500. Finally, we conclude and summarize our results in Sect.~\ref{s_conclusions}.

\section{Photometry}
\label{s_phot}

Here we present observations spanning days 100$-$732 in the optical, days 100$-$380 in the NIR and days 100$-$1061 in the MIR, and provide analysis and comparisons with SNe 1993J and 2008ax. The distance, extinction and the references for the photometric data used for SNe 1993J and 2008ax are the same as in \citetalias{Erg14a}. The lack of S-corrected photometry for SN 1993J complicates the comparison, whereas for SN 2008ax, the S-corrected Johnson-Cousins (JC) photometry by \citet{Tau11} agrees reasonably well with the JC photometry by \citet{Tsv09}. 

\begin{figure*}[tbp!]
\includegraphics[width=1.0\textwidth,angle=0]{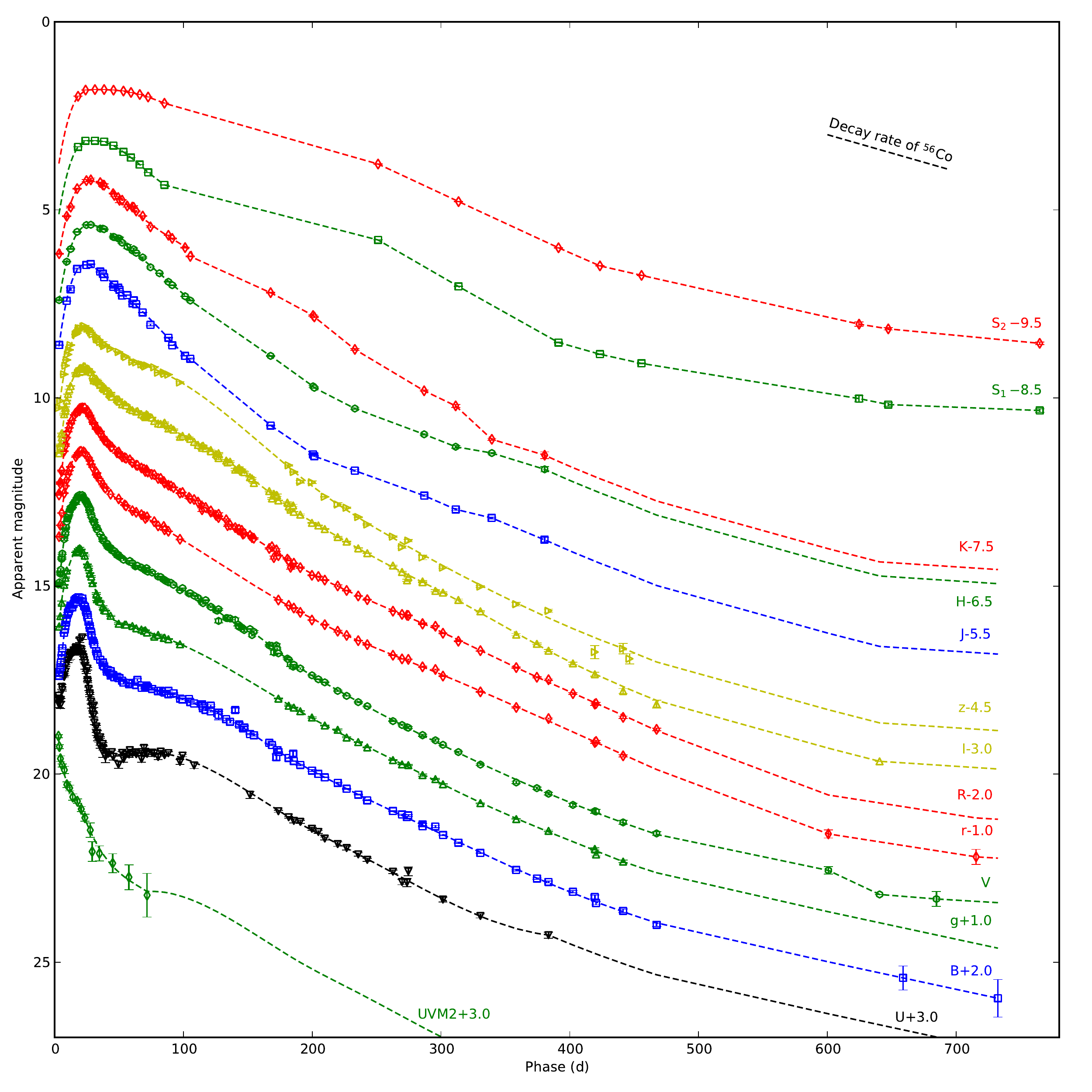}
\caption{The apparent UV, optical, NIR, and MIR broad-band lightcurves before day 800 for SN 2011dh. For clarity each band has been shifted in magnitude and the lightcurve annotated with the shift applied. We also show combinations of spline fits, interpolations, and extrapolations (dashed lines) used e.g.~in the calculations of the pseudo-bolometric lightcurves.}
\label{f_uv_opt_nir_mir}
\end{figure*}

\subsection{Observations}
\label{s_phot_obs}

The late-time data were obtained with the Liverpool Telescope (LT), the Nordic Optical Telescope (NOT), Telescopio Nazionale (TNG), the Calar Alto 3.5m (CA 3.5m) and 2.2m (CA 2.2m) telescopes, the Asiago 67/92cm Schmidt (AS Schmidt) and 1.82m Copernico (AS 1.82m) telescopes, the William Herschel Telescope (WHT), the Albanova Telescope\footnote{1.0 m Nasmyth-Cassegrain telescope at Albanova, Stockholm University. Equipped with an Andor DV438 CCD camera with an E2V 20C CCD chip and Bessel-like $UBVRI$ filters.} (AT), and the United Kingdom Infrared Telescope (UKIRT). The late-time dataset includes 61 (146 in total) and 9 (32 in total) epochs of optical and NIR imaging, respectively. Spitzer has systematically been observing SN 2011dh under programme ID 70207 (PI Helou), 90240 and 10136 (PI Kasliwal). \citet{Erg14a} and \citet{Hel13} presented Spitzer photometry up to days 85 and 625, respectively, and here we present observations up to day 1061. Throughout the paper, we label the 3.6 $\mu$m and 4.5 $\mu$m bands $S_1$ and $S_2$, respectively. The reduction and calibration procedures are described in \citetalias{Erg14a}, and issues specifically related to the late-time data are discussed in Appendix \ref{a_data_reductions}.

The S-corrected optical and NIR magnitudes and their corresponding errors are listed in Tables \ref{t_jc}, \ref{t_sloan}, and \ref{t_nir} and the Spitzer MIR magnitudes and their corresponding errors in Table~\ref{t_mir}. The UV, optical, NIR, and MIR broad-band lightcurves before day 800, including SWIFT \citepalias{Erg14a} and S-corrected HST \citep{Dyk13b} photometry, are shown in Fig.~\ref{f_uv_opt_nir_mir}. We also show cubic spline fits or, when the sampling is sparse, linear interpolations, as well as extrapolations assuming a constant colour to adjacent bands. These are used in a number of subsequent calculations, where the errors in the fitted splines are estimated by the standard deviation and then propagated. The extrapolation method used makes most sense in the early phase when the spectral energy distribution (SED) is dominated by the continuum. In particular, the NIR magnitudes are extrapolated between days $\sim$400 and $\sim$750, which introduces a considerable uncertainty.

\subsection{Photometric evolution}
\label{s_phot_evo}

\begin{figure}[tbp!]
\includegraphics[width=0.48\textwidth,angle=0]{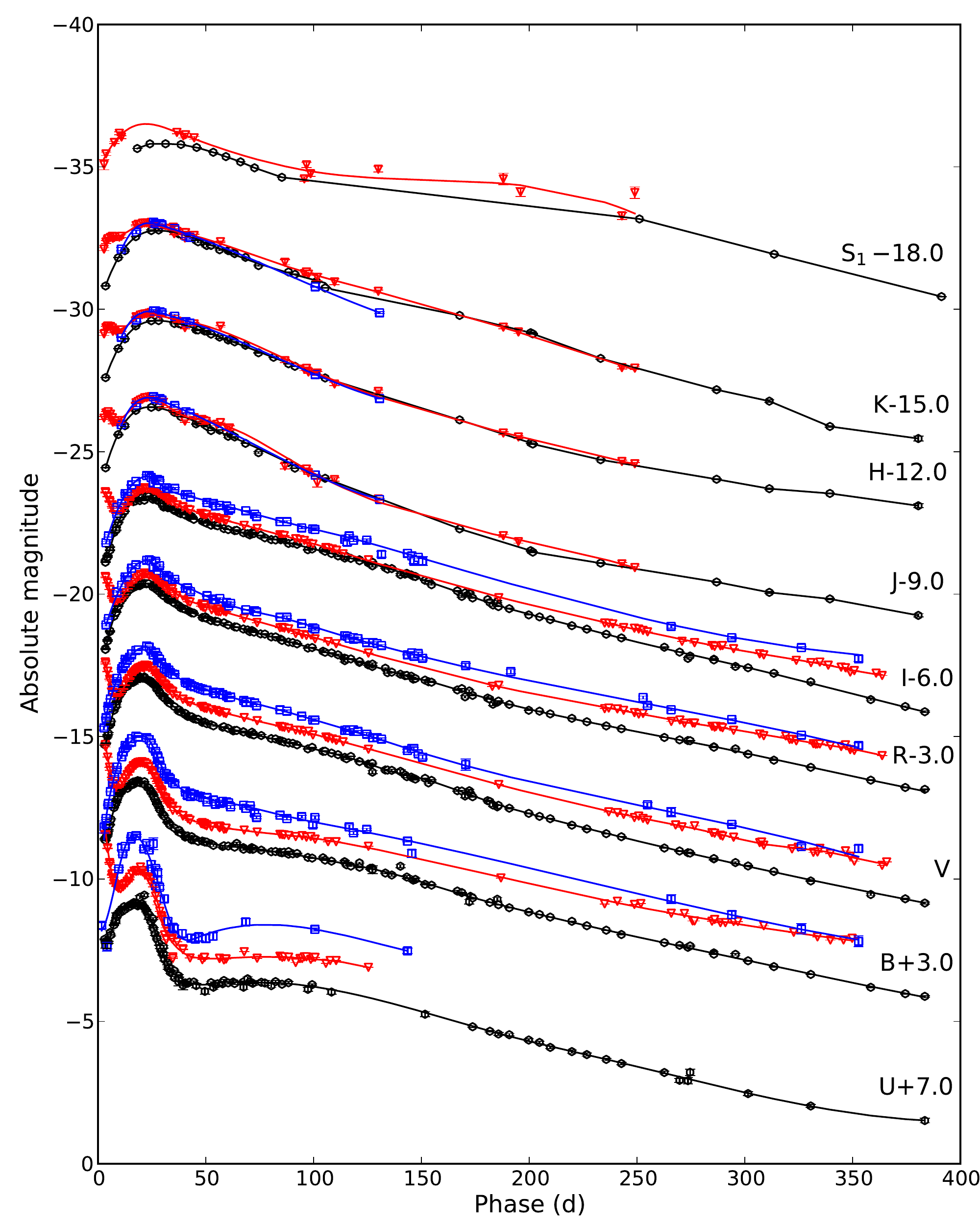}
\caption{The absolute optical, NIR, and MIR broad-band lightcurves before day 400 for SNe 2011dh (black circles), 1993J (red triangles), and 2008ax (blue squares). For clarity each band has been shifted in magnitude and the lightcurve annotated with the shift applied. We also show combinations of spline fits, interpolations, and extrapolations (solid lines) used e.g.~in the calculations of the pseudo-bolometric lightcurves.}
\label{f_opt_nir_comp}
\end{figure}

In Fig.~\ref{f_opt_nir_comp} we show absolute optical, NIR, and MIR broad-band lightcurves for SNe 2011dh, 1993J, and 2008ax, and in Table \ref{t_lc_char_comp_comb} we tabulate the tail decline rates at days 100, 200, and 300. Most striking is the similarity between the lightcurves, except for a shift towards higher luminosities for SNe 1993J and 2008ax, the shift being larger in bluer bands and negligible in the NIR, and most pronounced for SN 2008ax. As discussed in \citetalias{Erg14a}, this difference could be explained by errors in the adopted extinctions.

Given the caveat that SNe 1993J and 2008ax are only partly covered in $U$ and NIR, we find the following general trends. At day 100 the $V$-, $R$-, and $I$-band decline rates are roughly twice the decay rate of \element[ ][56]{Co}, and subsequently decrease towards day 300. The $U$- and $B$-band decline rates are significantly lower at day 100, after which they approach the other optical decline rates and then evolve similarly. The $J$- and $H$-band decline rates are considerably higher than the optical at day 100, subsequently approaches these rates and eventually become considerably lower. For SNe 2011dh and 1993J, the $K$ band behaves quite differently than the other NIR bands. At day 100 the decline rate is significantly lower, but as it remains roughly constant, it subsequently approaches the other NIR decline rates and eventually becomes considerably higher. As seen in Fig.~\ref{f_uv_opt_nir_mir}, the optical lightcurves of SN 2011dh flatten considerably after day $\sim$450, approaching a decline rate similar to, or lower than, the decay rate of \element[ ][56]{Co}.

SNe 2011dh and 1993J were also monitored in the MIR $S_1$ and $S_2$ and $L$ (similar to $S_1$) bands, respectively, and for both SNe a strong excess develops between days 100 and 250. For SN 1993J the MIR coverage ends at day $\sim$250, and for SN 2011dh the subsequent evolution is fairly similar to that in the optical. After day $\sim$400, a considerable flattening of the lightcurves is seen, similar to, but more pronounced than in the optical, and after day $\sim$750 (not shown in Fig.~\ref{f_uv_opt_nir_mir}) this trend continues and the MIR lightcurves become almost flat.

\begin{table}[tbp!]
\caption{Optical and NIR broad-band and optical-to-NIR pseudo-bolometric decline rates at days 100, 200, and 300 for SNe 2011dh, 1993J, and 2008ax.}
\begin{center}
\begin{tabular}{lllll}
\hline\hline \\ [-1.5ex]
SN & Band & Rate (100 d) & Rate (200 d) & Rate (300 d)\\ [0.5ex]
 & & (mag day$^{-1}$) & (mag day$^{-1}$) & (mag day$^{-1}$)\\
\hline \\ [-1.5ex]
2011dh & $U$ & 0.011 & 0.019 & 0.017\\
2011dh & $B$ & 0.014 & 0.018 & 0.016\\
2011dh & $V$ & 0.019 & 0.021 & 0.017\\
2011dh & $R$ & 0.021 & 0.018 & 0.016\\
2011dh & $I$ & 0.019 & 0.020 & 0.018\\
2011dh & $J$ & 0.036 & 0.018 & 0.011\\
2011dh & $H$ & 0.029 & 0.019 & 0.011\\
2011dh & $K$ & 0.020 & 0.020 & 0.024\\
2011dh & $U$-$K$ & 0.021 & 0.019 & 0.016\\
\hline \\ [-1.5ex]
1993J & $U$ & 0.006& ...& ...\\
1993J & $B$ & 0.011 & 0.017 & 0.012\\
1993J & $V$ & 0.019 & 0.018 & 0.017\\
1993J & $R$ & 0.022 & 0.014 & 0.013\\
1993J & $I$ & 0.022 & 0.018 & 0.013\\
1993J & $J$ & 0.041 & 0.017& ...\\
1993J & $H$ & 0.033 & 0.018& ...\\
1993J & $K$ & 0.022 & 0.024& ...\\
1993J & $U$-$K$ & 0.021 & 0.017 & 0.013\\
\hline \\ [-1.5ex]
2008ax & $U$ & 0.013& ...& ...\\
2008ax & $B$ & 0.015 & 0.018 & 0.016\\
2008ax & $V$ & 0.022 & 0.017 & 0.018\\
2008ax & $R$ & 0.023 & 0.015 & 0.016\\
2008ax & $I$ & 0.018 & 0.020 & 0.012\\
2008ax & $J$ & 0.035& ...& ...\\
2008ax & $H$ & 0.032& ...& ...\\
2008ax & $K$ & 0.032& ...& ...\\
2008ax & $U$-$K$ & 0.020 & 0.017 & 0.015\\ [0.5ex]
\hline
\end{tabular}

\end{center}
\label{t_lc_char_comp_comb}
\end{table}

\subsection {Bolometric evolution}
\label{s_bol_evo}

To calculate the bolometric lightcurves we use the spectroscopic and photometric methods described in \citetalias{Erg14a}, applied to wavelength regions with and without spectral information, respectively. Combinations of spline fits, interpolations, and extrapolations (Sect.~\ref{s_phot_obs}; Figs.~\ref{f_uv_opt_nir_mir} and \ref{f_opt_nir_comp}) have been used to calculate the magnitudes. Here and throughout the paper, the wavelength regions over which the luminosity is integrated are specified as follows; UV (1900$-$3300 \AA), optical (3300$-$10000 \AA), NIR (10000$-$24000 \AA), and MIR (24000$-$50000 \AA).

\begin{figure}[tbp!]
\includegraphics[width=0.48\textwidth,angle=0]{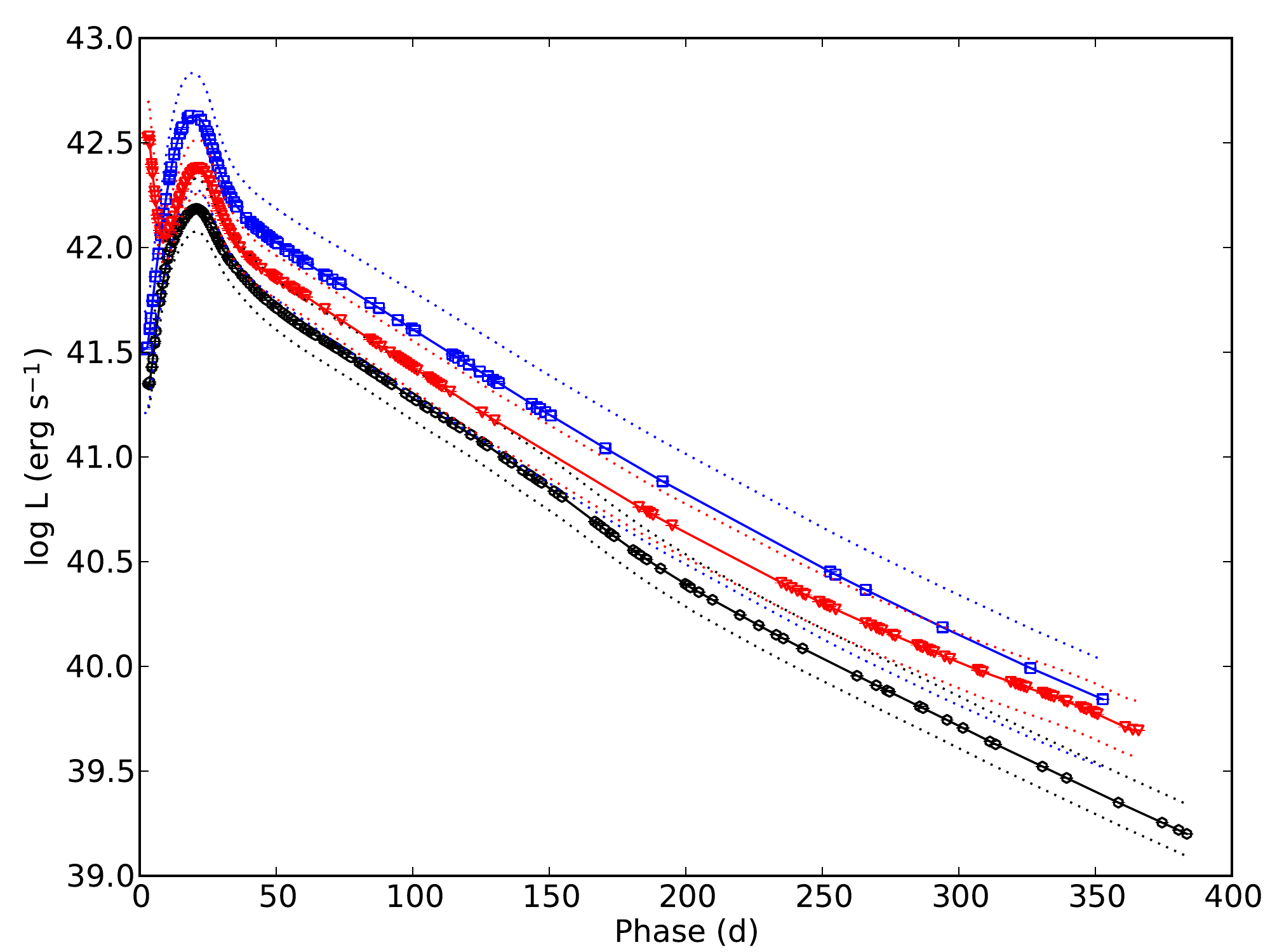}
\caption{The optical-to-NIR pseudo-bolometric lightcurve before day 400 for SNe 2011dh (black circles and solid line), 1993J (red triangles and solid line), and 2008ax (blue squares and solid line). The upper and lower error bars for the systematic error arising from extinction and distance (dotted lines) are also shown.}
\label{f_UK_bol_comp}
\end{figure}

Figure~\ref{f_UK_bol_comp} shows the optical-to-NIR pseudo-bolometric lightcurves before day 500 for SNe 2011dh, 1993J, and 2008ax, as calculated with the photometric method, and in Table \ref{t_lc_char_comp_comb} we tabulate the decline rates at days 100, 200, and 300. Given the caveat that SNe 1993J and 2008ax are not covered in NIR after days $\sim$250 and $\sim$150, respectively, their optical-to-NIR pseudo-bolometric lightcurves are remarkably similar to the one of SN 2011dh, except for the shift towards higher luminosities mentioned in Sect.~\ref{s_phot_evo}. The decline rates decrease from $\sim$0.020 mag day$^{-1}$, roughly twice the decay rate of \element[ ][56]{Co}, at day 100 to $\sim$0.015 mag day$^{-1}$ at day 300. There is, however, a significant increase in the decline rate to $\sim$0.025 mag day$^{-1}$ between days 150 and 200 for SN 2011dh, even more pronounced in the optical pseudo-bolometric lightcurve, not seen for SNe 1993J and 2008ax. For SN 1993J the decline rate becomes increasingly lower than for SNe 2011dh and 2008ax towards day 300, which is consistent with an increasing contribution from circumstellar medium (CSM) interaction in this phase.

\begin{figure}[tbp!]
\includegraphics[width=0.48\textwidth,angle=0]{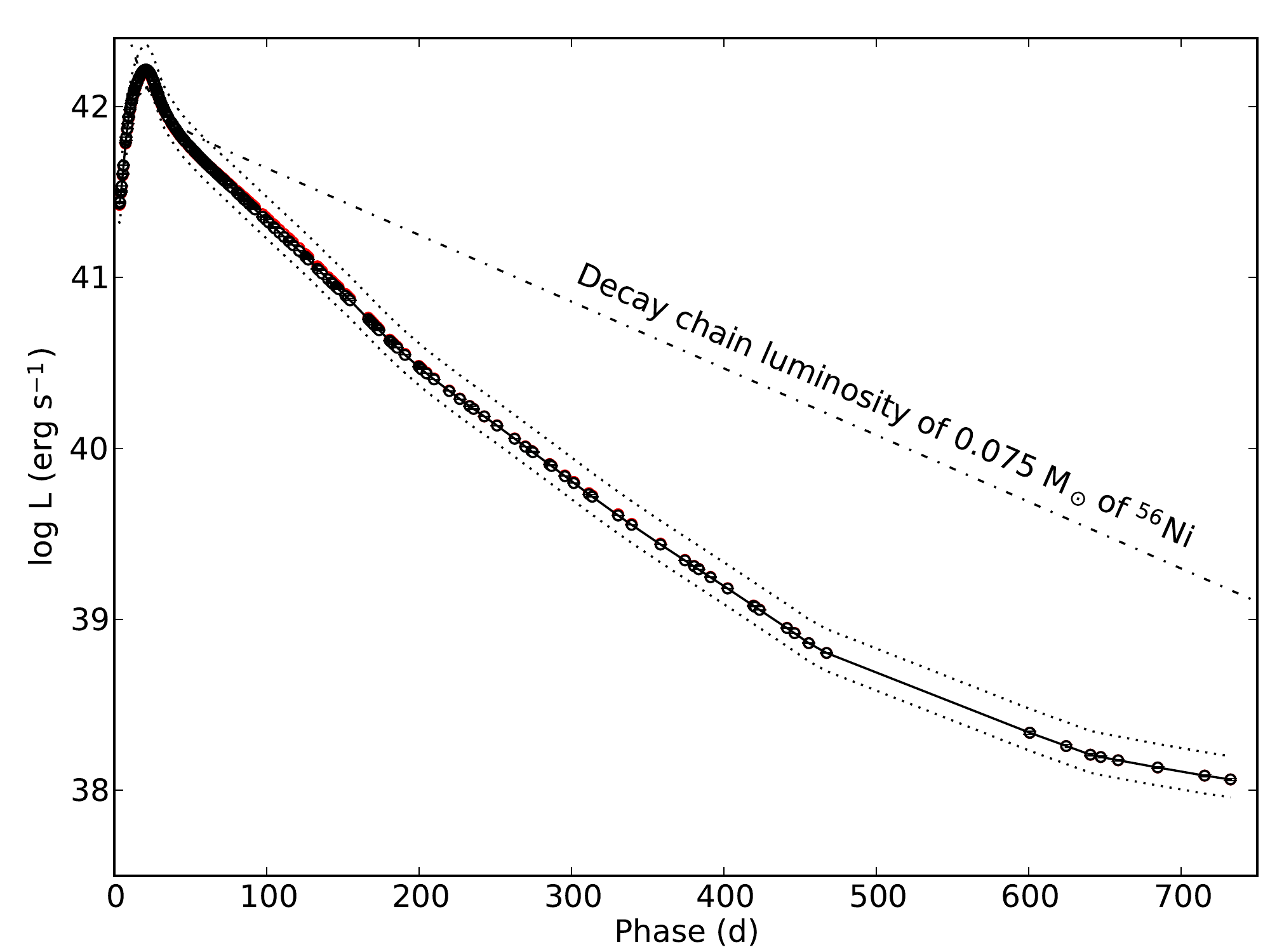}
\caption{The UV-to-MIR pseudo-bolometric lightcurve before day 750 for SN 2011dh (black circles and solid line). The upper and lower error bars for the systematic error arising from extinction and distance (black dotted lines) and the radioactive decay chain luminosity of 0.075 M$_\odot$ of \element[ ][56]{Ni} (black dash-dotted line) are also shown.}
\label{f_UV_MIR_bol}
\end{figure}

Figure~\ref{f_UV_MIR_bol} shows the UV-to-MIR pseudo-bolometric lightcurve before day 750 for SN 2011dh, as calculated with the combined spectroscopic and photometric methods, and in Table~\ref{t_UV_MIR_bol} we give the luminosity between days 3 and 400 (for which we have full $U$ to $S_2$ coverage) for reference. The decline rates are similar to those for the optical-to-NIR pseudo-bolometric lightcurve, but the increase between days 150 and 200 
is not as pronounced. Given the caveat that the NIR coverage ends at day $\sim$400, the UV-to-MIR pseudo-bolometric lightcurve shows a significant flattening after day $\sim$400, and the decline rates for days 467$-$601 and 601$-$732 are 0.0088 and 0.0061 mag day$^{-1}$, respectively. This flattening is also observed in the optical and MIR pseudo-bolometric lightcurves, for which we have full coverage, and after day $\sim$750 the MIR pseudo-bolometric lightcurve becomes almost flat. The optical decline rates are 0.0095 and 0.0069 mag day$^{-1}$ for days 467$-$601 and 601$-$732, respectively, whereas the MIR decline rates are 0.0069, 0.0031, and 0.0006 mag day$^{-1}$ for days 456$-$625, 625$-$765, and 765$-$1061, respectively. The flattening of the late-time lightcurves is discussed in more detail in Sect.~\ref{s_500_days_lightcurves}.

\begin{figure}[tbp!]
\includegraphics[width=0.48\textwidth,angle=0]{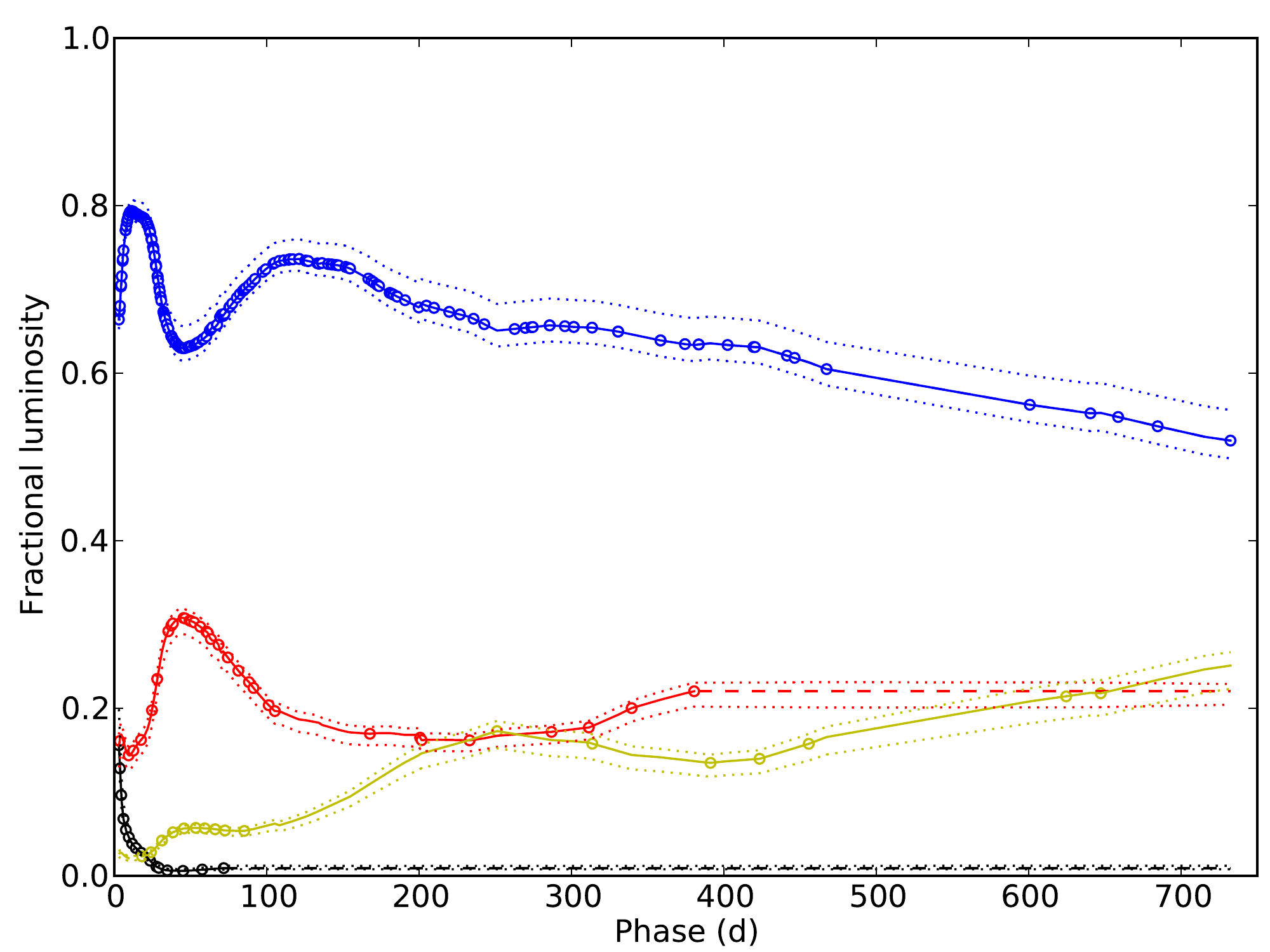}
\caption{The fractional UV (black dots), optical (blue dots), NIR (red dots), and MIR (yellow dots) luminosity before day 750 for SN 2011dh. Interpolations and extrapolations are shown as solid and dashed lines, respectively. The upper and lower error bars for the systematic error arising from extinction (dotted lines) are also shown.}
\label{f_bol_frac}
\end{figure}

Figure~\ref{f_bol_frac} shows the fractional UV, optical, NIR, and MIR luminosities for SN 2011dh. We assume the late-time extrapolated fractions to be constant and we do not use the adjacent colour based extrapolations applied elsewhere. After day 100 the most notable is the strong increase in the MIR fraction between days 100 and 250, together with a simultaneous decrease in the optical fraction. Also notable is the increase in the NIR fraction between days 200 and 400 caused by the evolution in the $J$ and $H$ bands. The evolution becomes quite uncertain after day $\sim$400 when the NIR coverage ends, but there seems to be a continuous increase in the  MIR fraction, at the expense of the optical fraction. Keeping the uncertainties in mind, it is worth noting the dominance of the optical luminosity even at day $\sim$750.

\begin{figure}[tbp!]
\includegraphics[width=0.48\textwidth,angle=0]{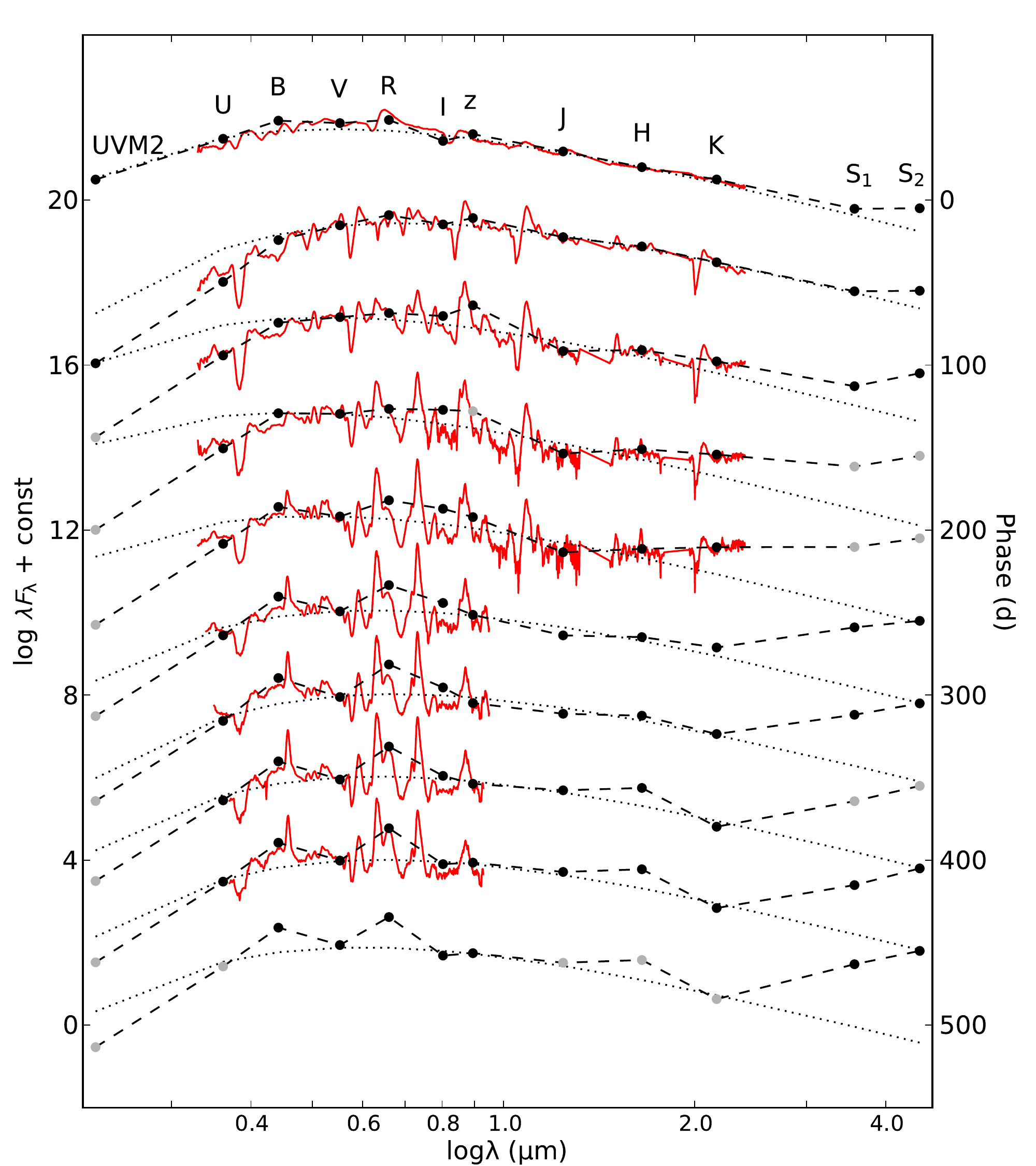}
\caption{The evolution of the SED (black dots and dashed lines) before day 500, overplotted with blackbody fits to the $V$, $I$, $J$, $H$, and $K$ photometry (black dotted lines) and the observed (interpolated; Sect.~\ref{s_spec_obs}) spectra (red solid lines). Fluxes based on extrapolations or interpolations over more than 25 days are shown in shaded colour.}
\label{f_sed_evo}
\end{figure}

Figure~\ref{f_sed_evo} shows the evolution of the SED before day 500 overplotted with blackbody fits to the $V$, $I$, $J$, $H$, and $K$ photometry, as well as the observed (interpolated) spectra. After day 100 the most notable is again the strong excess (relative to the blackbody fits) developing in the MIR between days 100 and 250, although an excess in the $S_2$ band is observed already during the first 100 days \citepalias{Erg14a}. There is also a similar excess (relative to the blackbody fits) developing in the $K$ band between days 100 and 200, that gradually fades away towards day 300. The evolution in the $K$ and MIR bands and the behaviour of the pseudo-bolometric lightcurves between days 150 and 200 are discussed in more detail in Sect.~\ref{s_modelling_molecules_dust} where we compare the observed lightcurves to results from modelling.

\section{Spectroscopy}
\label{s_spec}

Here we present observations spanning days 100$-$415 in the optical and days 100$-$206 in the NIR, and provide analysis and comparisons with SNe 1993J and 2008ax. Steady-state NLTE modelling of these data, as well as a detailed analysis of the formation of the identified lines and the evolution of their fluxes, are presented in \citetalias{Jer14}. Our analysis is complementary and focuses on the line profiles, and what can be learned about the distribution of the material from the nuclear burning zones. Following \citetalias{Jer14} these are the Fe/Co/He core, the oxygen-rich O/Si/S, O/Ne/Mg, and O/C zones, and the helium and hydrogen envelopes. This progression corresponds to the onion-like structure obtained in 1-D modelling \citep[e.g.][]{Woo07}, whereas in multi-dimensional modelling hydrodynamical instabilities result in macroscopic mixing of these zones \citep[e.g.][]{Iwa97}. We adopt a recession velocity of 515 km~s$^{-1}$ for SN 2011dh, as estimated from the H$\alpha$ velocity map in \citet{She07}. For SNe 1993J and 2008ax we adopt the systematic recession velocities for M81 \citep[$-$34 km~s$^{-1}$;][]{Tul08} and NGC 4490 \citep[565 km~s$^{-1}$;][]{Lav11}, respectively. The references for the spectroscopic data for SNe 1993J and 2008ax are the same as in \citetalias{Erg14a}, except that we also include spectra for SN 2008ax from \citet{Mil10}, and unpublished spectra for SN 1993J obtained with the WHT and the Isaac Newton Telescope (INT) (P. Meikle et al.; private communication), described in \citetalias{Jer14}.

\subsection{Observations}
\label{s_spec_obs}

The late-time data were obtained with the NOT, the TNG, the WHT, the INT, the CA 2.2m, the AS 1.82m, and the Gran Telescopio Canarias (GTC). The late-time dataset includes 26 (81 in total) optical spectra obtained at 21 (47 in total) epochs and 2 (20 in total) NIR spectra obtained at 2 (12 in total) epochs. The details of the late-time observations are given in Table~\ref{t_speclog}, and the reductions and calibration procedures are described in \citetalias{Erg14a}.

\begin{figure*}[tb]
\includegraphics[width=1.0\textwidth,angle=0]{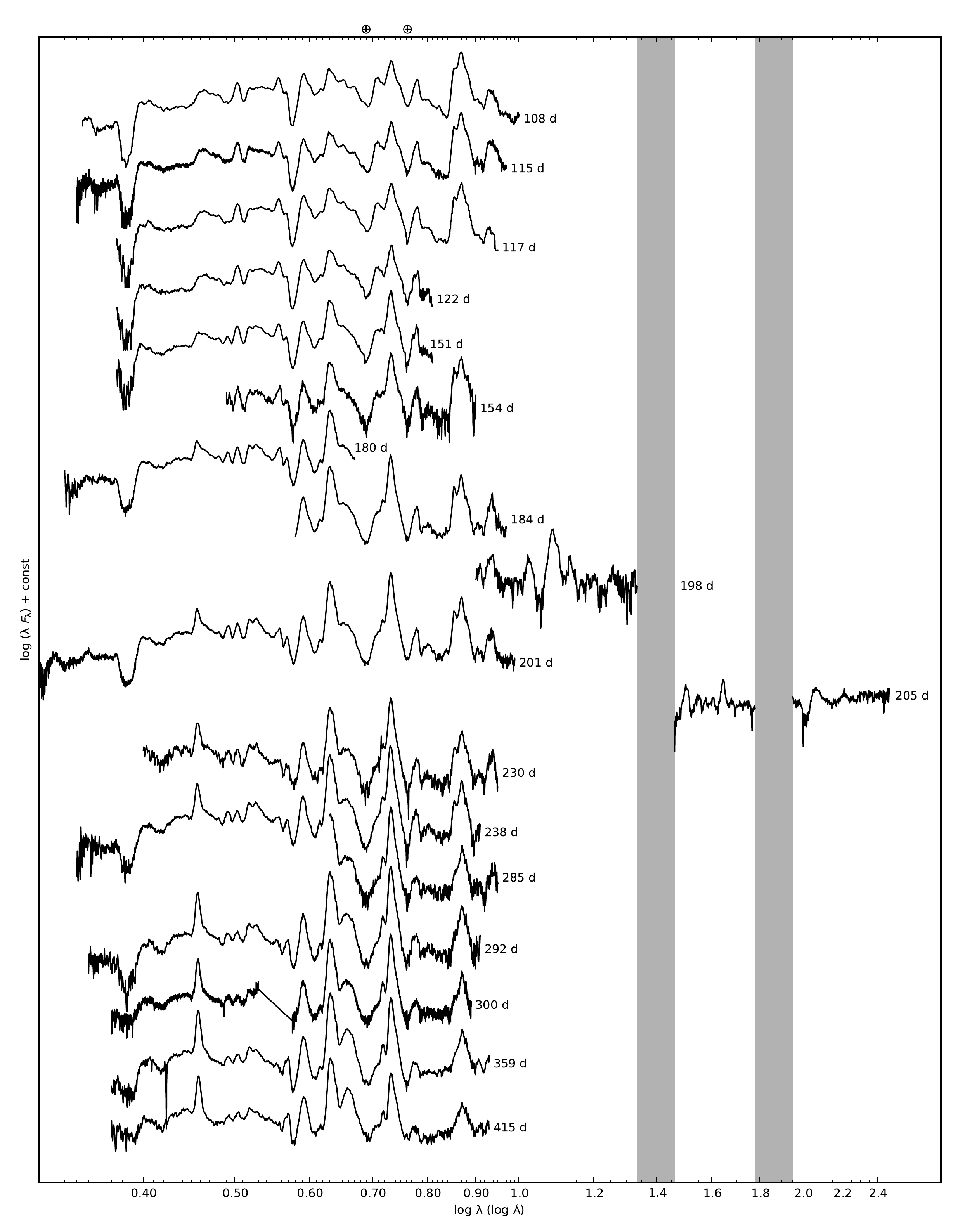}
\caption{Sequence of the observed late-time (day 100$-$415) spectra for SN 2011dh. Spectra obtained on the same night using the same telescope and instrument have been combined and each spectra have been labelled with the phase of the SN. Telluric absorption bands are marked with a $\oplus$ symbol in the optical and are shown as grey regions in the NIR.}
\label{f_spec_evo_opt_NIR_trad}
\end{figure*}

\begin{figure*}[tb]
\includegraphics[width=1.0\textwidth,angle=0]{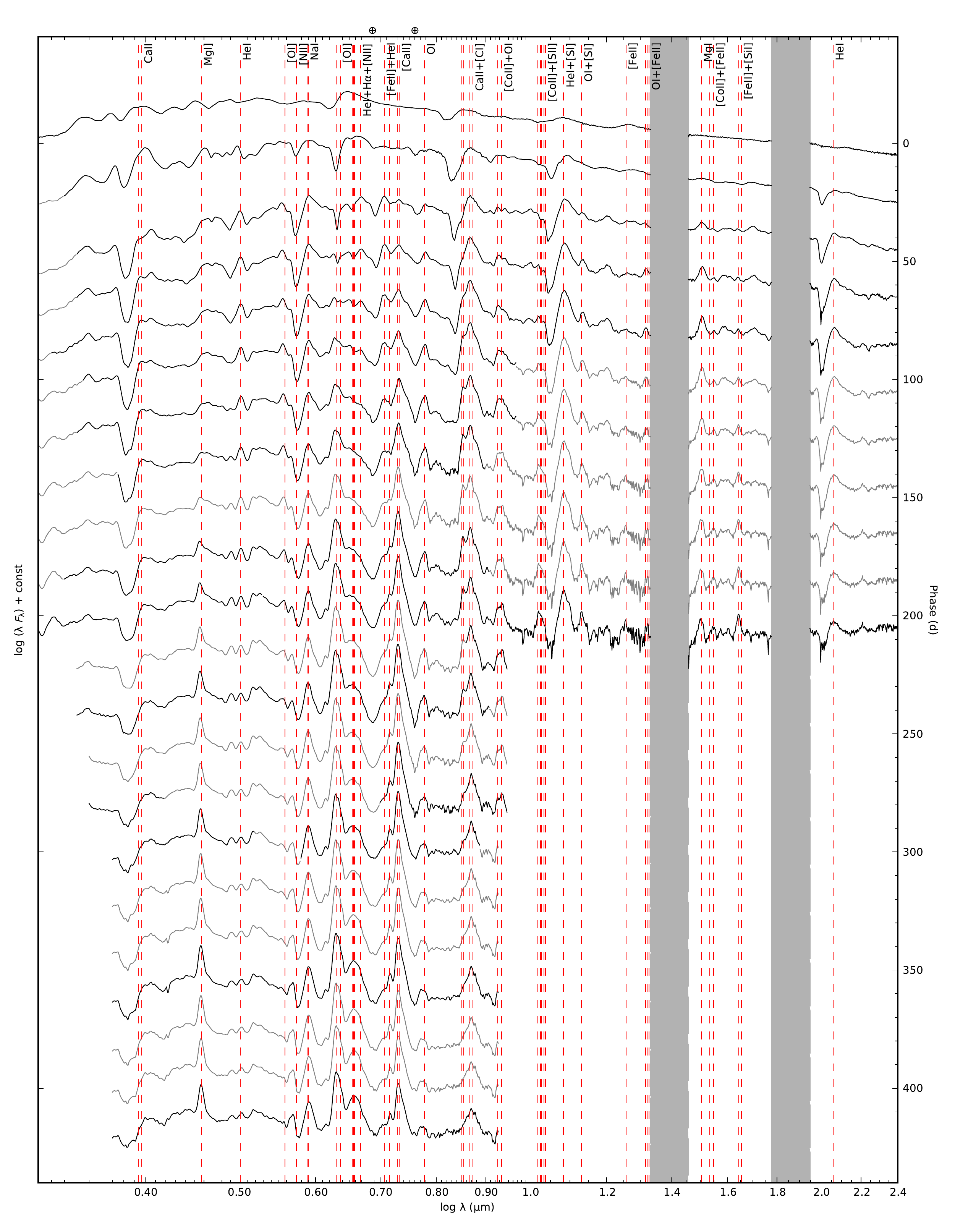}
\caption{The optical and NIR (interpolated) spectral evolution between days 5 and 425 for SN 2011dh with a 20-day sampling, where most of the lines identified in \citetalias{Jer14} have been marked with red dashed lines. Telluric absorption bands are marked with a $\oplus$ symbol in the optical and are shown as grey regions in the NIR.}
\label{f_spec_evo_opt_NIR}
\end{figure*}

All reduced and calibrated spectra will be made available for download from the Weizmann Interactive Supernova data REPository\footnote{http://www.weizmann.ac.il/astrophysics/wiserep/} (WISeREP) \citep{Yar12}.  Figure~\ref{f_spec_evo_opt_NIR_trad} shows the sequence of observed spectra, where those obtained on the same night using the same instrument have been combined. In addition to this, the NOT spectra obtained at days 289, 292, and 293 and the INT spectra obtained at days 228, 229, and 231 have been combined to increase the signal-to-noise ratio (SNR). For clarity some figures in this and the following sections are based on time-interpolations of the spectral sequence \citepalias{Erg14a}, and to further visualize the evolution the spectra have been aligned to a time-axis at the right border of the panels. Interpolated spectra are used in the calculations of S-corrections and the bolometric lightcurve (Sect.~\ref{s_bol_evo}). Figure~\ref{f_spec_evo_opt_NIR} shows the interpolated optical and NIR spectral evolution of SN 2011dh between days 5 and 425 with a 20-day sampling. All spectra in this and subsequent figures have been corrected for red-shift and interstellar extinction.

\subsection{Line evolution}
\label{s_spec_lines}

\begin{figure*}[tb]
\includegraphics[width=1.0\textwidth,angle=0]{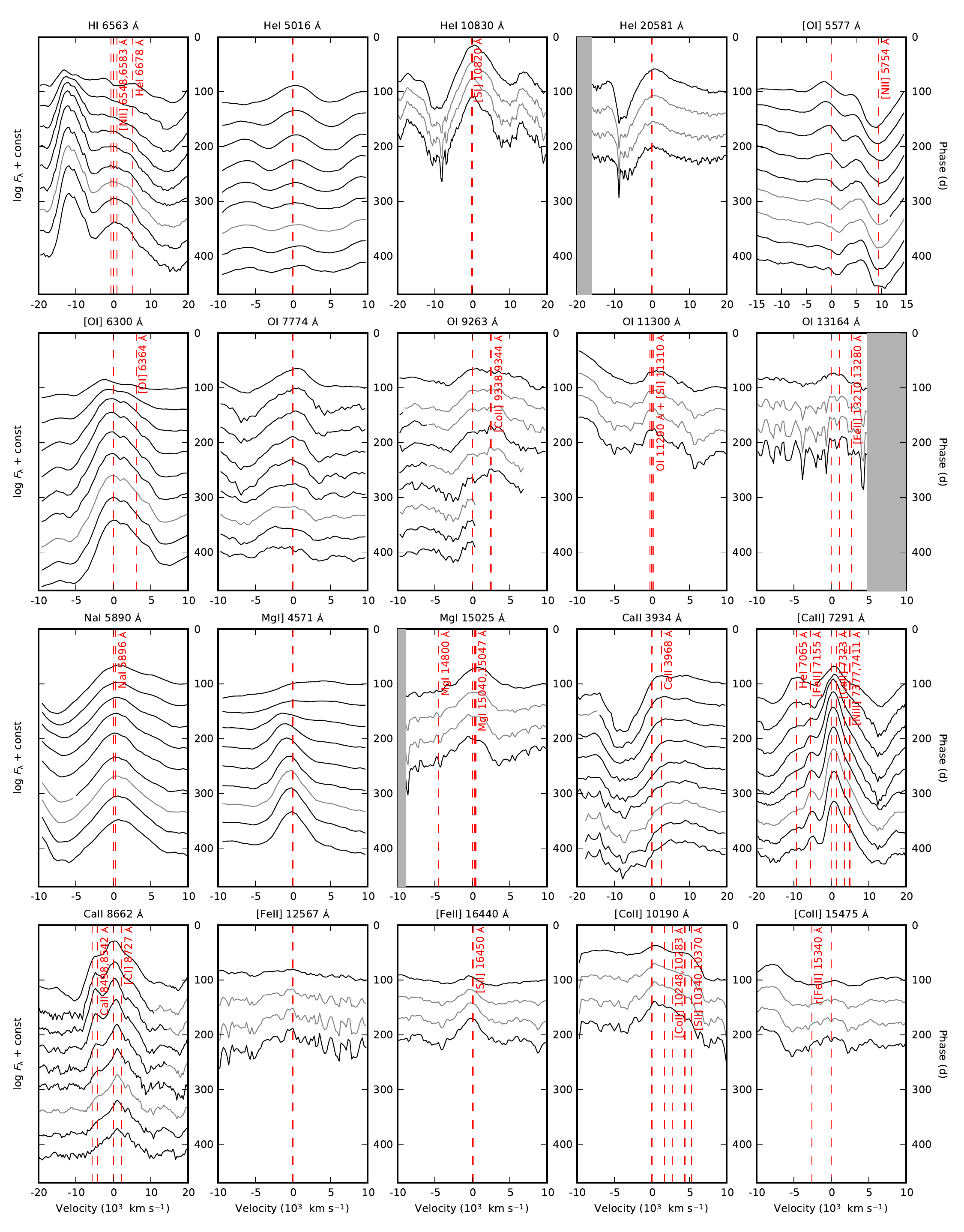}
\caption{The (interpolated) spectral evolution after day 100 for most of the lines identified in \citetalias{Jer14}. Multiple or blended lines are marked with red dashed lines and telluric absorption bands in the NIR are shown as grey regions.}
\label{f_spec_evo_lines}
\end{figure*}

Figure~\ref{f_spec_evo_lines} shows the (interpolated) evolution for most of the lines identified in \citetalias{Jer14}, and these are also marked in Fig.~\ref{f_spec_evo_opt_NIR}. Below we briefly discuss the evolution of the most important lines, and for a detailed discussion we refer to \citetalias{Jer14}. As mentioned, the optical coverage ends at day $\sim$400 and the NIR coverage at day $\sim$200. 

There is an emerging emission feature near the rest wavelength of H$\alpha$, increasing in strength towards day 450, but this feature is too narrow ($\sim$5500 km~s$^{-1}$ as fitted with the line profile model described in Appendix~\ref{a_line_measurements}) to be consistent with the $\gtrsim$11000 km~s$^{-1}$ observed for the hydrogen envelope \citepalias{Erg14a}. As discussed in \citetalias{Jer14} this feature is rather due to [\ion{N}{ii}] 6548,6583 \AA~emission originating from the helium envelope. There is also a dip in the [\ion{O}{i}] 6300,6364 \AA~line profile after day $\sim$150 (Fig.~\ref{f_line_fits}) that corresponds well to the early-time H$\alpha$ absorption minimum at $\sim$11000 km~s$^{-1}$ \citepalias{Erg14a}. However, as discussed in Sect.~\ref{s_spec_small_scale}, this feature repeats in a number of other lines, and is rather due to clumping/asymmetries in the ejecta.

The prominent \ion{He}{i} 10830 \AA~and \ion{He}{i} 20581 \AA~lines show P-Cygni like profiles \citep[][]{Fri12}, extending in absorption to at least $\sim$10000 km~s$^{-1}$, and are still strong when the NIR coverage ends, whereas the \ion{He}{i} 5016 \AA~line shows a P-Cygni like profile, extending in absorption to $\sim$5000 km~s$^{-1}$, and fades away at day $\sim$300. The \ion{He}{i} 5876 \AA~line is likely scattered away in the \ion{Na}{i} 5890,5896 \AA~line \citepalias{Jer14}, whereas the \ion{He}{i} 6678 \AA~and \ion{He}{i} 7065 \AA~lines are present at day $\sim$100, but then quickly disappear.

The [\ion{O}{i}] 6300,6364 \AA~line appears at day $\sim$100, and grows to become the most prominent line at day $\sim$400. The \ion{O}{i} 7774 \AA~line shows a P-Cygni like profile until day $\sim$300, extending in absorption to $\sim$8000 km~s$^{-1}$, whereas the [\ion{O}{i}] 5577 \AA~line fades away at day $\sim$300. The \ion{O}{i} 9263 \AA~line \citepalias[blended with the \text{[}\ion{Co}{ii}\text{]} 9338,9344 \AA~line;][]{Jer14} seems to be present at least until day $\sim$400, whereas the \ion{O}{i} 11290,11300 \AA~line is present until the NIR coverage ends. The \ion{Mg}{i}] 4571 \AA~line appears at day $\sim$150 and increase in strength towards day $\sim$400, whereas the \ion{Mg}{i} 15040 \AA~line is present until the NIR coverage ends.

The \ion{Na}{i} 5890,5896 line is present with a P-Cygni like profile until the optical coverage ends, extending in absorption to $\sim$10000 km~s$^{-1}$. The \ion{Ca}{ii} 3934,3968 \AA~line shows a strong absorption profile, extending to $\sim$15000 km~s$^{-1}$, and as the \ion{Ca}{ii} 8498,8542,8662 \AA~line \citepalias[blended with the \text{[}\ion{C}{i}\text{]} 8727 \AA~line;][]{Jer14} it is present until the optical coverage ends. The latter shows an absorption component at day $\sim$100, which quickly disappears, and the \ion{Ca}{ii} 8498,8542 \AA~contribution fades away towards day $\sim$400. The [\ion{Ca}{ii}] 7291,7323 \AA~line appears at day $\sim$100, soon becomes the most prominent line, and has a growing red wing extending to $\sim$10000 km~s$^{-1}$, which could be caused by [\ion{Ni}{ii}] 7378,7411 \AA~emission \citepalias{Jer14}.

The [\ion{Fe}{ii}] 7155 \AA~line appears at day $\sim$200 in the blue wing of the [\ion{Ca}{ii}] 7291,7323 \AA~line and is present until the optical coverage ends. A distinct feature near 16450 \AA, that could be either [\ion{Fe}{ii}] 16440 \AA~or [\ion{Si}{i}] 16450 \AA~is present until the NIR coverage ends. The [\ion{Co}{ii}] 9338,9344 \AA~line \citepalias[blended with the \ion{O}{i} 9263 \AA~line;][]{Jer14} seems to be present at least until day $\sim$300, whereas the [\ion{Co}{ii}] 10190,10248,10283 \AA~and [\ion{Co}{ii}] 15475 \AA~lines are present until the NIR coverage ends.

\subsection{CO emission}
\label{s_spec_CO}

\begin{figure}[tb!]
\includegraphics[width=0.48\textwidth,angle=0]{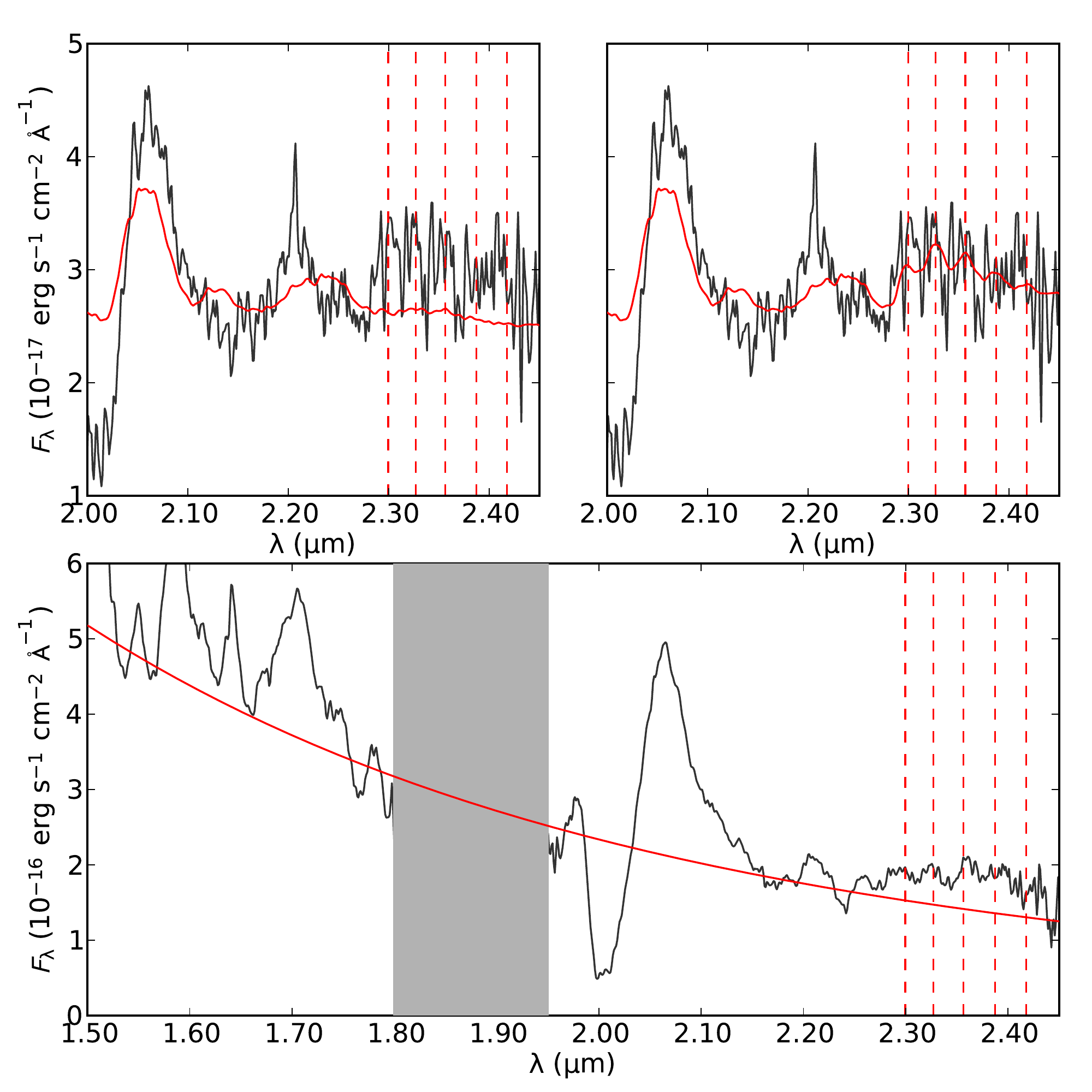}
\caption{Upper left panel: Observed (black solid line) and preferred steady-state NLTE model (red solid line) $K$-band spectra at days 206 and 200, respectively. Upper right panel: The same, but with the CO LTE model emission added to the steady-state NLTE model spectra. The CO overtone band heads have been marked by red dashed lines and the steady-state NLTE model spectra flux calibrated with the observed $K$-band flux at day 206. Lower panel: Observed (black solid line) $H$- and $K$-band spectrum at day 89 as compared to a blackbody fit to the continuum (red solid line).}
\label{f_spec_CO_flux}
\end{figure}

The upper left panel of Fig.~\ref{f_spec_CO_flux} shows the observed $K$-band spectrum at day 206 compared with the preferred steady-state NLTE model spectrum at day 200, flux calibrated with the observed $K$-band flux at day 206. The model (12F; see Sect.~\ref{s_modelling_optimal_model} and Appendix~\ref{a_nlte_modelling}) includes dust but not CO emission. The CO first overtone ($\Delta v$=2) emits between 2.25 and 2.45 $\mu$m and we have marked the location of the band heads, which are not sharp owing to the Doppler broadening of the individual transitions contributing. There is a strong excess in this region as compared to the model spectrum, the integrated flux being 6.2$\times$10$^{-15}$ erg s$^{-1}$ cm$^{-2}$. The contribution from this flux to the $K$-band flux is negligible though, because of the weak overlap with this band. We have used a simple LTE CO model described in \citet{Spy88} to provide a fit to the emission. The ratio of the band head emission is a diagnostic of the temperature of the CO, and the shape of the band heads a diagnostic of the expansion velocity of the CO gas. In the upper right panel of Fig.~\ref{f_spec_CO_flux} we show the combined CO model and preferred steady-state NLTE model spectrum. The shape of the observed emission and the agreement with the simple CO model provides strong support for the identification of this feature with emission from the first overtone of CO. The temperature in the CO model is 2300 K and the emission region extends to 1500 km~s$^{-1}$. The lower panel of Fig.~\ref{f_spec_CO_flux} shows the observed $H$- and $K$-band spectrum at day 89. As this spectrum was obtained before day 100, where the steady-state condition needed for the NLTE modelling is not satisfied, we instead use a blackbody fit to estimate the continuum. Although not as convincing as at day 206, there is a clear excess in the region where we expect CO first-overtone emission, the integrated flux being 7.5$\times$10$^{-14}$ erg s$^{-1}$ cm$^{-2}$. The contribution from this flux to the $K$-band flux is again negligible. In the absence of a model for the underlying emission we have not tried to fit the excess emission with the LTE CO model, but the structure of the feature shows a reasonable agreement with the expected positions of the band heads.

\subsection{Line emitting regions}
\label{s_spec_line_regions}

\begin{table*}[tbp!]
\caption{Radii of the line emitting regions as measured from line profile fits. The [\ion{O}{i}] 6300 \AA~line was decomposed with the method described in Appendix~\ref{a_line_measurements}. The continuum optical depths for the [\ion{O}{i}] 6300 \AA~and \ion{Mg}{i}] 4571 \AA~lines, and the [\ion{Ca}{ii}] 7293/7323 \AA~and [\ion{Co}{ii}] 10190/10248, 10190/10283 \AA~line ratios were also fitted, and are given in parentheses.}
\begin{center}
\scalebox{0.95}{
\begin{tabular}{l l l l l l l}
\hline\hline \\ [-1.5ex]
Phase & [\ion{O}{i}] 6300 \AA & \ion{Mg}{i}] 4571 \AA & \ion{Mg}{I} 15040 \AA & [\ion{Ca}{ii}] 7293,7323 \AA & [\ion{Fe}{ii}] 7155 \AA & [\ion{Co}{ii}] 10190,10248,10283 \AA \\ [0.5ex]
(d) & (10$^{3}$ km~s$^{-1}$) & (10$^{3}$ km~s$^{-1}$) & (10$^{3}$ km~s$^{-1}$) & (10$^{3}$ km~s$^{-1}$) & (10$^{3}$ km~s$^{-1}$) & (10$^{3}$ km~s$^{-1}$) \\ [0.5ex]
\hline \\ [-1.5ex]
202 & 3.3 (0.4) & 3.5 (1.1) & ...  & 2.5 (1.1) & ...  & ...  \\
206 & ...  & ...  & 3.3 & ...  & ...  & 1.8 (0.6,0.4) \\
300 & 3.0 (0.2) & 2.8 (0.7) & ...  & 2.2 (1.1) & 1.7 & ...  \\
415 & 2.9 (0.0) & 2.7 (0.2) & ...  & 2.3 (0.6) & 1.7 & ...  \\ [0.5ex]
\hline
\end{tabular}}
\end{center}
\label{t_line_emitting_regions}
\end{table*}

\begin{figure}[tbp!]
\includegraphics[width=0.48\textwidth,angle=0]{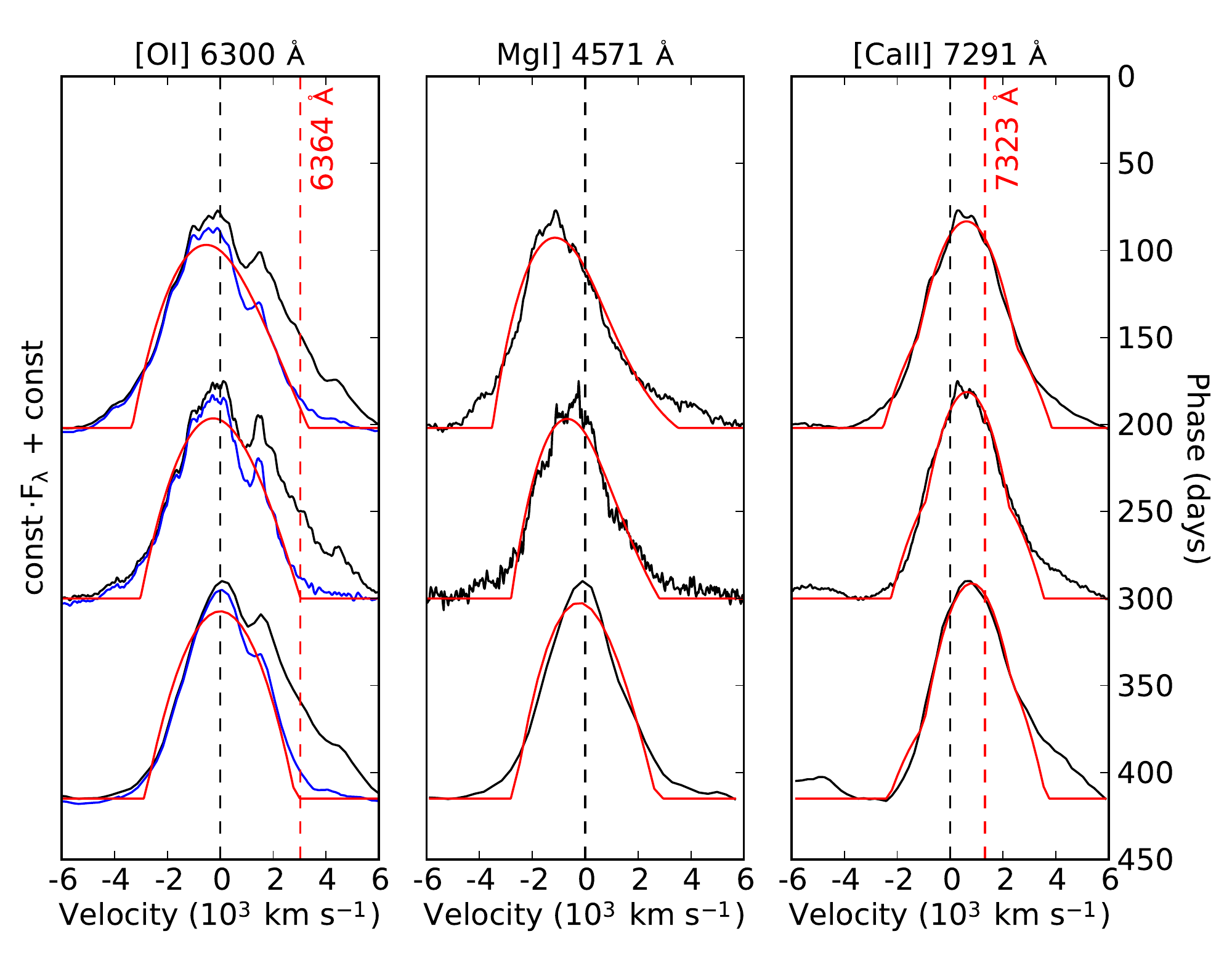}
\caption{The [\ion{O}{i}] 6300,6364 \AA~(left panel; black), decomposed [\ion{O}{i}] 6300 \AA~(left panel; blue), \ion{Mg}{i}] 4571 \AA~(middle panel; black), and [\ion{Ca}{ii}] 7291,7323 \AA~(right panel; black) lines at selected epochs, compared to line profile fits (red).}
\label{f_line_fits}
\end{figure}

\begin{figure}[tbp!]
\includegraphics[width=0.48\textwidth,angle=0]{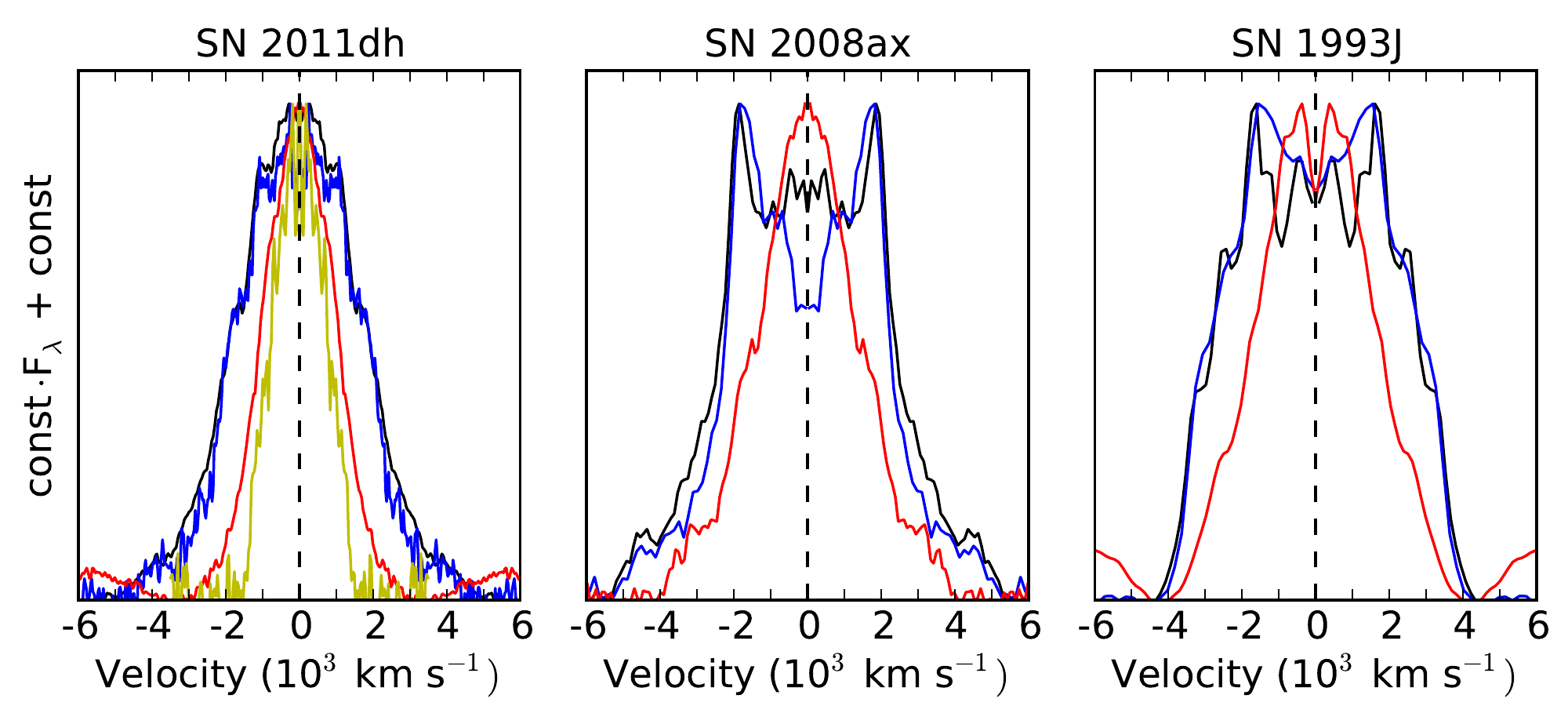}
\caption{Continuum subtracted mirrored blue-side profiles for the [\ion{O}{i}] 6300 \AA~(black), \ion{Mg}{i}] 4571 \AA~(blue), [\ion{Ca}{ii}] 7291 \AA~(red), and [\ion{Fe}{i}] 7155 \AA~(yellow) lines for SNe 2011dh, 2008ax, and 1993J at days 300, 307, and 283, respectively. The [\ion{Fe}{i}] 7155 \AA~line is only shown for SN 2011dh.}
\label{f_spec_line_blue_side_comp}
\end{figure}

The sizes of the line emitting regions are related to the distribution of the material from the different nuclear burning zones, determined by hydrodynamical instabilities in the explosion. In \citetalias{Jer14} we find that the \ion{Mg}{i} lines arise from the O/Ne/Mg zone, and the \ion{O}{i} lines from the O/Ne/Mg zone and, depending on the amount of molecular (CO and SiO) cooling, the O/C and O/Si/S zones. We also find that the [\ion{Ca}{ii}] 7291,7323 \AA~line arises mainly from the Si/S zone, and the \ion{Fe}{ii} and \ion{Co}{ii} lines from the Fe/Co/He zone. To measure the sizes of the line emitting regions, we fit a simple model for optically thin line emission to the observed line profiles. The model has a constant emissivity, and includes an absorptive continuum opacity to mimic blue-shifts caused by obscuration. The details are given in Appendix \ref{a_line_measurements}, where we also describe our method to decompose the [\ion{O}{i}] 6300 \AA~line. The fits have been applied to lines for which the scattering contribution appears to be small (no absorption component observed), the SNR is good, and contamination from other lines is likely to be small (based on the results in \citetalias{Jer14}). The results are given in Table~\ref{t_line_emitting_regions} and a selection of fits are shown in Fig.~\ref{f_line_fits}. The estimated radii of the \ion{O}{i} and \ion{Mg}{i} line emitting regions of 2900$-$3300 and 2700$-$3500 km~s$^{-1}$, respectively, are similar, whereas the estimated radii of the [\ion{Ca}{ii}] 7291,7323 \AA~and the \ion{Fe}{ii} and \ion{Co}{ii} line emitting regions of 2200$-$2500 and 1700$-$1800 km~s$^{-1}$, respectively, are progressively smaller. This progression is consistent with the (unmixed) onion-like structure of the nuclear burning zones, and suggests incomplete mixing of the oxygen, Si/S, and Fe/Co/He material. This is in agreement with results from 2-D hydrodynamical modelling of SN 1993J by \citet{Iwa97}, which shows significant differences in the distribution of the oxygen and Fe/Co/He material (but see \citealp{Hac91} for somewhat different results).

We have also compared the radii of the line emitting regions to those of SNe 1993J and 2008ax. For SN 1993J the estimated radii are 4000$-$4100 and 3800$-$3900 km~s$^{-1}$ for the [\ion{O}{i}] 6300 \AA~and \ion{Mg}{i}] 4571 \AA~lines, respectively, and 3200$-$3400 km~s$^{-1}$ for the [\ion{Ca}{ii}] 7291,7323 \AA~line. For SN 2008ax the estimated radii are 3900$-$4000 and 3500$-$3600 km~s$^{-1}$ for the [\ion{O}{i}] 6300 \AA~and \ion{Mg}{i}] 4571 \AA~lines, respectively, and 3000$-$3200 km~s$^{-1}$ for the [\ion{Ca}{ii}] 7291,7323 \AA~line. These radii are larger than for SN 2011dh, and larger for SN 1993J than for SN 2008ax. The radii of the \ion{Mg}{i}] 4571 \AA~and the [\ion{O}{i}] 6300 \AA~line emitting regions are again similar, and the radius of the [\ion{Ca}{ii}] 7291,7323 \AA~line emitting region is again smaller, suggesting incomplete mixing of the oxygen and Si/S material also for SNe 1993J and 2008ax. Figure~\ref{f_spec_line_blue_side_comp} shows mirrored blue-side line profiles for the [\ion{O}{i}] 6300 \AA, \ion{Mg}{i}] 4571 \AA, [\ion{Ca}{ii}] 7291 \AA, and [\ion{Fe}{i}] 7155 \AA~lines for SNe 1993J, 2008ax, and 2011dh at day $\sim$300. The blue side is less affected by obscuration, and contamination from the [\ion{O}{i}] 6364 \AA~and [\ion{Ca}{ii}] 7323 \AA~lines to the [\ion{O}{i}] 6300 \AA~and [\ion{Ca}{ii}] 7291 \AA~lines is probably modest. This figure illustrates the different sizes of the line emitting regions discussed, and also shows a remarkable similarity of the blue-side [\ion{O}{i}] 6300 \AA~and \ion{Mg}{i}] 4571 \AA~line profiles for all SNe. For SN 2011dh this similarity persists even in small scale fluctuations (Sect.~\ref{s_spec_small_scale}), which suggests that these lines arise mainly from the O/Ne/Mg zone, and that the contributions from the O/C and O/Si/S zones to the [\ion{O}{i}] 6300 \AA~flux are modest.

\subsection{Line asymmetries}
\label{s_spec_line_asymmetries}

\begin{figure}[tb!]
\includegraphics[width=0.48\textwidth,angle=0]{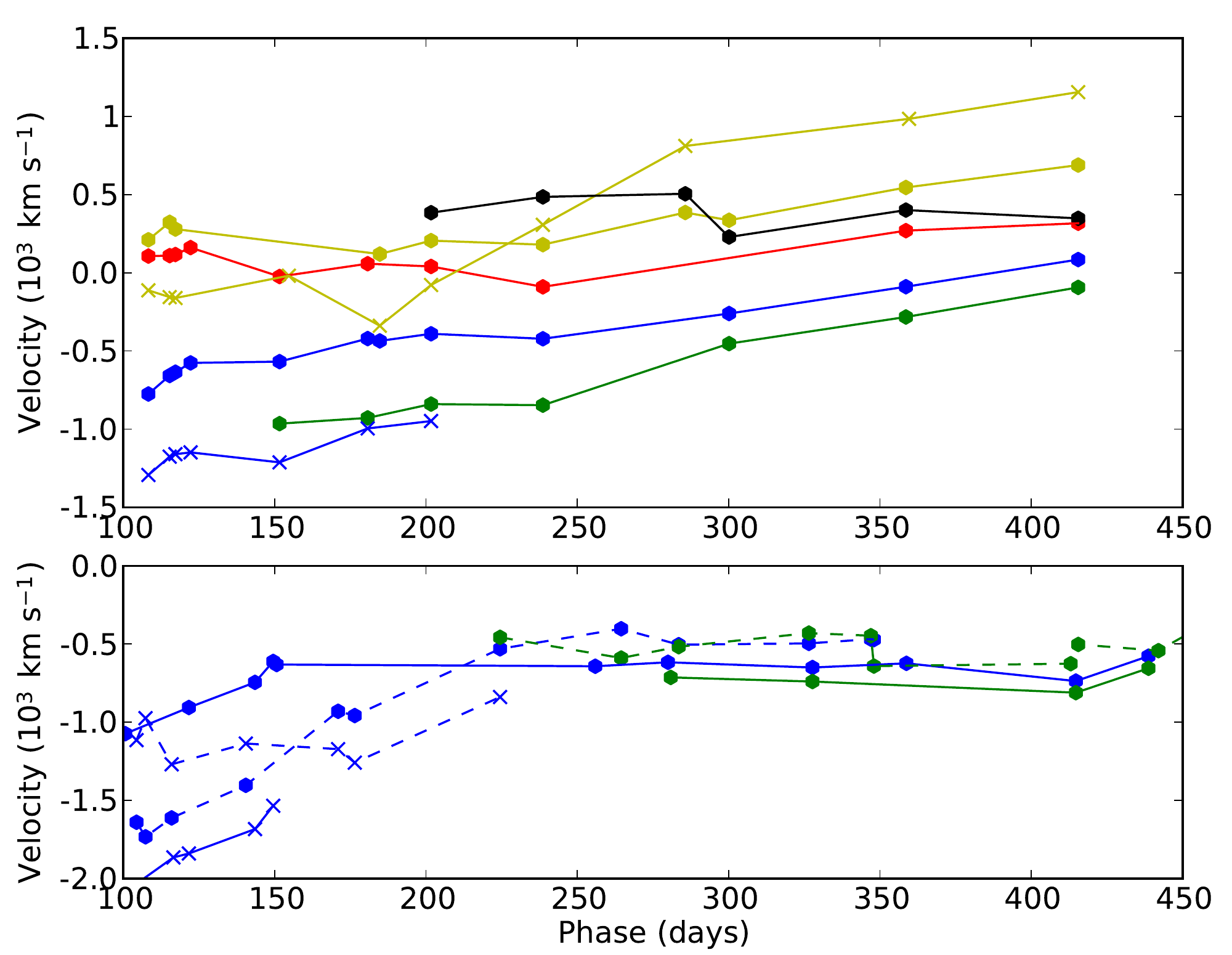}
\caption{Upper panel: Center of flux velocities for the [\ion{O}{i}] 6300,6364 \AA~(blue circles), [\ion{O}{i}] 5577 \AA~(blue crosses), \ion{Mg}{i}] 4571 \AA~(green circles), [\ion{Ca}{ii}] 7291,7323 \AA~(yellow circles), \ion{Ca}{ii} 8498,8542,8662 \AA~(yellow crosses), \ion{Na}{i} 5890,5896 \AA~(red circles), and [\ion{Fe}{ii}] 7155 \AA~(black circles) lines for SN 2011dh. Lower panel: Center of flux velocities for the [\ion{O}{i}] 6300,6364 \AA~(blue circles), [\ion{O}{i}] 5577 \AA~(blue crosses), and \ion{Mg}{i}] 4571 \AA~(green circles) lines for SNe 1993J (dashed lines) and 2008ax (solid lines).}
\label{f_line_vel_evo_comp}
\end{figure}

Line asymmetries may be caused by intrinsic asymmetries in the ejecta, radiative transfer effects, or blends with other lines. To measure these we calculate the first wavelength moment of the flux (center of flux; Appendix \ref{a_line_measurements}). The upper panel of Fig.~\ref{f_line_vel_evo_comp} shows the center of flux velocities for a selection of lines between days 108 and 415 for SN 2011dh. At early times the [\ion{O}{i}] 6300,6364 \AA, [\ion{O}{i}] 5577 \AA~and \ion{Mg}{i}] 4571 \AA~lines all show significant blue-shifts, which gradually disappear towards day 400. As shown in the lower panel of Fig.~\ref{f_line_vel_evo_comp}, such blue-shifts are also present, and even more pronounced for SNe 1993J and 2008ax, but in this case the blue-shifts saturate at $\sim$500 km~s$^{-1}$ after day 200. In \citetalias{Jer14} we provide a thorough discussion of these blue-shifts, and suggest the cause to be obscuration of receding-side emission by line-blocking in the core. We find no significant blue-shifts in the \ion{O}{i} 11300 \AA, \ion{O}{i} 13164 \AA, and \ion{Mg}{i} 15040 \AA~lines for SNe 2011dh and 2008ax in support of this hypothesis, as line-blocking is less effective in the NIR (\citetalias{Jer14}). After day $\sim$250 we instead see increasing red-shifts in the [\ion{Ca}{ii}] 7291,7323 \AA~and \ion{Ca}{ii} 8498,8542,8662 \AA~lines for SN 2011dh, likely produced by blends with the [\ion{Ni}{ii}] 7378,7411 \AA~and [\ion{C}{i}] 8727 \AA~lines, respectively \citepalias{Jer14}. Given our interpretation of the line-shifts, the systematic redward trend seen in most lines between days 250 and 400 for SN 2011dh is coincidental.

Obscuration of receding-side emission may also be caused by dust, and it is not obvious how to disentangle the effects of line-blocking and dust. In order to explain the behaviour of the lightcurves, our steady-state NLTE models have a homogeneous dust opacity in the core (Sect.~\ref{s_modelling_day_100_500}). The preferred model used in this work (12F) has $\tau_\mathrm{dust}$=0.44, which is found to best reproduce the observed lightcurves, whereas the original version presented in \citetalias{Jer14} (12C) has $\tau_\mathrm{dust}$=0.25. This lower value gives a less accurate reproduction of the lightcurves, but produces smaller blue-shifts, which seems to be in better agreement with observations. Using our line profile model, we find blue-shifts of the center of flux of 150 and 250 km~s$^{-1}$ for optical depths of 0.25 and 0.44, respectively, for a line-emitting region with a radius of 3000 km~s$^{-1}$ (Sect.~\ref{s_spec_line_regions}). At day 415, when the effect from line-blocking would be the weakest, the center of flux for the [\ion{O}{i}] 6300,6364 \AA~and \ion{Mg}{i}] 4571 \AA~lines, which arise solely from the core \citepalias{Jer14}, shows a red-shift of $\sim$100 km~s$^{-1}$ and a blue-shift of $\sim$100 km~s$^{-1}$, respectively. The absence of significant blue-shifts in these lines at day 415 suggests a small optical depth of the dust and indicate that, if the optical depth is as high as 0.44, the dust cannot be homogeneously distributed within the core. It also suggests that the relatively large blue-shifts seen in these lines at earlier epochs can be attributed to the line-blocking effect.

\subsection{Small scale fluctuations}
\label{s_spec_small_scale}

Small scale fluctuations in the line profiles may provide evidence for a clumpy ejecta, as previously demonstrated for SNe 1987A \citep{Sta91,Chu94} and 1993J \citep{Spy94,Mat00}. In a simplified way we can represent the material of some nuclear burning zone by a number of randomly distributed clumps, having a typical size and occupying some fraction of the ejecta volume (filling factor). The small scale fluctuations then arise from statistical fluctuations in the distribution of the clumps, the root mean square (RMS) of the fluctuations increasing with decreasing number of clumps or filling factor, or increasing size of the clumps. 

\begin{figure}[tb]
\includegraphics[width=0.48\textwidth,angle=0]{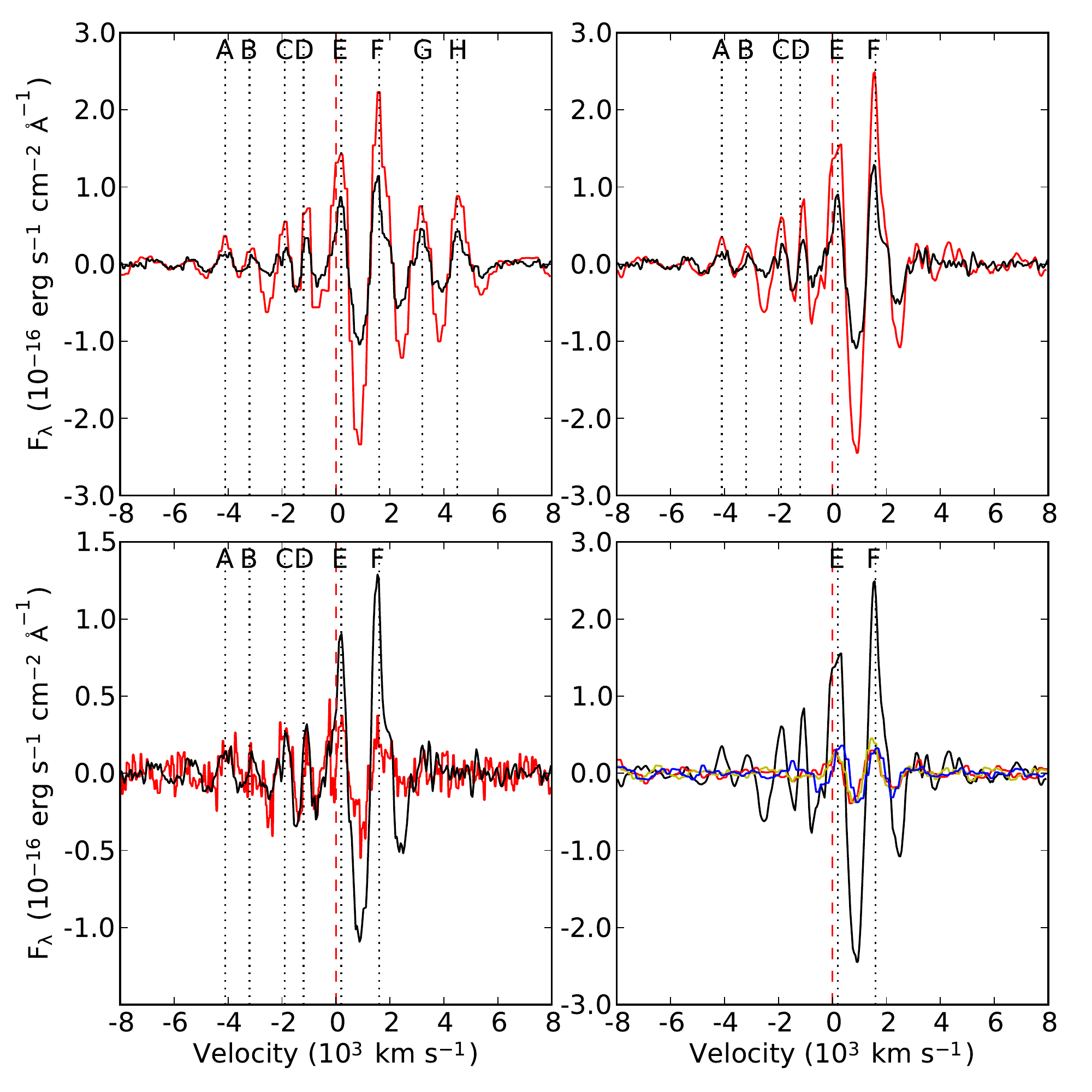}
\caption{Comparison of large scale subtracted line profiles. The upper left panel shows the [\ion{O}{i}] 6300,6364 \AA~line at days 202 (red) and 300 (black). The upper right panel shows the decomposed [\ion{O}{i}] 6300 \AA~line at days 202 (red) and 300 (black). The lower left panel shows the decomposed [\ion{O}{i}] 6300 \AA~line (black) and the \ion{Mg}{i}] 4571 \AA~line (red) at day 300. The lower right panel shows the decomposed [\ion{O}{i}] 6300 \AA~line (black) and the [\ion{O}{i}] 5577 \AA~(red), \ion{O}{i} 7774 \AA~(green), and \ion{Na}{i} 5890,5896 \AA~(blue) lines at day 202.}
\label{f_spec_line_profiles_comp}
\end{figure}

Figure~\ref{f_spec_line_profiles_comp} shows small scale fluctuations in the [\ion{O}{i}] 6300,6364 \AA, [\ion{O}{i}] 5577 \AA, \ion{O}{i} 7774 \AA, \ion{Mg}{i}] 4571 \AA, and \ion{Na}{i} 5890,5896 \AA~lines at days 202 and 300. The resolution is $\sim$600 and $\sim$250 km~s$^{-1}$ in the spectra obtained at days 202 and 300, respectively, and the large scale (>1000 km~s$^{-1}$) structure has been subtracted using the box-averaged line profile as described in Appendix~\ref{a_line_measurements}. In the upper left panel we show a comparison of the [\ion{O}{i}] 6300 \AA~line at days 202 and 300, and there is not much evolution in the small scale structure during this period. We identify eight features marked A-H with a FWHM between 300 and 600 km~s$^{-1}$ present at both epochs. The G and H features interpreted as belonging to the [\ion{O}{i}] 6364 \AA~line, match very well the E and F features interpreted as belonging to the [\ion{O}{i}] 6300 \AA~line, so these are likely to be repetitions. Subtracting the [\ion{O}{i}] 6364 \AA~flux (Appendix~\ref{a_line_measurements}) and minimizing the RMS of the small scale fluctuations redwards 3000 km~s$^{-1}$, we find that a line ratio of 2.9 at days 202 and 300 gives a complete removal of features G and H. This ratio is close to the 3.1 expected for optically thin emission and is consistent with the results in \citetalias{Jer14}. 

In the upper right panel we show the decomposed [\ion{O}{i}] 6300 \AA~line at days 202 and 300, and in the lower left panel we show a comparison to the \ion{Mg}{i}] 4571 \AA~line at day 300. All features except B are clearly identified in both lines and the agreement is good. The features on the red side are weaker for the \ion{Mg}{i}] 4571 \AA~line, which is consistent with the larger red-side flux deficit for this line (Fig.~\ref{f_line_fits}), but the relative (normalized with the large scale flux) strength of all features are similar. As mentioned in Sect.~\ref{s_spec_line_regions}, the good agreement suggests that the [\ion{O}{i}] 6300 \AA~and \ion{Mg}{i}] 4571 \AA~lines arise mainly from the O/Ne/Mg zone, and that the contributions from the O/Si/S and O/C zones to the [\ion{O}{i}] 6300 \AA~flux are modest. 

In the lower right panel we show a comparison of the decomposed [\ion{O}{i}] 6300 \AA~line and the [\ion{O}{i}] 5577 \AA, \ion{O}{i} 7774 \AA, and \ion{Na}{i} 5890,5896 \AA~lines at day 202. The E and F features are clearly identified in all of these lines, whereas the other features are only seen in the [\ion{O}{i}] 6300 \AA~line. However, as the A, B, C and D features are weaker, they may just be too faint to be seen in the other lines. The small scale fluctuations in the [\ion{Ca}{ii}] 7291,7323 \AA~line (not shown) do not match very well with those in the [\ion{O}{i}] 6300 \AA~line, and the relative strength of the features is weaker. These results are consistent with the results in \citetalias{Jer14}, where the oxygen lines are found to arise from the oxygen-rich zones and the \ion{Na}{i} 5890,5896 \AA~line partly from the O/Ne/Mg zone, whereas the [\ion{Ca}{ii}] 7291,7323 \AA~line is found to arise mainly from the Si/S zone.

\citet{Shi13} presented an analysis of the [\ion{O}{i}] 6300,6364 \AA, \ion{O}{i} 7774 \AA, and \ion{Mg}{i}] 4571 \AA~lines at day 268, where by Gaussian decomposition they find two narrow features, likely corresponding to our E and F features, repeating in all these lines, in agreement with our analysis. \citet{Mat00} presented an analysis of small scale fluctuations in the spectra of SN 1993J, and found a good agreement between the fluctuations in the [\ion{O}{i}] 6300 \AA, [\ion{O}{i}] 5577 \AA, and \ion{O}{i} 7774 \AA~lines, in agreement with our results for SN 2011dh. Surprisingly, they did not find a good agreement between the fluctuations in the [\ion{O}{i}] 6300 \AA~and \ion{Mg}{i}] 4571 \AA~lines. One possible explanation is that the [\ion{O}{i}] 6300 \AA~line is dominated by flux from the O/Ne/Mg zone for SN 2011dh, but not for SN 1993J (see above). \citet{Fil89} presented an analysis of small scale fluctuations in the spectra of the Type Ib SN 1985F. Similar to our analysis they found repetitions of the identified features in the [\ion{O}{i}] 6300 \AA~and 6364 \AA~lines, and a line ratio close to 3 using the strongest feature.

\citet{Mat00} applied the statistical model by \citet{Chu94} to their spectra of SN 1993J, giving a filling factor of $\sim$0.06 for oxygen zone material, distributed within a sphere with 3800 km s$^{-1}$ radius. Using their estimated typical clump size of 300 km s$^{-1}$, this corresponds to $\sim$900 clumps. The model requires the radius of the sphere containing the clumps, the typical size of the clumps and the RMS of relative flux fluctuations in lines originating from the clumps. In the case of SN 2011dh, we adopt a radius of the sphere containing the bulk of the oxygen zone material of $\sim$3500 km s$^{-1}$, based on the estimates of the \ion{O}{i} and \ion{Mg}{i} line emitting regions in Sect.~\ref{s_spec_lines}. For SN 1987A a typical clump size of 120 km s$^{-1}$ was estimated from the power spectrum of the [\ion{O}{i}] 6300 \AA~line by \citet{Sta91}, using high-resolution spectroscopy, but it is not clear how this was done by \citet{Mat00}. As we do not have high-resolution spectroscopy for SN 2011dh, we can only estimate an upper limit on the typical clump size, taken to be 300 km s$^{-1}$, the smallest size of the features seen. The RMS of the relative flux fluctuations in the inner part \citep[$\pm{2000}$ km s$^{-1}$, see][]{Chu94} of the sphere is $\sim$0.13 and $\sim$0.09 at day 300 for the decomposed [\ion{O}{i}] 6300 \AA~line and the \ion{Mg}{i}] 4571 \AA~line, respectively. Taking the latter value as a lower limit and applying \citet[Eq.~11]{Chu94}, we find an upper limit on the filling factor of oxygen zone material (within the sphere) of $\sim$0.07, and a lower limit on the number of oxygen zone clumps of $\sim$900. These values are similar to the values estimated by \citet{Mat00} for the clumping of oxygen-rich material in SN 1993J.

\section{Lightcurve modelling}
\label{s_modelling}

Modelling of nebular-phase pseudo-bolometric or broad-band lightcurves is rare, and it is therefore of interest to examine this case using the high quality observations obtained for SN 2011dh. To accomplish this we use the nebular-phase spectral synthesis code described in \citet{Jer11,Jer12}, hereafter referred to as the steady-state NLTE code. In \citetalias{Jer14} we present spectra for a set of models, and here we present pseudo-bolometric and broad-band lightcurves for the same models. As this code assumes steady-state, the models begin at day 100, and because of the complexity of the physics, a restricted volume of parameter space is covered. Consistent modelling of the diffusion-phase \emph{and} nebular-phase lightcurves is desirable, and to facilitate this we make use of HYDE, a hydrodynamical code described in \citet[][hereafter \citetalias{Erg15}]{Erg15} and summarized in Appendix~\ref{a_hyde}. HYDE is capable of evolving the SN from the explosion into the nebular phase, but is limited to bolometric lightcurve modelling, so in practice it can only be used before day 100 when the bolometric correction (BC) is small. Because of its simplicity it is fast, and in \citetalias{Erg15} we construct a grid of SN models covering a large volume of parameter space. Feeding all these models into the steady-state NLTE code is not feasible, but by combining the advantages of each code we are able to stretch the limitations. First, to extend the coverage of the \citetalias{Jer14} models to early times, we calculate bolometric lightcurves after day 100 for these with HYDE. Secondly, to extend the coverage of the \citetalias{Erg15} model grid to late times, we determine the BC with the steady-state NLTE code, and use it to fit the optical-to-MIR pseudo-bolometric lightcurve before day 400 to the model grid.

\subsection{Steady-state NLTE modelling of the pseudo-bolometric and broad-band lightcurves between days 100 and 500}
\label{s_modelling_day_100_500}

In \citetalias{Jer14} we present a set of spectral models for Type IIb SNe, with the specific aim of modelling SNe 1993J, 2008ax, and in particular SN 2011dh. Here we present synthetic pseudo-bolometric and broad-band lightcurves between days 100 and 500 for these models, constructed using the standard system filter response functions specified in \citetalias{Erg14a}. The preferred model for SN 2011dh, presented in \citetalias{Jer14} and refined in this paper with respect to dust, has been chosen to give the best agreement with both nebular spectra and the pseudo-bolometric and broad-band lightcurves. In \citetalias{Jer14} we discuss the constraints on the model parameters provided by the nebular spectra, and here we discuss the constraints provided by the lightcurves. In this work we restrict the set of models \citepalias[][Table 3]{Jer14} to the model families differing in a single parameter \citepalias[][Table 4]{Jer14}, and two additional models (12E and 12F) described in Appendix \ref{a_nlte_modelling}. These models differ from model 12C only in the absence (12E) and the properties (12F) of the dust. All models have been evaluated at days 100, 150, 200, 300, 400, and 500.

\subsubsection{Model parameters}
\label{s_modelling_parameters}

The models vary in the following parameters: initial mass (12, 13, or 17 M$_\odot$), macroscopic mixing (medium or strong), density contrast (low or high), positron trapping (local or non-local), molecular cooling (complete or none), and dust ($\tau_{\mathrm{dust}}$=0, 0.25, or 0.44). The first four parameters are described in detail in \citetalias{Jer14}. In short, the initial mass parameter determines the overall properties of the ejecta model (e.g.~mass and composition), whereas the macroscopic mixing and contrast factor parameters mimic the mixing of the nuclear burning zones and subsequent expansion of the Fe/Co/He clumps arising in multi-dimensional modelling. The positron trapping parameter mimics the effect of a magnetic field (on the positron trapping). We note that the strong and medium mixing models differ only in that the former has 50 per cent of the Fe/Co/He material mixed into the helium envelope, and that the mass of the $^{56}$Ni is fixed at 0.075 M$_\odot$ (Sect.~\ref{s_modelling_day_0_300}) for all models.

The modelling does not include a treatment of the formation of molecules and dust in the ejecta, but the effects of these are included in a simplified way as described in Appendix~\ref{a_nlte_modelling}. The treatment of molecules only supports two extremes, no emission or complete dominance of the cooling in the O/C and O/Si/S zones by CO and SiO emission. This is motivated by the fact that molecules are efficient coolers and tend to dominate the cooling once formed. Dust absorption is represented as a grey absorptive opacity in the core, and the emission as blackbody emission. Absorption is turned on at day 200, and the optical depth chosen to reproduce the optical pseudo-bolometric lightcurve, whereas the temperature of the blackbody emission is determined from fits to the $K$ and $S_1$ bands.

\subsubsection{The preferred model.}
\label{s_modelling_optimal_model}

\begin{figure}[tb]
\includegraphics[width=0.48\textwidth,angle=0]{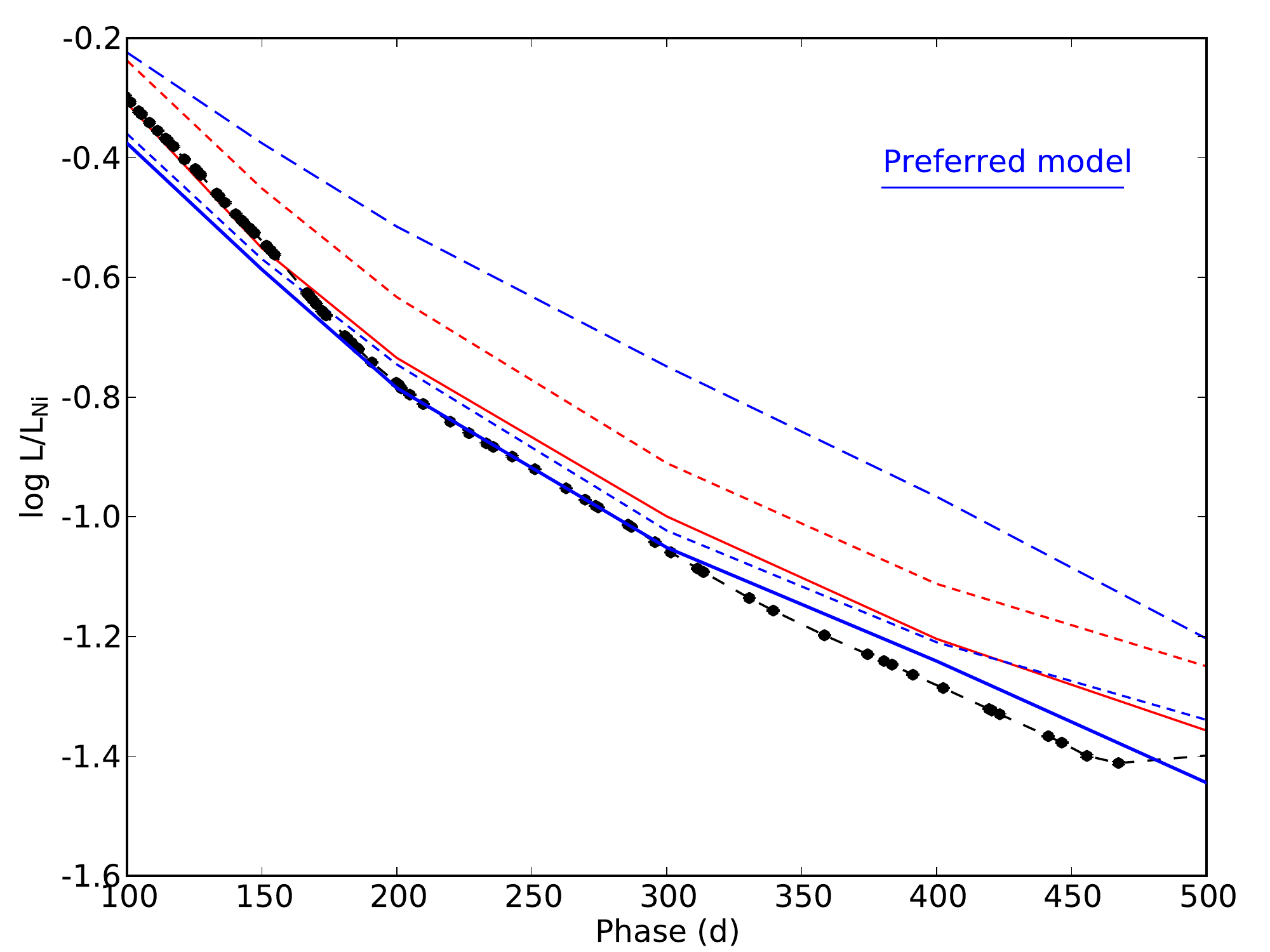}
\caption{The observed (black circles) and \citetalias{Jer14} model optical-to-MIR pseudo-bolometric lightcurves between days 100 and 500 normalized to the radioactive decay chain luminosity of 0.075 M$_{\odot}$ of \element[ ][56]{Ni}. Selected representatives (12A,12F,13A,13C,17A) of the model families differing in initial mass (12 M$_\odot$; solid, 13 M$_\odot$; short-dashed, 17 M$_\odot$; long-dashed) and macroscopic mixing (medium; red, strong; blue) are shown. Following this coding, the preferred model is shown as a blue solid line.}
\label{f_bol_opt_mir_mcomp}
\end{figure}

\begin{figure}[tb]
\includegraphics[width=0.48\textwidth,angle=0]{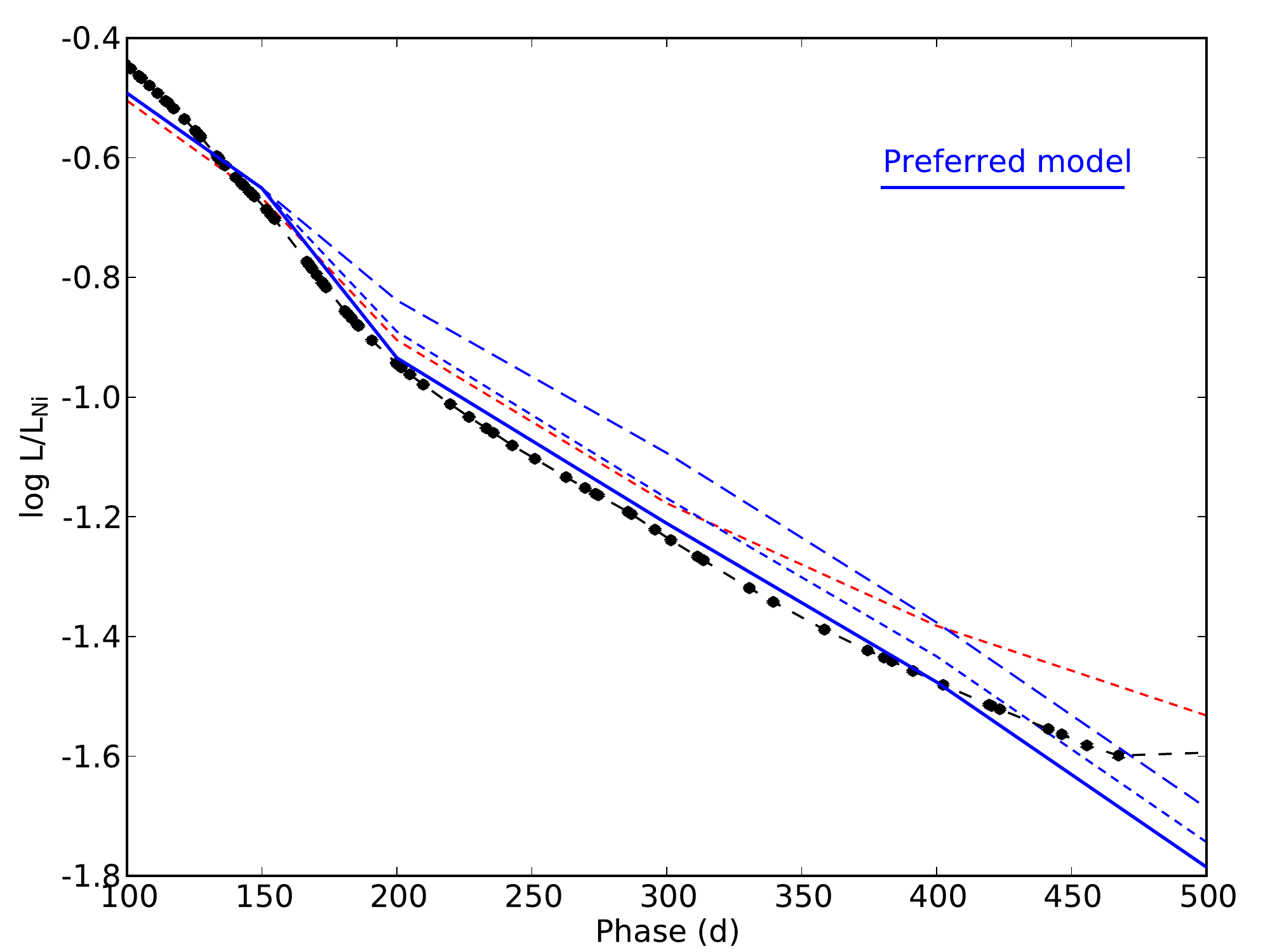}
\caption{The observed (black circles) and \citetalias{Jer14} model optical pseudo-bolometric lightcurves between days 100 and 500 normalized to the radioactive decay chain luminosity of 0.075 M$_{\odot}$ of \element[ ][56]{Ni}. The preferred model (12F; blue solid line), model 12C (blue short-dashed line) and model 12E (blue long-dashed line), which differ only in the optical depth (12F; 0.44, 12C; 0.25, 12E; 0) and temperature of the dust, are shown together with model 12B (red short-dashed line), which differs from model 12C only in the positron trapping (12C; local, 12B; non-local).}
\label{f_bol_opt_mcomp}
\end{figure}

\begin{figure*}[tb]
\includegraphics[width=1.0\textwidth,angle=0]{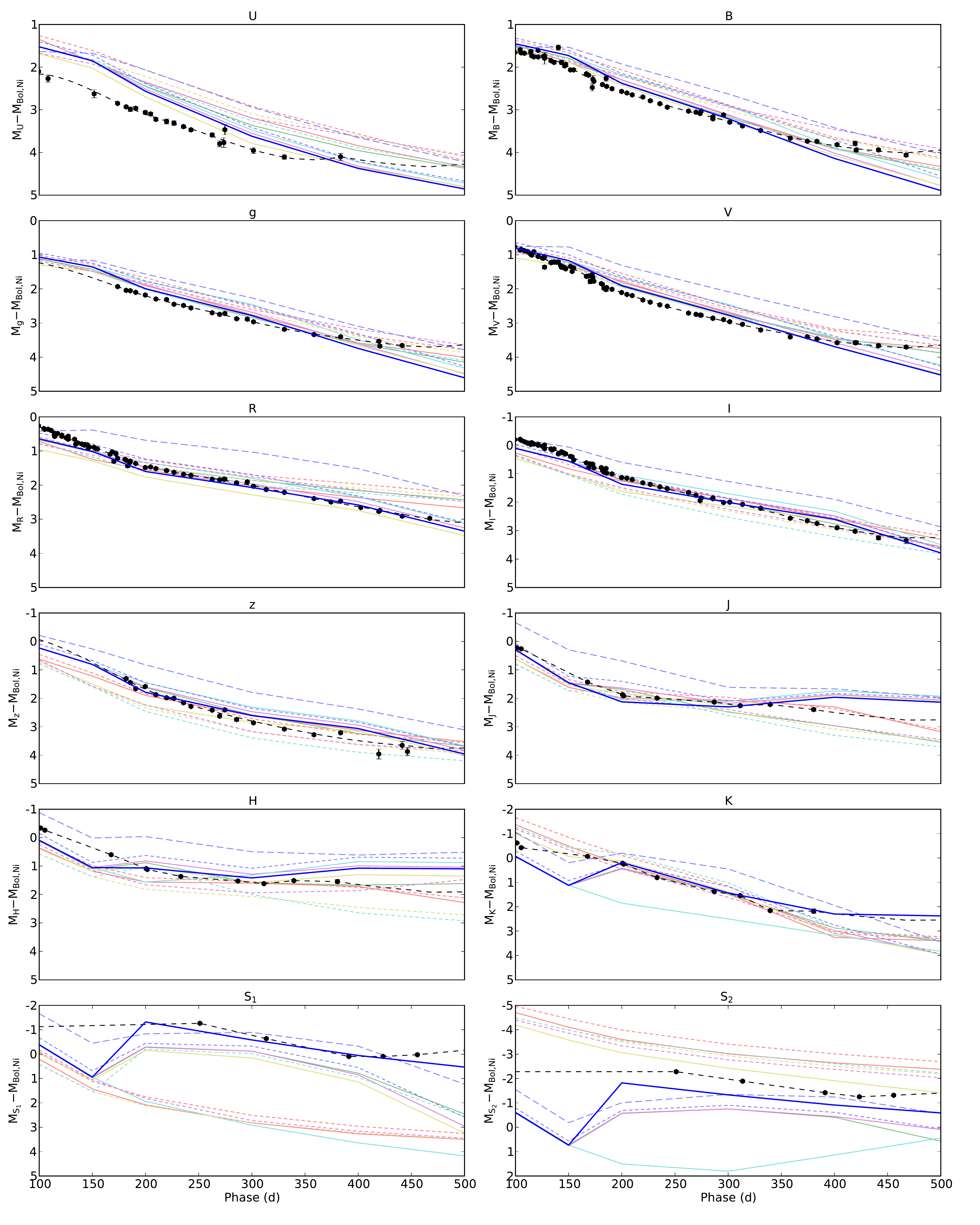}
\caption{The observed (black circles) and \citetalias{Jer14} model optical, NIR, and MIR (absolute) magnitudes between days 100 and 500 normalized to the radioactive decay chain luminosity of 0.075 M$_{\odot}$ of \element[ ][56]{Ni}. The preferred model (12F) is shown as a blue solid line. The other models are shown in shaded colour as follows; 12A (red solid line), 12B (green solid line), 12C (magenta solid line), 12D (yellow solid line), 12E (cyan solid line), 13A (red short-dashed line), 13C (yellow short-dashed line), 13D (cyan short-dashed line), 13E (magenta short-dashed line), 13G (blue short-dashed line), 17A (blue long-dashed line).}
\label{f_opt_nir_mir_mcomp}
\end{figure*}

The preferred model presented in \citetalias{Jer14} (12C) has an initial mass of 12 M$_\odot$, strong macroscopic mixing, local positron trapping, no molecular cooling, dust with $\tau_{\mathrm{dust}}$=0.25, and a high density contrast. However, in this work we use a refined version of the preferred model (12F), which differs only in the optical depth ($\tau_{\mathrm{dust}}$=0.44) and the temperature of the dust (Appendix \ref{a_nlte_modelling}). It should be noted that the good spectroscopic agreement found for model 12C in \citetalias{Jer14} does not necessarily apply to model 12F, but the differences are likely to be small (Appendix \ref{a_nlte_modelling}). Figures~\ref{f_bol_opt_mir_mcomp} and \ref{f_bol_opt_mcomp} show the observed and \citetalias{Jer14} model, optical and optical-to-MIR pseudo-bolometric lightcurves between days 100 and 500. In these figures, and in a number of figures that follows we have normalized the luminosity with the decay chain luminosity of the mass of \element[ ][56]{Ni}, which (to first order) removes the dependence on this quantity. The preferred model shows a good agreement with observations, the differences being $\lesssim$10 and $\lesssim$15 per cent for the optical and optical-to-MIR pseudo-bolometric lightcurves, respectively. However, after day $\sim$400 the observed optical pseudo-bolometric lightcurve flattens and starts to diverge from that of the preferred model. This flattening continues after day 500 and is discussed in Sect.~\ref{s_500_days_lightcurves}. Figure~\ref{f_opt_nir_mir_mcomp} shows the observed and \citetalias{Jer14} model broad-band lightcurves between days 100 and 500. The preferred model shows an overall good agreement with observations, the differences being mostly $\lesssim$0.3 mag, but there are some notable exceptions, such as the $U$ and $S_2$ bands. The discrepancy seen in the optical pseudo-bolometric lightcurve after day $\sim$400 is also reflected in the $B$, $g$, and $V$ bands. The discrepancy in the $S_2$ band is discussed further in Sect.~\ref{s_modelling_molecules_dust}.

\subsubsection{Calculation of the bolometric lightcurves before day 100 with HYDE}
\label{s_modelling_day_3_100_j14}

\begin{figure}[tb]
\includegraphics[width=0.48\textwidth,angle=0]{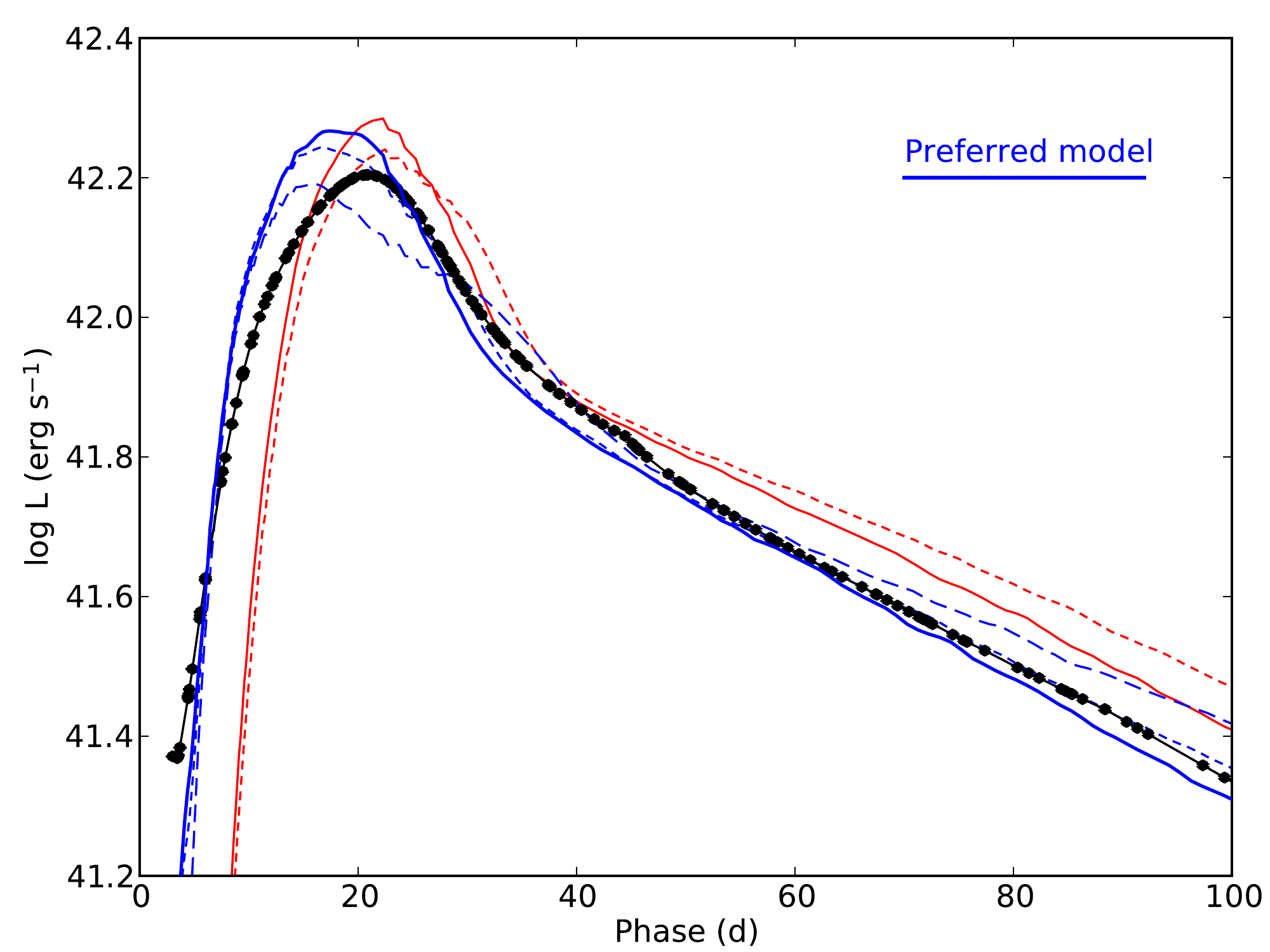}
\caption{The \citetalias{Jer14} model bolometric lightcurves between days 3 and 100 calculated with HYDE, compared to the observed optical-to-MIR pseudo-bolometric lightcurve (black circles). The model families differing in initial mass (12 M$_\odot$; solid, 13 M$_\odot$; short-dashed, 17 M$_\odot$; long-dashed) and macroscopic mixing (medium; red; strong; blue) are shown. Following this coding, the preferred model is shown as a blue solid line.}
\label{f_bol_opt_mir_mcomp_100d}
\end{figure}

To investigate the behaviour of the models in the early phase, when steady-state is not satisfied and the NLTE code cannot be used, we use HYDE in homologous mode to produce bolometric lightcurves between days 3 and 100. The \citetalias{Jer14} models are first (homologously) rescaled to day 1, and then evolved with HYDE until day 100. The initial temperature profile at day 1 is adopted from the best-fit hydrodynamical model for SN 2011dh (Sect.~\ref{s_modelling_day_0_300}), with the mixing of the \element[ ][56]{Ni} adjusted to match that of the \citetalias{Jer14} models. The subsequent evolution is not sensitive to the choice of initial temperature profile, as it is powered by the continuous injection of radioactive decay energy. Figure~\ref{f_bol_opt_mir_mcomp_100d} shows the bolometric lightcurves between days 3 and 100 for the \citetalias{Jer14} model families differing in initial mass and macroscopic mixing, compared to the observed optical-to-MIR pseudo-bolometric lightcurve. The other \citetalias{Jer14} model parameters have negligible influence on the bolometric lightcurve (Appendix~\ref{a_nlte_modelling}), and the optical-to-MIR BC is likely >$-$0.15 mag during this period (Appendix~\ref{a_nlte_modelling}), so the comparison is justified. The preferred model shows an overall agreement with observations, although the peak occurs a few days earlier and is overproduced by $\sim$0.2 mag.

\subsubsection{Constraints on the model parameters}
\label{s_modelling_nlte_parameters}

\begin{figure}[tb]
\centering
\includegraphics[width=0.36\textwidth,angle=0]{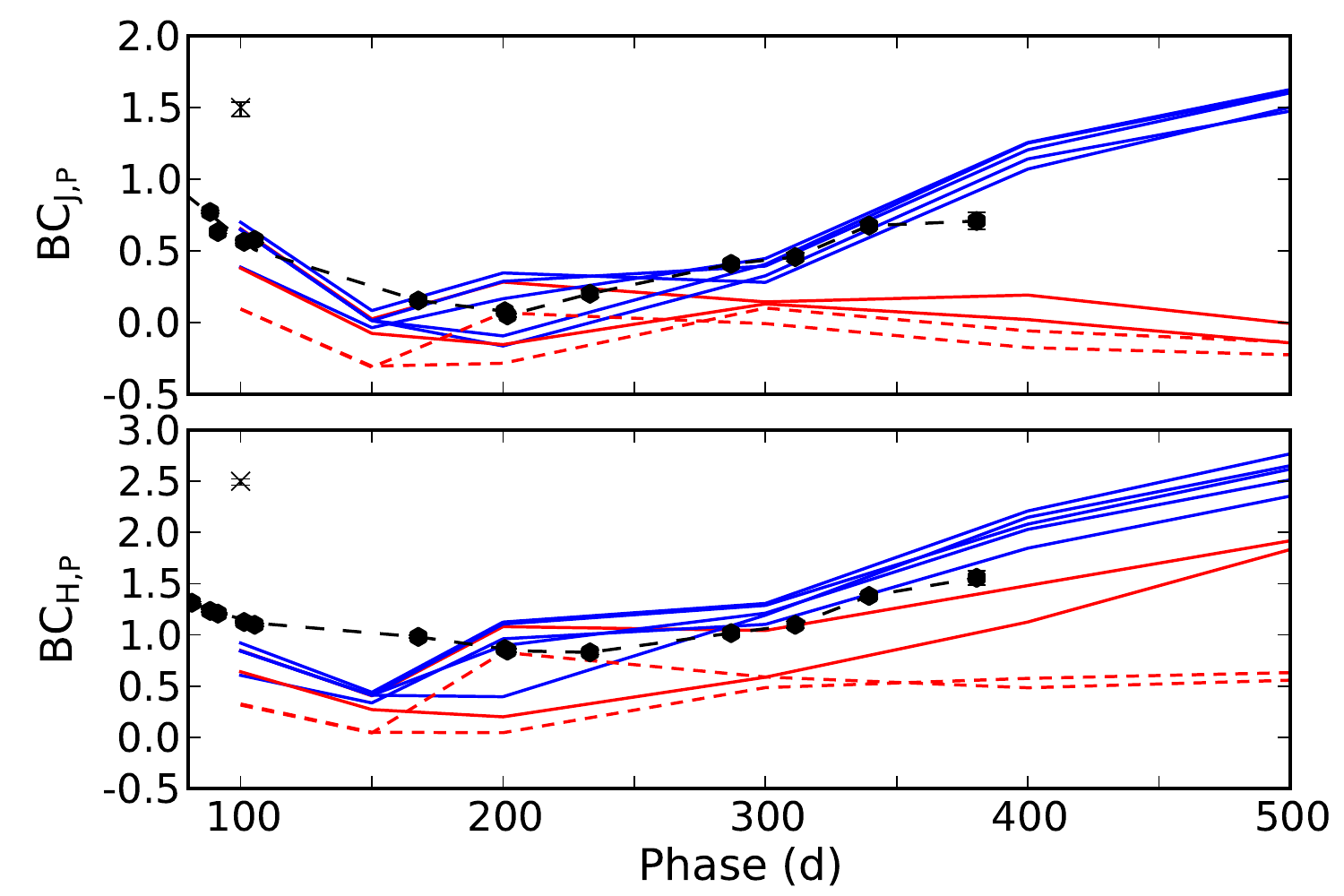}
\caption{The observed (black circles) and \citetalias{Jer14} model pseudo-BCs between days 100 and 500 for the $J$ and $H$ bands. The model families differing in contrast factor (high; solid, low; short-dashed) and positron trapping (local; blue, non-local; red) are shown. 17 M$_\odot$ and medium mixing models are not shown. The error bars arising from the extinction are marked in the upper left or right corner.}
\label{f_pseudo_bc_opt_nir_mir_mcomp_1}
\end{figure}

Here we discuss the constraints obtained from the lightcurves on the model parameters, except for the molecular cooling and dust parameters, which are discussed separately in Sect.~\ref{s_modelling_molecules_dust}. As explained in Appendix \ref{a_nlte_modelling}, a factorization of the model lightcurves into bolometric lightcurves and BCs is useful for the analysis, as the bolometric lightcurve depends significantly only on the initial mass and macroscopic mixing, whereas the other parameters only significantly affect the BCs. However, these quantities cannot be compared with observations, and observational counterparts are the optical-to-MIR pseudo-bolometric lightcurve and the (optical-to-MIR) pseudo-BCs. The pseudo-BC for band $X$ is abbreviated $\mathrm{BC}_{X,\mathrm{P}}$, and differs from BC$_{X}$ only in that it is based on the optical-to-MIR pseudo-bolometric luminosity. Figures~\ref{f_bol_opt_mir_mcomp_100d} and \ref{f_bol_opt_mir_mcomp} show the observed and \citetalias{Jer14} model optical-to-MIR pseudo-bolometric lightcurves\footnote{The model lightcurves between days 3 and 100 are bolometric (Sect.~\ref{s_modelling_day_3_100_j14})} for days 3$-$100 and 100$-$500, respectively, and from those constraints on the initial mass and macroscopic mixing can be obtained. Figures~\ref{f_pseudo_bc_opt_nir_mir_mcomp_1} and ~\ref{f_pseudo_bc_opt_nir_mir_mcomp_2} show the observed and \citetalias{Jer14} model pseudo-BCs for those bands from which constraints can be obtained on the contrast factor and positron trapping ($J$ and $H$) and the molecular cooling and optical depth of the dust ($I$, $z$, $K$, $S_1$, and $S_2$), respectively. Additional constraints can be obtained from the broad-band lightcurves (Fig.~\ref{f_opt_nir_mir_mcomp}; all models) and the optical pseudo-bolometric lightcurves (Fig.~\ref{f_bol_opt_mcomp}; selected models differing in the positron trapping and the properties of the dust). 

\paragraph{Initial mass and macroscopic mixing}

These parameters affect both the bolometric lightcurve and the BCs. The constraints from the optical-to-MIR pseudo-bolometric lightcurve are likely better obtained with the procedure used in Sect.~\ref{s_modelling_day_0_300}, but in agreement with those results, 17 M$_\odot$ and medium mixing models seem to be excluded. In particular, the tail luminosity becomes too high for 17 M$_\odot$ models and the rise to peak luminosity begins too late for medium mixing models (Figs.~\ref{f_bol_opt_mir_mcomp} and \ref{f_bol_opt_mir_mcomp_100d}). The effects on the BCs are generally too weak, or degenerate with the effects of other parameters, to provide useful constraints. On the other hand, the $R$-band lightcurve provides a constraint on the initial mass, at least as strong as the one obtained from the bolometric lightcurve, and seems to exclude 17  M$_\odot$ models (Fig.~\ref{f_opt_nir_mir_mcomp}). This is because the [\ion{O}{i}] 6300,6364 \AA~line, which depends strongly on the initial mass \citepalias{Jer14}, contributes $\sim$50 per cent of the flux in this band. Furthermore, as the fractional oxygen mass increases with initial mass \citepalias{Jer14}, both the $R$-band BC and the bolometric luminosity increase with initial mass, making $R$-band photometry particularly useful to constrain the initial mass. The choice of initial mass is motivated by the agreement with nebular spectra \citepalias{Jer14} and the optical-to-MIR pseudo-bolometric and $R$-band lightcurves, whereas the choice of macroscopic mixing is motivated by the optical-to-MIR pseudo-bolometric lightcurve.

\paragraph{Density contrast}

This parameter only (significantly) affects the BCs. In general the effect on the BCs is small, but there is a strong effect on BC$_{H}$ after day 300, caused by the [\ion{Si}{i}] 16450 \AA~line. However, there is a similar effect on BC$_{H}$ caused by the positron trapping (see below), and it is not possible to constrain the density contrast alone. Our NIR coverage ends at day $\sim$400, but the observed evolution of BC$_{H,\mathrm{P}}$ does not favour models with low density contrast and non-local positron trapping (Fig.~\ref{f_pseudo_bc_opt_nir_mir_mcomp_1}). The choice of a high density contrast is mainly motivated by the agreement with nebular spectra \citepalias{Jer14}, but is consistent with the oxygen zone filling factor of $\lesssim$0.07, derived from small scale fluctuations in the [\ion{O}{i}] 6300,6364 \AA~and \ion{Mg}{i}] 4571 \AA~lines (Sect.~\ref{s_spec_small_scale}).

\paragraph{Positron trapping}

This parameter only (significantly) affects the BCs. The effect is prominent at late times, when the positrons start to dominate the energy deposition, because locally trapped positrons deposit all their energy in the low temperature Fe/Co/He zone. This results in redder emission, and the luminosity of lines originating from this zone is boosted. Due to this, the optical decline rate is higher after day 300 for models with local trapping, which is in better agreement with observations (Fig.~\ref{f_bol_opt_mcomp}). After day 400, the observed optical decline rate starts to decrease, and models with local positron trapping start to diverge from observations. This could be a sign of additional energy sources (Sect.~\ref{s_500_days_lightcurves}), and is not necessarily in conflict with locally trapped positrons. We find particularly strong line effects in BC$_{J}$ and BC$_{H}$ after day 300, caused by the [\ion{Fe}{ii}] 12567 \AA~and [\ion{Fe}{ii}] 16440 \AA~lines, respectively. As discussed above, the effect on BC$_{H}$ is degenerate with a similar effect caused by the density contrast. Our NIR coverage ends at $\sim$400 days, but the observed evolution of BC$_{J,\mathrm{P}}$ seems to be in better agreement with models with local positron trapping (Fig.~\ref{f_pseudo_bc_opt_nir_mir_mcomp_1}). The choice of local positron trapping is motivated by the optical pseudo-bolometric lightcurve and the evolution of BC$_{J,\mathrm{P}}$ between days 300 and 400, although the decreasing optical decline rate observed after day 400 is a caveat. The constraints obtained from nebular spectra are not conclusive \citepalias{Jer14}.

\subsubsection{Dust, molecules, and the MIR evolution}
\label{s_modelling_molecules_dust}

Some of the most interesting results obtained from the pseudo-bolometric and broad-band lightcurves are related to the MIR evolution and the molecular cooling and dust parameters. There is a strong increase in the fractional MIR luminosity between days 100 and 250, during which an increase in the decline rate of the optical pseudo-bolometric lightcurve is also observed (Sect.~\ref{s_phot}). This behaviour is reminiscent of dust formation in the ejecta, and has previously been observed in SN 1987A \citep{Sun90}. From \citetalias{Erg14a} we know that there is an excess in the $S_2$ band, as compared to blackbody fits, already developing during the first 100 days. The $S_2$ band overlaps with the CO fundamental and SiO first-overtone bands, so in this band molecule emission provides an alternative or complementary explanation. Carbon-monoxide first-overtone emission is observed at day 206, and possibly at day 89 (Sect.~\ref{s_spec_CO}), which implies at least some contribution from CO fundamental band emission to the $S_2$ flux. Below we discuss the constraints on the molecular cooling and dust parameters obtained from the lightcurves, but also additional constraints obtained from spectra not discussed in \citetalias{Jer14}.

\begin{figure}[tb]
\centering
\includegraphics[width=0.36\textwidth,angle=0]{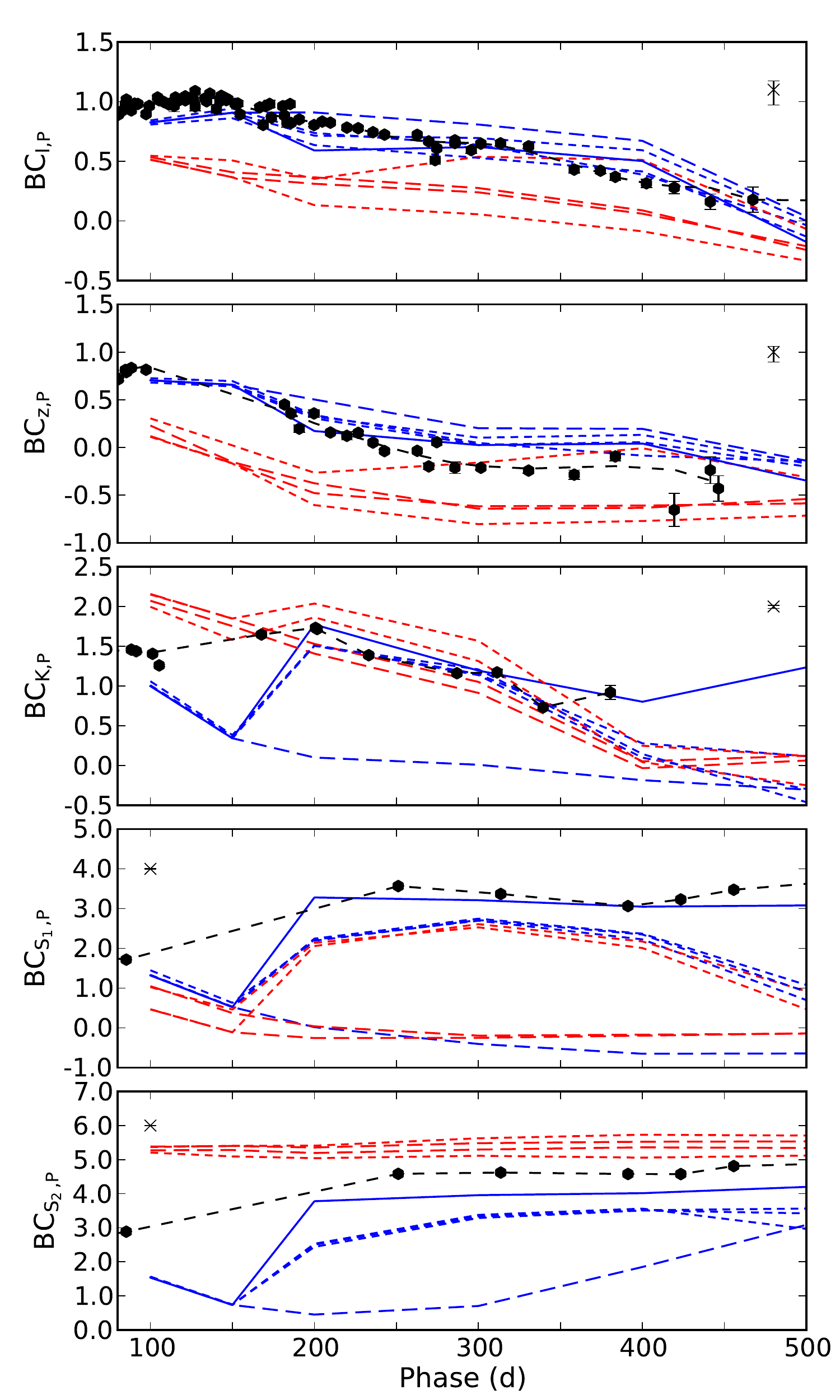}
\caption{The observed (black circles) and \citetalias{Jer14} model pseudo-BCs between days 100 and 500 for the $I$, $z$, $K$, $S_1$, and $S_2$ bands. The model families differing in molecular cooling (complete; red, none; blue) and the optical depth of the dust (0; long-dashed, 0.25; short-dashed, 0.44; solid) are shown. 17 M$_\odot$ and medium mixing models are not shown. The error bars arising from the extinction are marked in the upper left or right corner.}
\label{f_pseudo_bc_opt_nir_mir_mcomp_2}
\end{figure}

\paragraph{Dust} 

\begin{figure}[tb!]
\includegraphics[width=0.48\textwidth,angle=0]{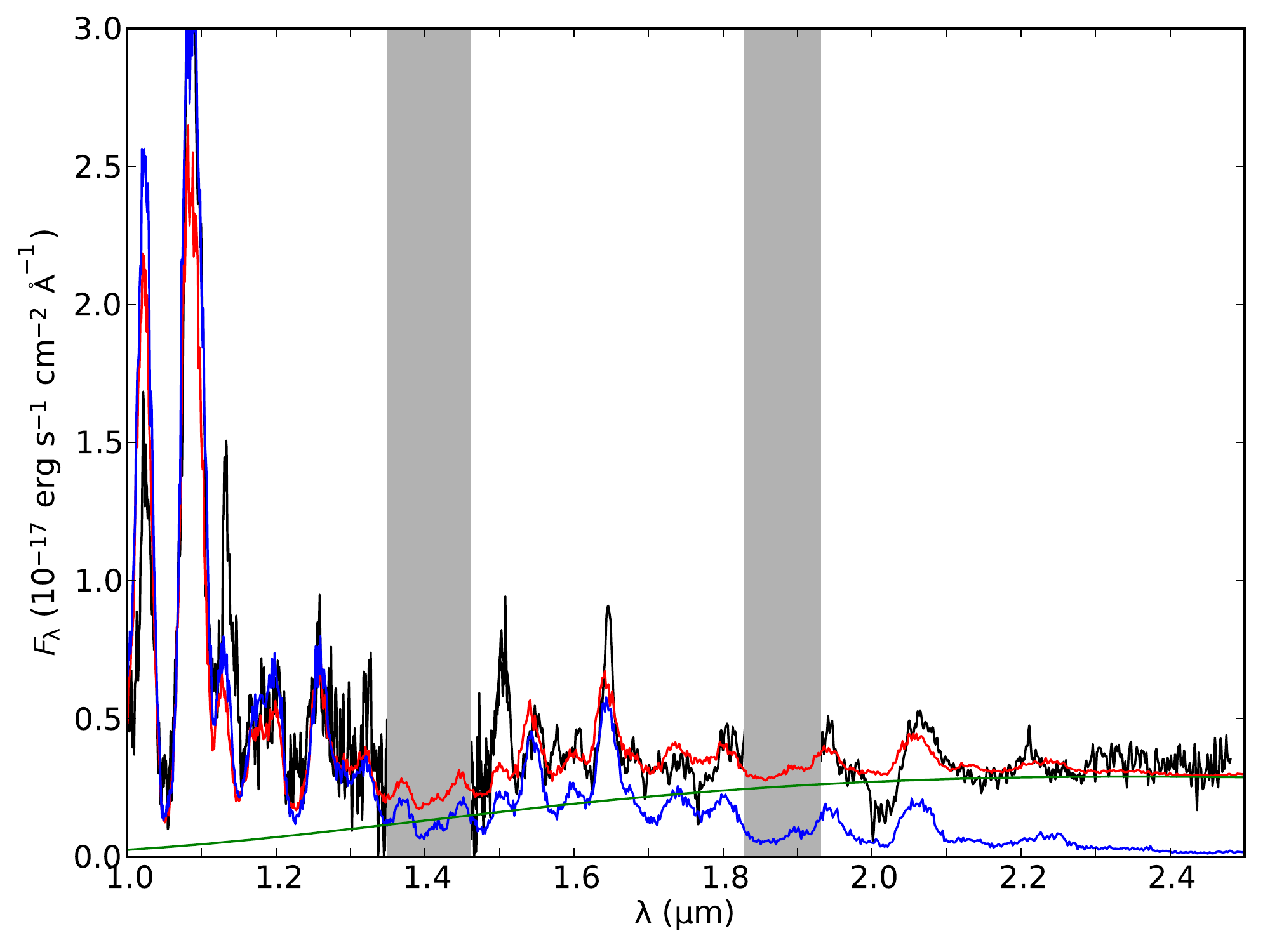}
\caption{Observed zJ and HK spectra at days 198 and 206 (black) compared to NIR spectra for the preferred model (12F; red) and model 12E (blue), which differs only in that $\tau_{\mathrm{dust}}$=0. The observed spectra were flux calibrated with the interpolated observed $J$, $H$, and $K$ magnitudes at day 200 and we also show the dust emission component for the preferred model (green).}
\label{f_spec_nir_200d_obs_12F_12E_comp}
\end{figure}

This parameter only affects the BCs and tends to decrease the optical emission and increase the $K$-band and MIR emission, as the hot radiation from the SN is absorbed and re-emitted at a lower temperature. Carbon-monoxide first-overtone emission contributes negligible to the $K$-band flux (Sect.~\ref{s_spec_CO}), so the observed excess in this band can be solely attributed to dust emission. The preferred model (with $\tau_{\mathrm{dust}}$=0.44) reproduces well both the drop in the optical pseudo-bolometric lightcurve (Fig.~\ref{f_bol_opt_mcomp}) and the increase in BC$_{K,\mathrm{P}}$ and BC$_{S_1,\mathrm{P}}$ (Fig.~\ref{f_pseudo_bc_opt_nir_mir_mcomp_2}), whereas the original \citetalias{Jer14} models (with  $\tau_{\mathrm{dust}}$=0.25) only qualitatively reproduce this behaviour. This shows, most importantly, that the absorbed and emitted luminosities are in good agreement, and suggests a scenario where the emission arises from nearby, newly formed dust, presumably in the ejecta. We note that the dust contribution is insufficient to reproduce the excess in the $S_2$ band (Fig.~\ref{f_pseudo_bc_opt_nir_mir_mcomp_2}), suggesting a contribution from molecule emission in this band (see below). 

Further support for a local origin of the emission is gained from the spectra. In Fig.~\ref{f_spec_nir_200d_obs_12F_12E_comp} we show the observed NIR spectrum at day $\sim$200, as compared to NIR spectra for the preferred model and model 12E (same but without dust) at day 200, as well as the dust emission component for the preferred model. The observed flux at the absorption minimum of the \ion{He}{i} 20581 \AA~line is a factor of $\sim$2 below the model dust emission level\footnote{The model dust emission has not been subject to radiative transfer (Appendix~\ref{a_nlte_modelling}), so the depth of the model \ion{He}{i} 20581 \AA~absorption corresponds to that expected for a thermal echo.}, which suggests that this emission originates from $\lesssim$10000 km~s$^{-1}$, and disfavours a thermal echo from heated CSM dust (but see \citealp{Hel13} with respect to the period before day 100). The small blue-shifts in the [\ion{O}{i}] 6300,6364 \AA~and \ion{Mg}{i}] 4571 \AA~lines at day 415 (Sect.~\ref{s_spec_line_asymmetries}) and the lack of a physical model for the temperature evolution are caveats though. Regardless of the location, dust emission seems to be required to explain the strong discrepancy between observations and models without dust in the $K$ and $S_1$ bands (Fig.~\ref{f_pseudo_bc_opt_nir_mir_mcomp_2}), further illustrated by the spectral comparison in Fig.~\ref{f_spec_nir_200d_obs_12F_12E_comp}. A MIR excess developing between days 100 and 250 was also observed for SN 1993J (Sect.~\ref{s_bol_evo}). The cause of this excess was suggested by \citet{Mat02} to be dust, but the absence of an increase in the optical pseudo-bolometric decline rate indicates that, if this emission is due to dust, it would rather arise from heated CSM dust. Except for SNe 2011dh and 1993J and the peculiar Type Ibn SN 2008jc \citep{Smi08}, we have found no reports of dust emission in Type IIb or stripped envelope SNe in the literature.

\paragraph{Molecular cooling} 

This parameter only affects the BCs and results in a redistribution of cooling emission from the O/C and O/Si/S zones to the CO and SiO molecular bands. The effect is strong in the $S_2$ band, which overlaps with the CO fundamental and SiO first-overtone bands, and there is also a significant effect in the $K$ band, which overlaps partly with the CO first-overtone band. The observed CO first-overtone emission implies some amount of CO cooling at day 206, and possibly at day 89, but is $\sim$3 and $\sim$3.5 mag too faint, respectively, as compared to models with complete molecular cooling. As mentioned, there is an excess in the $S_2$ band as compared to the preferred model (without molecular cooling), which shows the best agreement for a conceivable dust component. Compared to this model, BC$_{S_2,\mathrm{P}}$ is 0.5$-$1.0 mag too large, but compared to models with complete molecular cooling, BC$_{S_2,\mathrm{P}}$ is $\sim$2 and 0.5$-$1.0 mag too small at day 85 and after day 251, respectively (Fig.~\ref{f_pseudo_bc_opt_nir_mir_mcomp_2}). As molecule emission in the $S_2$ band is likely to be dominated by the CO fundamental band (Appendix~\ref{a_nlte_modelling}), this suggests a small amount of CO cooling at day 85, increasing to an intermediate amount after day 251, whereas the observed CO first-overtone emission suggests a small amount of CO cooling at days 89 and 206. 

The interpretation of the CO first-overtone emission and the $S_2$ excess depends on the fundamental to first-overtone band ratio, which is assumed to be the same as observed in SN 1987A in the models (Appendix~\ref{a_nlte_modelling}). The observed CO first-overtone and $S_2$ fluxes give an upper limit on this ratio, which can be improved by subtracting a model for the underlying emission from the $S_2$ fluxes, and if we again assume the excess in the $S_2$ band to be dominated by CO fundamental band emission, we can estimate it. Using the preferred model for the underlying emission, keeping in mind that the $S_2$ magnitudes have to be interpolated between days 85 and 251, we get upper limits of 6.6 and 28 and estimates of 3.9 and 9.9 at days 89 and 206, respectively. These values are significantly larger than for SN 1987A ($\sim$1 and $\sim$2, respectively), and would naively suggest a lower temperature for the CO. So, either the fundamental to first-overtone band ratios are significantly higher than observed in SN 1987A, or the excess in the $S_2$ band is not caused by CO (but e.g.~by SiO or heated CSM dust).

Further constraints on the amount of molecular cooling is gained from thermal lines originating in the O/C an O/Si/S zones, in particular the blended \ion{Ca}{ii} 8498,8542,8662 \AA~and [\ion{C}{i}] 8727 \AA~lines \citepalias{Jer14}. The evolution of this blend is reflected in the evolution of the $I$ and $z$ bands, which seems to exclude models with complete molecular cooling before day $\sim$250 \citepalias[Fig.~\ref{f_pseudo_bc_opt_nir_mir_mcomp_2}; but see also][]{Jer14}, Such models also seem to be excluded at later times because of the (likely) continued presence of the [\ion{C}{i}] 8727 \AA~line (Sect.~\ref{s_spec_line_asymmetries}). Finally, the remarkable similarity between the [\ion{O}{i}] 6300 \AA~and \ion{Mg}{i}] 4571 \AA~lines suggests the contributions to the [\ion{O}{i}] 6300 \AA~flux from the O/C and O/Si/S zones to be modest (Sect.~\ref{s_spec_line_regions}), in turn suggesting a considerable amount of molecular cooling after day $\sim$200. 

We have chosen no molecular cooling for the preferred model, but neither this scenario nor complete molecular cooling shows a satisfactory agreement with observations. However, the constraints obtained are not necessarily inconsistent, and at least in the O/C zone, a small amount of molecular (CO) cooling at day $\sim$100, increasing to an intermediate amount after day $\sim$250, seems reasonable. Carbon-monoxide emission has been reported for the Type Ic SNe 2002ew \citep{Ger02} and 2007gr \citep{Hun09}, but for Type Ib and IIb SNe we have found no reports of CO emission in the literature. However, there is a feature near 23000 \AA~in the NIR spectra of SN 1993J \citep{Mat02} that could be interpreted as a modest amount of CO first-overtone emission.

\subsection{Hydrodynamical modelling of the optical-to-MIR pseudo-bolometric lightcurve before day 400}
\label{s_modelling_day_0_300}

\citet{Ber12} presented hydrodynamical modelling of SN 2011dh, that well reproduced the UV-to-MIR pseudo-bolometric lightcurve before day 80 presented in \citetalias{Erg14a}. In \citetalias{Erg15} we use a grid of SN models constructed with HYDE and MESA STAR \citep{Pax11,Pax13} to fit the UV-to-MIR pseudo-bolometric lightcurve before day 100, and here we use an extended version of this grid to fit the optical-to-MIR pseudo-bolometric lightcurve before day 400. This is the period for which we have full $U$ to $S_2$ coverage and, except for the additional $\sim$300 days of data, the method allows us to quantify the errors in the derived quantities and the degeneracy of the solution. The stellar models consist of bare helium cores without a hydrogen envelope, which is a sound approximation as long as the optical depth of the hydrogen envelope is $\ll$1, and the mass low enough not to (appreciable) decelerate the helium core \citepalias{Erg15}. The original model grid is described in \citetalias{Erg15} and the extended grid is summarized in Appendix~\ref{a_hyde}, and the model parameters are the helium core mass (M$_{\mathrm{He}}$), the explosion energy (E), and the mass (M$_{\mathrm{Ni}}$) and distribution (Mix$_{\mathrm{Ni}}$) of the \element[ ][56]{Ni}. The fitting is done by minimization of the square of the relative residuals, giving equal weights to the diffusion phase lightcurve, the early tail lightcurve, the late tail lightcurve, and the diffusion phase photospheric velocities. 

The optical-to-MIR BC, which is required to fit the observations, is determined from the \citetalias{Jer14} steady-state NLTE models, and is discussed in more detail in Appendix~\ref{a_nlte_modelling}. Before day 100 the BC is likely >$-$0.15 mag and is assumed to be negligible, and between days 100 and 400 we take advantage of the small spread (<$\pm$0.1 mag) in the model BCs during this period, and use the BC for the preferred steady-state NLTE model. As discussed in Appendix~\ref{a_nlte_modelling} this is motivated for all hydrodynamical models that could possibly give a good fit to the lightcurve.

\begin{figure}[tb!]
\includegraphics[width=0.5\textwidth,angle=0]{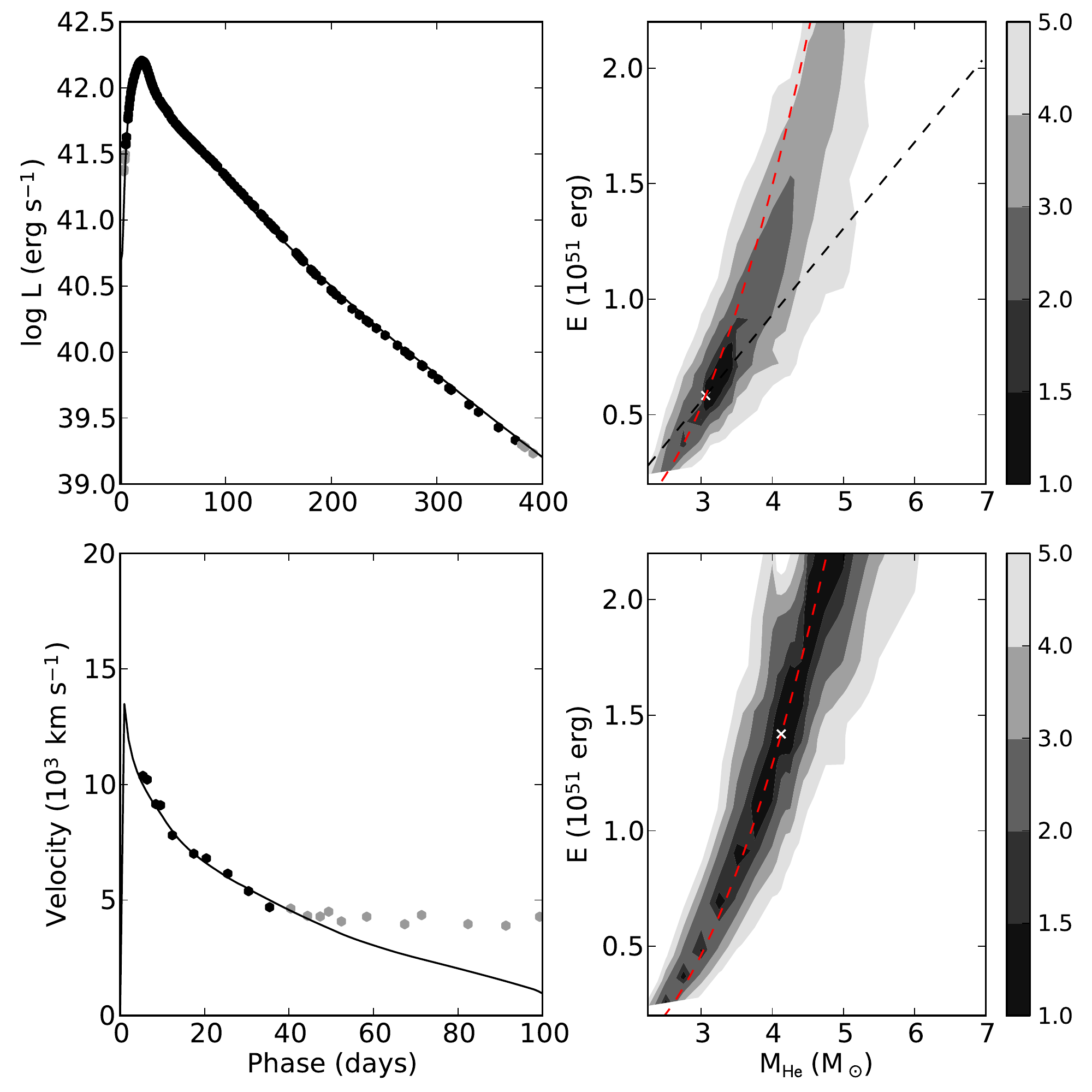}
\caption{Left panels: Optical-to-MIR pseudo-bolometric lightcurve (upper left panel) and photospheric velocities (lower left panel) for the best-fit model (solid line) compared to observations (circles), where the photospheric velocities have been estimated from the absorption minimum of the \ion{Fe}{ii} 5169 \AA~line. Observations not included in the fit are shown in grey. Right panels: Contour plots showing the (grey-scale coded) standard deviation in the fit, normalized to that of the best-fit model, projected onto the E-M$_{\mathrm{He}}$ plane for the case where the photospheric velocities were used (upper right panel) and not used (lower right panel). We also show the constraints M$_{\mathrm{ej}}$/E=const (blue line) and M$_{\mathrm{ej}}^{2}$/E=const (red line) provided by the photospheric velocities and the bolometric lightcurve, respectively \citepalias{Erg15}.}
\label{f_mg_fit_mc}
\end{figure}

\begin{table*}[tb!]
\caption{Helium core mass, explosion energy and mass and distribution of the $^{56}$Ni as derived from the optical-to-MIR pseudo-bolometric lightcurve before day 400 in this work, and from the UV-to-MIR pseudo-bolometric lightcurves before days 80 and 100 in \citetalias{Ber12} and \citetalias{Erg15}, respectively. The errors are given in parenthesis and include those arising from the distance, extinction, and photospheric velocities (assumed to be 15 per cent), but do not take the degeneracy of the solution into account. The mass of the \element[ ][56]{Ni} given in \citetalias{Ber12} has been adjusted with respect to the distance and extinction \citepalias{Erg14a}.}
\begin{center}
\begin{tabular}{l l l l l}
\hline\hline \\ [-1.5ex]
Reference & M$_{\mathrm{He}}$ & E & M$_{\mathrm{Ni}}$ & Mix$_{\mathrm{Ni}}$ \\ [0.5ex]
 & (M$_\odot$) & (10$^{51}$ erg) & (M$_\odot$) & \\ [0.5ex]
\hline \\ [-1.5ex]
This work & 3.06 (+0.68,-0.44) & 0.58 (+0.45,-0.25) & 0.075 (+0.028,-0.020) & 0.95 (+0.06,-0.04) \\
\citetalias{Erg15} & 3.31 (+0.54,-0.57) & 0.64 (+0.38,-0.30) & 0.075 (+0.028,-0.020) & 1.05 (+0.08,-0.00) \\
\citetalias{Ber12} & 3.3-4.0 & 0.6-1.0 & 0.06-0.10 & 0.95 \\ [0.5ex]
\hline
\end{tabular}
\end{center}
\label{t_prog_sn_parameters}
\end{table*}

The left panels of Fig.~\ref{f_mg_fit_mc} show the optical-to-MIR pseudo-bolometric lightcurve and the photospheric velocities for the best-fit model as compared to observations, where the photospheric velocities have been estimated from the absorption minimum of the \ion{Fe}{ii} 5169 \AA~line 
\citepalias{Erg14a}. The right panels of Fig.~\ref{f_mg_fit_mc} show contour plots of the standard deviation in the fit, normalized to that of the best-fit model. The fits to the lightcurve and the photospheric velocities are both good, and the solution is well constrained in the E-M$_{\mathrm{He}}$ plane (upper right panel). When only the lightcurve is used in the fit (lower right panel), the solution is completely degenerate along the M$_{\mathrm{ej}}^{2}$/E=const curve, in agreement with the results in \citetalias{Erg15} for the period before day 100. The good fit to the lightcurve suggests that other isotopes than \element[ ][56]{Ni} and \element[ ][56]{Co}, or additional energy sources like those discussed in Sect.~\ref{s_500_days_lightcurves}, do not contribute substantially to the luminosity before day 400.

Table~\ref{t_prog_sn_parameters} lists the best-fit values for the  helium core mass, explosion energy, and mass and distribution of the $^{56}$Ni. These results are consistent within error bars with those for the lightcurves before days 80 and 100 obtained in \citetalias{Ber12} and \citetalias{Erg15} (Table~\ref{t_prog_sn_parameters}). The addition of the late-time data seems to decrease the explosion energy and helium core mass slightly, and these are $\sim$10 per cent lower than the results obtained in \citetalias{Erg15}. The strong mixing of the $^{56}$Ni found (95 per cent in mass, see Appendix~\ref{a_hyde}) is in good agreement with the results obtained in \citetalias{Ber12} and \citetalias{Erg15}, and seems to be required to fit the rise to peak luminosity (see also Sect.~\ref{s_modelling_nlte_parameters}). The fraction of the $^{56}$Ni outside 3500 km~s$^{-1}$ is $\sim$50 per cent in the best-fit model, and naively this is in contradiction with the small size of the Fe/Co/He line emitting region estimated in Sect.~\ref{s_spec_line_regions}. However, the amount of high velocity Fe/Co/He material does not necessarily need to be large to reproduce the rise to peak luminosity, and further modelling is needed to resolve this issue. The errors in the distance, extinction, and photospheric velocities (where we have assumed a systematic error of 15 per cent) have been propagated to the derived quantities, but without taking the degeneracy of the solution into account. However, as is evident from the contour plots, the constraint on the quantity M$_{\mathrm{ej}}^{2}$/E is strong, and even if the photospheric velocities were considerably \emph{underestimated}, which seems unlikely, a helium core mass of $\gtrsim$4 M$_\odot$ would be excluded.

\section{The lightcurves after day 500}
\label{s_500_days_lightcurves}

\begin{figure}[tb]
\includegraphics[width=0.48\textwidth,angle=0]{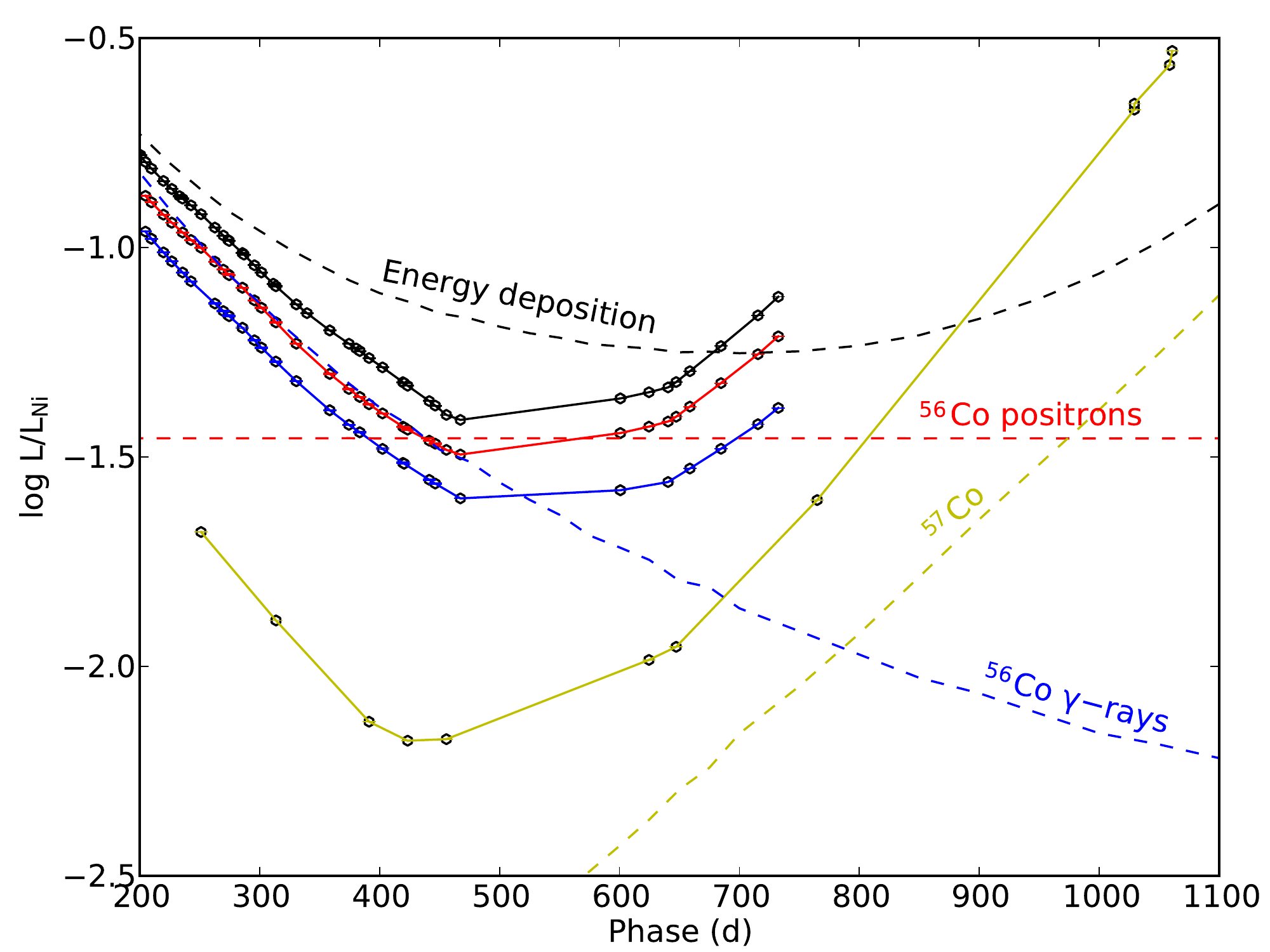}
\caption{The optical-to-MIR (black dots), optical-plus-MIR (red dots), optical (blue dots), and MIR (yellow dots) pseudo-bolometric lightcurves 
after day 200, compared to the bolometric lightcurve (black dashed line), deposited \element[ ][56]{Co} $\gamma$-ray (blue dashed line) and positron (red dashed line) luminosity, and deposited \element[ ][57]{Co} luminosity (yellow dashed line) for the preferred steady-state NLTE model (12F). The lightcurves have been normalized to the radioactive decay chain luminosity of 0.075 M$_{\odot}$ of \element[ ][56]{Ni}.}
\label{f_bol_late}
\end{figure}

As discussed in Sect.~\ref{s_bol_evo} there is a strong flattening of the observed pseudo-bolometric lightcurves\footnote{There is an observational gap between days 467 and 601.} after day $\sim$450, which is also when these start to diverge from the preferred steady-state NLTE model (Sect.~\ref{s_modelling_optimal_model}). Figure~\ref{f_bol_late} shows the observed optical-to-MIR, optical-plus-MIR (excluding NIR), optical, and MIR pseudo-bolometric lightcurves 
after day 200. In the figure we also show the bolometric lightcurve (energy deposition), deposited \element[ ][56]{Co} $\gamma$-ray and positron luminosity, and deposited \element[ ][57]{Co} luminosity for the preferred steady-state NLTE model\footnote{These were calculated with HYDE using the same ejecta model as the preferred steady-state NLTE model.}. \citet{Shi13} suggested that the SN entered a phase powered by the positrons emitted in the \element[ ][56]{Co} decay after day 300-350. As seen in  Figure~\ref{f_bol_late}, this suggestion seems to be roughly correct in the sense that the positron contribution dominates the deposited luminosity after day $\sim$450. However, as we discuss below, there is evidence for additional energy sources, and the positron contribution does not seem to dominate the emitted luminosity.

The \citetalias{Jer14} models were not evolved beyond day 500 as the steady-state condition is no longer valid (see below). However, a similar set of models, which were evolved to day 700, all show a decrease in the optical and optical-to-MIR BCs after day 500, corresponding to redder emission. This decrease is particularly strong for models with local positron trapping because of the increasing contribution from the low temperature Fe/Co/He zone. Assuming that the actual BCs behave similarly and do not increase, the pseudo-bolometric decline rates after day $\sim$600, which becomes significantly lower than the decay rate of \element[ ][56]{Co} (Sect.~\ref{s_bol_evo}), are not consistent with a model powered by this decay. Furthermore, assuming again that the actual BCs do not increase, the actual bolometric luminosity at day 732 would be $\gtrsim$0.6 mag higher than the radioactive energy deposition in the preferred model. Even if we ignore the BCs, the observed optical-plus-MIR luminosity at day 732 equals the radioactive energy deposition in the preferred model, and the observed MIR luminosity at day 1061 is 1.1 mag higher. In summary, both the observed decline rates and the observed luminosities suggest that additional energy sources are required to power the lightcurves after day $\sim$600.

If the recombination time scales become longer than the time scale of the \element[ ][56]{Co} decay, some fraction of the energy is deposited in a growing ionization reservoir, which through recombination emission could eventually dominate the emitted luminosity. This process, called freeze-out, is an example of time-dependent effects that might violate the steady-state assumption in late phases. We use a time-dependent NLTE code \citep{Koz92,Koz98a,Koz98b} to test this assumption, and find that time-dependent effects begin to become important at day $\sim$600, and after day $\sim$800 they provide a dominant and increasing contribution to the optical flux. In the particular model tested, which is similar to the preferred model, this contribution is not sufficient to reproduce the observed lightcurves. However, because of clumping or asymmetries, low density components may exist in the ejecta, for which freeze-out would occur earlier. The day 678 spectrum of SN 2011dh presented by \citet{Shi13}, shows features not present in our last optical spectra that \emph{could} be identified as the \ion{He}{i} 6678 \AA~and 7065 \AA~lines, whereas the strong feature identified as \ion{Na}{i} 5890,5896 \AA~by the authors \emph{could} have a significant contribution from, or be fed by, the \ion{He}{i} 5876 \AA~line. This is consistent with a contribution from recombination emission due to freeze-out in the helium envelope. 

For SN 1993J, CSM interaction became the dominant energy source after day $\sim$300, giving rise to broad box-like H$\alpha$ and \ion{Na}{i} 5890,5896 \AA~lines and a considerable flattening of the lightcurves. The day 678 spectrum of SN 2011dh shows a feature that \citet{Shi13} interpret as broad box-like H$\alpha$ emission, but no broad box-like \ion{Na}{i} 5890,5896 \AA~emission is seen. The interpretation of the broad feature as H$\alpha$ emission is far from clear, as a number of other lines may contribute in this wavelength range (including the [\ion{N}{ii}] 6548,6583 \AA~line discussed in Sect.~\ref{s_spec_lines} and the \ion{He}{i} 6678 \AA~line mentioned above), and the feature is much weaker than for SN 1993J at a similar epoch. Additional energy could also be provided by the decay of radioactive isotopes other than \element[ ][56]{Co}. In the preferred steady-state NLTE model, the fractional bolometric luminosity deposited by the \element[ ][57]{Co} decay is $\sim$10 per cent at day 700 and increasing. A higher mass of \element[ ][57]{Co} would help explain the observed evolution and cannot be excluded. Another possibility is that the observed flux does not solely originate from the SN, but this seems to be ruled out by the day 678 spectrum, which is line dominated and has a minimum flux level close to zero. 

In summary, we find  observational and theoretical evidence that the \element[ ][56]{Co} decay does not dominate the observed luminosity after day $\sim$600. A substantial contribution from time-dependent effects is likely, whereas contributions from CSM interaction or other radioactive isotopes than \element[ ][56]{Co} cannot be excluded. A contribution from coincident sources seems to be ruled out.

\section{Conclusions}
\label{s_conclusions}

We present the late-time part of our optical and NIR, photometric and spectroscopic dataset for the Type IIb SN 2011dh. The dataset spans two years, and together with SWIFT and Spitzer observations it covers the UV to MIR wavelength range. Particular attention is paid to the pseudo-bolometric and broad-band lightcurves, where we use steady-state NLTE modelling and hydrodynamical modelling to put constraints on the SN and progenitor parameters. 

We analyse the optical-to-MIR pseudo-bolometric lightcurve before day 400 using a grid of hydrodynamical SN models and an optical-to-MIR BC determined with steady-state NLTE modelling. Using this method we find model parameters consistent within error bars with those obtained in \citetalias{Ber12} using the UV-to-MIR pseudo-bolometric lightcurve before day 80. In particular, we find a helium core mass of 3.1$^{+0.7}_{-0.4}$ M$_\odot$, and upper limits on the helium core and initial mass of $\lesssim$4 M$_\odot$ and $\lesssim$15 M$_\odot$, respectively.

We analyse the pseudo-bolometric and broad-band lightcurves between days 100 and 500 using the set of steady-state NLTE models presented in \citetalias{Jer14}. The preferred 12 M$_\odot$ (initial mass) model, refined in this work with respect to the dust, shows an overall agreement with the observed broad-band and pseudo-bolometric lightcurves. In particular, the simultaneous increase in the optical pseudo-bolometric decline rate and the fractional $K$ and $S_1$ luminosities between days 100 and 250 is reproduced by this model, which has a modest dust opacity in the core ($\tau_{\mathrm{dust}}=0.44$), turned on at day 200. A local origin of the excess emission is supported by the depth of the \ion{He}{i} 20581 \AA~absorption. We find the dust contribution insufficient to reproduce the $S_2$ magnitudes and, assuming this additional excess to be dominated by CO fundamental band emission, an intermediate amount of CO cooling in O/C zone is likely. This is consistent with the detected CO first-overtone emission, although the fundamental to first-overtone band ratios need to be significantly higher than observed in SN 1987A.

The lightcurves after day 500 have not been systematically modelled, but there is both observational and theoretical evidence that the SN becomes powered by additional energy sources in this phase. After day 400 the optical and MIR pseudo-bolometric lightcurves flatten to fast, and after day 600 their luminosities become too high to be powered by the \element[ ][56]{Co} decay. Modelling with a time-dependent NLTE code \citep{Koz92,Koz98a,Koz98b} shows that time-dependent effects become important at day $\sim$600, after which a steady-state assumption is no longer valid, and after day $\sim$800 they provide a dominant and increasing contribution to the optical luminosity. We find substantial contributions from CSM interaction or other radioactive isotopes than \element[ ][56]{Co} less likely, whereas a substantial contribution from coincident sources is ruled out.

We estimate progressively smaller sizes, ranging from $\sim$3000 to $\sim$1500 km~s$^{-1}$, for the line emitting regions of oxygen, magnesium, [\ion{Ca}{ii}] 7291,7323 \AA, and iron. These regions correspond to the oxygen, O/Ne/Mg, Si/S, and Fe/Co/He nuclear burning zones \citepalias{Jer14}, and suggest incomplete mixing of the core material. The profiles of the [\ion{O}{i}] 6300 \AA~and \ion{Mg}{i}] 4571 \AA~lines show a remarkable similarity, suggesting that these lines arise from the O/Ne/Mg zone. We use repetitions of small scale fluctuations in the [\ion{O}{i}] 6300 \AA~and 6364 \AA~lines to find a line ratio close to 3, consistent with optically thin emission, from day 200 and onwards. Applying the method of \citet{Chu94} to these small scale fluctuations, we find an upper limit on the filling factor for the [\ion{O}{i}] 6300 \AA~and \ion{Mg}{i}] 4571 \AA~line emitting material of $\sim$0.07, and a lower limit on the number of clumps of $\sim$900.

This paper concludes our work on SN 2011dh presented in \citetalias{Mau11}, \citetalias{Erg14a}, and \citetalias{Jer14}. In addition, modelling of our data have been presented by \citetalias{Ber12}. We have applied stellar evolutionary progenitor analysis, hydrodynamical modelling, SN atmosphere modelling, and steady-state NLTE modelling to our extensive set of observational data. Although a number of issues remain unsolved, the main characteristics of the SN and its progenitor star found by the different methods are consistent. The progenitor star appears to have been of moderate ($\lesssim$15 M$_\odot$) initial mass, and the 3$-$4 M$_\odot$ helium core surrounded by a low-mass ($\sim$0.1 M$_\odot$) and extended (200$-$300 R$_\sun$) hydrogen envelope. Given the upper bound on the initial mass of $\sim$15 M$_\odot$, the mass-loss rate is probably not strong enough to expel the hydrogen envelope before core-collapse, and a binary origin for SN 2011dh is suggested. \citet{Fol14} find a blue source they suggest to be the companion star in HST UV imaging obtained at $\sim$1200 days, but it is unclear how they can exclude a contribution from the SN, and further work is likely needed to settle this issue.

\section{Acknowledgements}

This work is based on observations obtained with the Nordic Optical Telescope, operated by the Nordic Optical Telescope Scientific Association at the Observatorio del Roque de los Muchachos, La Palma, Spain, of the Instituto de Astrofisica de Canarias; the German-Spanish Astronomical Center, Calar Alto, jointly operated by the Max-Planck-Institut für Astronomie Heidelberg and the Instituto de Astrofísica de Andalucía (CSIC); the United Kingdom Infrared Telescope, operated by the Joint Astronomy Centre on behalf of the Science and Technology Facilities Council of the U.K.; the William Herschel Telescope and its service programme (proposals SW2011b21 and SW2012a02), operated on the island of La Palma by the Isaac Newton Group in the Spanish Observatorio del Roque de los Muchachos of the Instituto de Astrofisica de Canarias; the Copernico 1.82m Telescope and Schmidt 67/92 Telescope operated by INAF - Osservatorio Astronomico di Padova at Asiago, Italy; by the 3.6m Italian Telescopio Nazionale Galileo operated by the Fundaci\'on Galileo Galilei - INAF on the island of La Palma; the Liverpool Telescope, operated on the island of La Palma by Liverpool John Moores University in the Spanish Observatorio del Roque de los Muchachos of the Instituto de Astrofisica de Canarias with financial support from the UK Science and Technology Facilities Council; the AlbaNova telescope operated by the Department of Astronomy at Stockholm University and funded by a grant from the Knut and Alice Wallenberg Foundation; the Gran Telescopio Canarias (GTC), installed in the Spanish Observatorio del Roque de los Muchachos of the Instituto de Astrofísica de Canarias, in the island of La Palma. We acknowledge the exceptional support we got from the NOT staff throughout this campaign, we thank the Calar Alto Observatory for the allocation of the director's discretionary time and we thank Philip Dufton, Paul Dunstall, Darryl Wright and Lindsay Magill for assistance with the WHT observations.

The Oskar Klein Centre is funded by the Swedish Research Council. J.S. acknowledge support by the Swedish Research Council. S.J.S. thanks European Research Council under the European Union's Seventh Framework Programme (FP7/2007-2013)/ERC Grant agreement n$^{\rm o}$ [291222] and STFC. A.P., L.T. and S.B. are partially supported by the PRIN-INAF 2011 with the project “Transient Universe: from ESO Large to PESSTO”. M.F. acknowledges support by the European Union FP7 programme through ERC grant number 320360. N.E.R. acknowledges the support from the European Union Seventh Framework Programme (FP7/2007-2013) under grant agreement n$^{\rm o}$ 267251 “Astronomy Fellowships in Italy” (AstroFIt). R.K. acknowledges observing time at the LT, NOT, TNG, and WHT via programme CCI-04. M.M.K. acknowledges generous support from the Hubble Fellowship and Carnegie-Princeton Fellowship.

We thank Melina Bersten for providing the post-explosion density profile for the He4R270 model \citepalias{Ber12}, for inspiration and a great contribution to the understanding of SN 2011dh. We thank Peter Meikle and Dan Milisavljevic for providing spectra on SN 1993J and SN 2008ax, respectively.

\begin{table*}[p]
\caption{Late-time (after day 100) optical colour-corrected JC $U$ and S-corrected JC $BVRI$ magnitudes for SN 2011dh. Errors are given in parentheses.}
\begin{center}
\scalebox{1.00}{
\begin{tabular}{l l l l l l l l }
\hline\hline \\ [-1.5ex]
JD (+2400000) & Phase & $U$ & $B$ & $V$ & $R$ & $I$ & Telescope (Instrument)\\ [0.5ex]
(d) & (d) & (mag) & (mag) & (mag) & (mag) & (mag) & \\
\hline \\ [-1.5ex]
55817.35 & 104.35 & ... & 16.00 0.03 & 15.20 0.02 & 14.64 0.02 & 14.04 0.02 &  AS-Schmidt (SBIG) \\
55818.33 & 105.33 & ... & 16.08 0.02 & 15.19 0.02 & 14.68 0.01 & 14.09 0.01 &  AS-Schmidt (SBIG) \\
55821.31 & 108.31 & 16.77 0.08 & 16.12 0.02 & 15.26 0.01 & 14.70 0.01 & 14.17 0.01 &  CA-2.2m (CAFOS) \\
55824.32 & 111.32 & ... & ... & 15.31 0.02 & 14.77 0.03 & 14.25 0.03 &  AS-Schmidt (SBIG) \\
55827.33 & 114.33 & ... & 16.14 0.03 & 15.42 0.01 & 14.88 0.02 & 14.28 0.01 &  AS-Schmidt (SBIG) \\
55827.48 & 114.48 & ... & 16.15 0.06 & 15.41 0.03 & 14.97 0.05 & 14.32 0.05 &  AT (ANDOR) \\
55828.27 & 115.27 & ... & 16.25 0.03 & 15.46 0.02 & 14.91 0.01 & 14.27 0.01 &  AT (ANDOR) \\
55830.28 & 117.28 & ... & 16.30 0.02 & 15.38 0.01 & 14.91 0.01 & 14.34 0.01 &  AS-1.82m (AFOSC) \\
55834.26 & 121.26 & ... & 16.18 0.03 & 15.55 0.03 & 15.03 0.01 & 14.38 0.02 &  AT (ANDOR) \\
55834.31 & 121.31 & ... & 16.34 0.02 & 15.56 0.01 & 15.00 0.02 & 14.42 0.02 &  AS-Schmidt (SBIG) \\
55838.34 & 125.34 & ... & ... & 15.64 0.02 & 15.12 0.03 & 14.49 0.01 &  AS-Schmidt (SBIG) \\
55839.28 & 126.28 & ... & 16.37 0.03 & 15.65 0.02 & 15.13 0.02 & 14.52 0.02 &  AS-Schmidt (SBIG) \\
55840.26 & 127.26 & ... & 16.37 0.03 & 15.62 0.02 & 15.18 0.02 & 14.47 0.02 &  AT (ANDOR) \\
55840.30 & 127.30 & ... & 16.42 0.14 & 15.93 0.06 & 15.09 0.04 & 14.60 0.04 &  AS-Schmidt (SBIG) \\
55846.26 & 133.26 & ... & 16.54 0.03 & 15.85 0.03 & 15.24 0.02 & 14.66 0.02 &  AT (ANDOR) \\
55847.30 & 134.30 & ... & ... & 15.85 0.03 & 15.39 0.02 & 14.71 0.02 &  AT (ANDOR) \\
55849.26 & 136.26 & ... & 16.61 0.04 & 15.86 0.03 & 15.38 0.02 & 14.69 0.02 &  AT (ANDOR) \\
55853.27 & 140.27 & ... & 16.30 0.06 & 15.90 0.05 & 15.45 0.05 & 14.91 0.04 &  AS-Schmidt (SBIG) \\
55855.38 & 142.38 & ... & ... & 16.02 0.03 & 15.50 0.03 & 14.89 0.03 &  AT (ANDOR) \\
55856.24 & 143.24 & ... & 16.68 0.05 & 16.08 0.03 & 15.49 0.02 & 14.86 0.02 &  AT (ANDOR) \\
55858.29 & 145.29 & ... & ... & 16.10 0.03 & 15.52 0.02 & 14.92 0.03 &  AT (ANDOR) \\
55859.23 & 146.23 & ... & 16.79 0.04 & 16.14 0.03 & 15.62 0.03 & 14.96 0.03 &  AT (ANDOR) \\
55860.22 & 147.22 & ... & 16.76 0.04 & 16.17 0.04 & 15.60 0.02 & 14.97 0.02 &  AT (ANDOR) \\
55864.69 & 151.69 & 17.55 0.09 & 16.94 0.02 & 16.14 0.01 & 15.65 0.01 & 15.10 0.01 &  AS-1.82m (AFOSC) \\
55866.28 & 153.28 & ... & ... & 16.30 0.03 & 15.71 0.02 & 15.14 0.02 &  AT (ANDOR) \\
55867.70 & 154.70 & ... & 16.97 0.02 & 16.22 0.03 & 15.73 0.02 & 15.26 0.02 &  CA-2.2m (CAFOS) \\
55879.66 & 166.66 & ... & 17.17 0.03 & 16.58 0.02 & 16.00 0.02 & 15.48 0.02 &  AS-Schmidt (SBIG) \\
55881.74 & 168.74 & ... & 17.23 0.02 & 16.59 0.02 & 15.95 0.01 & 15.67 0.03 &  CA-2.2m (CAFOS) \\
55883.24 & 170.24 & ... & ... & 16.76 0.06 & 16.24 0.04 & 15.55 0.03 &  AT (ANDOR) \\
55885.21 & 172.21 & ... & 17.55 0.08 & 16.58 0.04 & 16.04 0.04 & 15.58 0.03 &  AT (ANDOR) \\
55885.73 & 172.73 & ... & 17.36 0.08 & 16.67 0.02 & ... & 15.59 0.03 &  AS-1.82m (AFOSC) \\
55886.75 & 173.75 & 17.99 0.03 & 17.42 0.01 & 16.79 0.01 & 16.19 0.01 & 15.72 0.01 &  NOT (ALFOSC) \\
55893.71 & 180.71 & ... & ... & 16.93 0.02 & 16.29 0.01 & 15.78 0.02 &  AS-Schmidt (SBIG) \\
55894.76 & 181.76 & 18.15 0.02 & 17.57 0.01 & 16.96 0.01 & 16.33 0.01 & 15.88 0.01 &  NOT (ALFOSC) \\
55896.20 & 183.20 & ... & ... & 17.08 0.06 & 16.50 0.05 & 15.97 0.04 &  AT (ANDOR) \\
55898.19 & 185.19 & ... & 17.46 0.06 & 17.14 0.05 & 16.43 0.04 & 15.86 0.03 &  AT (ANDOR) \\
55898.73 & 185.73 & 18.24 0.05 & 17.65 0.01 & 17.09 0.01 & 16.39 0.01 & 16.03 0.01 &  NOT (ALFOSC) \\
55903.76 & 190.76 & 18.27 0.04 & 17.76 0.02 & 17.19 0.02 & 16.51 0.02 & 16.10 0.02 &  NOT (ALFOSC) \\
55912.79 & 199.79 & 18.46 0.04 & 17.91 0.02 & 17.37 0.02 & 16.72 0.01 & 16.32 0.01 &  NOT (ALFOSC) \\
55917.79 & 204.79 & 18.54 0.04 & 17.99 0.01 & 17.48 0.01 & 16.75 0.01 & 16.39 0.01 &  NOT (ALFOSC) \\
55922.76 & 209.76 & 18.71 0.04 & 18.09 0.01 & 17.56 0.01 & 16.84 0.01 & 16.48 0.01 &  NOT (ALFOSC) \\
55932.79 & 219.79 & 18.86 0.06 & 18.23 0.01 & 17.79 0.02 & 17.00 0.02 & 16.70 0.02 &  NOT (ALFOSC) \\
55939.73 & 226.73 & 18.97 0.05 & 18.39 0.01 & 17.92 0.02 & 17.12 0.01 & 16.82 0.01 &  NOT (ALFOSC) \\
55948.73 & 235.73 & 19.14 0.04 & 18.55 0.01 & 18.09 0.02 & 17.26 0.01 & 17.00 0.01 &  NOT (ALFOSC) \\
55955.76 & 242.76 & 19.28 0.04 & 18.70 0.01 & 18.19 0.01 & 17.36 0.01 & 17.13 0.01 &  NOT (ALFOSC) \\
55975.69 & 262.69 & 19.60 0.05 & 18.98 0.01 & 18.59 0.01 & 17.67 0.01 & 17.45 0.01 &  NOT (ALFOSC) \\
55982.74 & 269.74 & 19.87 0.07 & 19.08 0.02 & 18.69 0.02 & 17.75 0.01 & 17.62 0.01 &  NOT (ALFOSC) \\
55986.62 & 273.62 & 19.88 0.11 & 19.15 0.03 & 18.75 0.03 & 17.78 0.02 & 17.85 0.03 &  CA-2.2m (CAFOS) \\
55987.62 & 274.62 & 19.59 0.12 & 19.10 0.02 & 18.77 0.02 & 17.78 0.01 & 17.76 0.01 &  LT (RATCam) \\
55998.66 & 285.66 & ... & 19.34 0.04 & 18.97 0.04 & 18.00 0.02 & 17.88 0.02 &  LT (RATCam) \\
55998.67 & 285.67 & ... & 19.39 0.02 & 18.96 0.02 & 18.01 0.01 & 17.90 0.01 &  NOT (ALFOSC) \\
56008.66 & 295.66 & ... & 19.39 0.03 & 19.10 0.03 & 18.07 0.02 & 18.13 0.03 &  LT (RATCam) \\
\hline
\end{tabular}}

\end{center}
\label{t_jc}
\end{table*}

\setcounter{table}{3}
\begin{table*}[p]
\caption{Continued.}
\begin{center}
\scalebox{1.00}{
\begin{tabular}{l l l l l l l l }
\hline\hline \\ [-1.5ex]
JD (+2400000) & Phase & $U$ & $B$ & $V$ & $R$ & $I$ & Telescope (Instrument)\\ [0.5ex]
(d) & (d) & (mag) & (mag) & (mag) & (mag) & (mag) & \\
\hline \\ [-1.5ex]
56014.51 & 301.51 & 20.34 0.07 & 19.62 0.01 & 19.22 0.02 & 18.25 0.01 & 18.17 0.01 &  NOT (ALFOSC) \\
56026.49 & 313.49 & ... & 19.83 0.02 & 19.41 0.02 & 18.47 0.01 & 18.37 0.02 &  NOT (ALFOSC) \\
56043.59 & 330.59 & 20.77 0.05 & 20.10 0.02 & 19.75 0.02 & 18.72 0.01 & 18.67 0.02 &  NOT (ALFOSC) \\
56071.42 & 358.42 & ... & 20.55 0.02 & 20.22 0.03 & 19.17 0.02 & 19.29 0.03 &  NOT (ALFOSC) \\
56087.43 & 374.43 & ... & 20.78 0.02 & 20.37 0.03 & 19.42 0.02 & 19.54 0.02 &  NOT (ALFOSC) \\
56096.48 & 383.48 & 21.28 0.08 & 20.87 0.03 & 20.52 0.03 & 19.49 0.02 & 19.72 0.03 &  NOT (ALFOSC) \\
56115.44 & 402.44 & ... & 21.13 0.03 & 20.82 0.04 & 19.86 0.02 & 20.06 0.04 &  NOT (ALFOSC) \\
56132.43 & 419.43 & ... & 21.27 0.05 & 20.99 0.05 & 20.14 0.03 & 20.35 0.05 &  NOT (ALFOSC) \\
56133.40 & 420.40 & ... & 21.43 0.05 & 20.99 0.05 & 20.12 0.03 & ... &  NOT (ALFOSC) \\
56154.39 & 441.39 & ... & 21.63 0.04 & 21.28 0.05 & 20.49 0.03 & 20.79 0.06 &  NOT (ALFOSC) \\
56180.37 & 467.37 & ... & 22.01 0.04 & 21.58 0.05 & 20.81 0.03 & 21.14 0.11 &  NOT (ALFOSC) \\
56313.73 & 600.73 & ... & ... & 22.44 0.10 & ... & ... &  NOT (ALFOSC) \\
56353.50 & 640.50 & ... & ... & 23.02 0.00 & ... & 22.58 0.00 &  HST (ACS) \\
56371.69 & 658.69 & ... & 23.42 0.32 & ... & ... & ... &  NOT (ALFOSC) \\
56397.64 & 684.64 & ... & ... & 23.20 0.20 & ... & ... &  NOT (ALFOSC) \\
56445.43 & 732.43 & ... & 23.96 0.50 & ... & ... & ... &  NOT (ALFOSC) \\
\hline
\end{tabular}}

\end{center}
\end{table*}

\begin{table*}[p]
\caption{Late-time (after day 100) optical colour-corrected SDSS $u$ and S-corrected SDSS $griz$ magnitudes for SN 2011dh. Errors are given in parentheses.}
\begin{center}
\scalebox{1.00}{
\begin{tabular}{l l l l l l l l }
\hline\hline \\ [-1.5ex]
JD (+2400000) & Phase & $u$ & $g$ & $r$ & $i$ & $z$ & Telescope (Instrument)\\ [0.5ex]
(d) & (d) & (mag) & (mag) & (mag) & (mag) & (mag) & \\
\hline \\ [-1.5ex]
55886.75 & 173.75 & 18.76 0.04 & 16.99 0.01 & 16.37 0.01 & 16.14 0.01 & ... &  NOT (ALFOSC) \\
55894.76 & 181.76 & 18.92 0.03 & 17.18 0.02 & 16.51 0.01 & 16.27 0.01 & 16.29 0.02 &  NOT (ALFOSC) \\
55898.73 & 185.73 & 19.02 0.04 & 17.23 0.01 & 16.58 0.01 & 16.39 0.01 & 16.47 0.02 &  NOT (ALFOSC) \\
55903.76 & 190.76 & 19.13 0.08 & 17.33 0.02 & 16.70 0.01 & 16.45 0.01 & 16.73 0.04 &  NOT (ALFOSC) \\
55912.79 & 199.79 & 19.29 0.06 & 17.49 0.02 & 16.90 0.01 & 16.64 0.01 & 16.74 0.04 &  NOT (ALFOSC) \\
55922.77 & 209.77 & 19.47 0.02 & 17.71 0.01 & 17.03 0.01 & 16.79 0.01 & 17.13 0.01 &  NOT (ALFOSC) \\
55932.79 & 219.79 & 19.62 0.05 & 17.82 0.02 & 17.20 0.02 & 16.96 0.02 & 17.33 0.04 &  NOT (ALFOSC) \\
55939.74 & 226.74 & 19.76 0.05 & 18.03 0.01 & 17.31 0.01 & 17.07 0.01 & 17.42 0.02 &  NOT (ALFOSC) \\
55948.73 & 235.73 & 19.93 0.03 & 18.15 0.01 & 17.45 0.01 & 17.22 0.01 & 17.67 0.01 &  NOT (ALFOSC) \\
55955.76 & 242.76 & 20.08 0.03 & 18.29 0.01 & 17.56 0.01 & 17.34 0.01 & 17.86 0.02 &  NOT (ALFOSC) \\
55975.69 & 262.69 & 20.32 0.04 & 18.63 0.01 & 17.84 0.01 & 17.65 0.01 & 18.19 0.02 &  NOT (ALFOSC) \\
55982.74 & 269.74 & 20.68 0.06 & 18.74 0.01 & 17.94 0.01 & 17.83 0.01 & 18.47 0.03 &  NOT (ALFOSC) \\
55987.62 & 274.62 & 20.49 0.11 & 18.76 0.01 & 17.95 0.01 & 17.94 0.01 & 18.29 0.03 &  LT (RATCam) \\
55998.66 & 285.66 & ... & 19.02 0.03 & 18.15 0.02 & 18.06 0.02 & 18.74 0.06 &  LT (RATCam) \\
56008.66 & 295.66 & ... & 19.13 0.02 & 18.22 0.01 & 18.31 0.03 & 19.26 0.10 &  LT (RATCam) \\
56014.52 & 301.52 & 21.08 0.05 & 19.27 0.01 & 18.39 0.01 & 18.41 0.01 & 19.01 0.02 &  NOT (ALFOSC) \\
56043.60 & 330.60 & 21.53 0.04 & 19.77 0.01 & 18.81 0.01 & 18.91 0.01 & 19.51 0.03 &  NOT (ALFOSC) \\
56071.43 & 358.43 & ... & 20.20 0.02 & 19.23 0.02 & 19.56 0.03 & 19.98 0.05 &  NOT (ALFOSC) \\
56096.49 & 383.49 & 22.07 0.07 & 20.51 0.02 & 19.52 0.02 & 20.00 0.03 & 20.16 0.05 &  NOT (ALFOSC) \\
56132.43 & 419.43 & ... & 20.99 0.03 & 20.16 0.02 & 20.65 0.04 & 21.25 0.17 &  NOT (ALFOSC) \\
56133.41 & 420.41 & ... & 21.14 0.04 & 20.13 0.03 & 20.78 0.05 & ... &  NOT (ALFOSC) \\
56154.39 & 441.39 & ... & 21.32 0.04 & 20.51 0.03 & 21.10 0.06 & 21.17 0.14 &  NOT (ALFOSC) \\
56159.38 & 446.38 & ... & ... & ... & ... & 21.43 0.13 &  NOT (ALFOSC) \\
56313.75 & 600.75 & ... & ... & 22.46 0.11 & ... & ... &  NOT (ALFOSC) \\
56428.46 & 715.46 & ... & ... & 23.10 0.20 & ... & ... &  NOT (ALFOSC) \\
\hline
\end{tabular}}

\end{center}
\label{t_sloan}
\end{table*}

\begin{table*}[p]
\caption{Late-time (after day 100) NIR S-corrected 2MASS $JHK$ magnitudes for SN 2011dh. Errors are given in parentheses.}
\begin{center}
\scalebox{0.80}{
\begin{tabular}{l l l l l l }
\hline\hline \\ [-1.5ex]
JD (+2400000) & Phase & $J$ & $H$ & $K$ & Telescope (Instrument)\\ [0.5ex]
(d) & (d) & (mag) & (mag) & (mag) & \\
\hline \\ [-1.5ex]
55814.32 & 101.32 & 14.38 0.01 & 13.80 0.01 & 13.50 0.01 &  CA-3.5m (O2000) \\
55818.36 & 105.36 & 14.45 0.02 & 13.91 0.01 & 13.74 0.01 &  NOT (NOTCAM) \\
55880.72 & 167.72 & 16.23 0.01 & 15.38 0.01 & 14.70 0.01 &  CA-3.5m (O2000) \\
55913.68 & 200.68 & 17.00 0.01 & 16.19 0.02 & 15.31 0.02 &  CA-3.5m (O2000) \\
55914.66 & 201.66 & 17.05 0.01 & 16.23 0.02 & 15.35 0.02 &  CA-3.5m (O2000) \\
55946.13 & 233.13 & 17.43 0.02 & 16.78 0.02 & 16.21 0.02 &  UKIRT (WFCAM) \\
55999.91 & 286.91 & 18.10 0.02 & 17.47 0.02 & 17.31 0.02 &  UKIRT (WFCAM) \\
56024.38 & 311.38 & 18.46 0.03 & 17.80 0.03 & 17.71 0.04 &  WHT (LIRIS) \\
56052.47 & 339.47 & 18.69 0.02 & 17.96 0.02 & 18.60 0.03 &  WHT (LIRIS) \\
56093.48 & 380.48 & 19.27 0.06 & 18.40 0.07 & 19.02 0.09 &  WHT (LIRIS) \\
\hline
\end{tabular}}

\end{center}
\label{t_nir}
\end{table*}

\begin{table*}[p]
\caption{Late-time (after day 100) MIR Spitzer $S_1$ and $S_2$ magnitudes for SN 2011dh. Errors are given in parentheses.}
\begin{center}
\scalebox{0.80}{
\begin{tabular}{l l l l l }
\hline\hline \\ [-1.5ex]
JD (+2400000) & Phase & $S_1$ & $S_2$ & Telescope (Instrument)\\ [0.5ex]
(d) & (d) & (mag) & (mag) & \\
\hline \\ [-1.5ex]
55964.14 & 251.14 & 14.300 0.002 & 13.280 0.002 &  SPITZER (IRAC) \\
56026.63 & 313.63 & 15.536 0.007 & 14.280 0.003 &  SPITZER (IRAC) \\
56104.23 & 391.23 & 17.025 0.012 & 15.507 0.005 &  SPITZER (IRAC) \\
56136.41 & 423.41 & 17.337 0.014 & 15.988 0.006 &  SPITZER (IRAC) \\
56168.69 & 455.69 & 17.581 0.014 & 16.241 0.008 &  SPITZER (IRAC) \\
56337.59 & 624.59 & 18.517 0.083 & 17.537 0.056 &  SPITZER (IRAC) \\
56360.27 & 647.27 & 18.675 0.045 & 17.663 0.030 &  SPITZER (IRAC) \\
56477.83 & 764.83 & 18.833 0.047 & 18.047 0.039 &  SPITZER (IRAC) \\
56742.28 & 1029.28 & ... & 18.506 0.091 &  SPITZER (IRAC) \\
56742.30 & 1029.30 & ... & 18.410 0.075 &  SPITZER (IRAC) \\
56771.67 & 1058.67 & 18.858 0.077 & 18.543 0.070 &  SPITZER (IRAC) \\
56773.84 & 1060.84 & ... & 18.364 0.058 &  SPITZER (IRAC) \\ [0.5ex]
\hline
\end{tabular}}

\end{center}
\label{t_mir}
\end{table*}

\begin{table*}[p]
\caption{List of late-time (after day 100) optical and NIR spectroscopic observations.}
\begin{center}
\scalebox{0.8}{
\begin{tabular}{l l l l l l l}
\hline\hline \\ [-1.5ex]
JD (+2400000) & Phase & Grism & Range & Resolution & Resolution & Telescope (Instrument)\\ [0.5ex]
(d) & (d) & & (\AA) & & (\AA) &\\ [0.5ex]
\hline \\ [-1.5ex]
55821.33 & 108.33 & b200 & 3300-8700 & ... & 12.0 & CA-2.2m (CAFOS) \\
55821.33 & 108.33 & r200 & 6300-10500 & ... & 12.0 & CA-2.2m (CAFOS) \\
55828.35 & 115.35 & R300B & 3200-5300 & ... & 4.1 & WHT (ISIS) \\
55828.35 & 115.35 & R158R & 5300-10000 & ... & 7.7 & WHT (ISIS) \\
55830.25 & 117.25 & Grism 4 & 3500-8450 & 292 & 19.9 & AS 1.82m (AFOSC) \\
55830.28 & 117.28 & Grism 2 & 3720-10200 & 191 & 37.6 & AS 1.82m (AFOSC) \\
55835.25 & 122.25 & Grism 4 & 3500-8450 & 292 & 19.9 & AS 1.82m (AFOSC) \\
55864.65 & 151.65 & Grism 4 & 3500-8450 & 292 & 19.9 & AS 1.82m (AFOSC) \\
55867.71 & 154.71 & g200 & 4900-9900 & ... & 12.0 & CA-2.2m (CAFOS) \\
55893.76 & 180.76 & Grism 3 & 3200-6700 & 345 & 12.4 & NOT (ALFOSC) \\
55897.76 & 184.76 & Grism 5 & 5000-10250 & 415 & 16.8 & NOT (ALFOSC) \\
55911.20 & 198.20 & zJ & 8900-15100 & 700 & ... & WHT (LIRIS) \\
55914.70 & 201.70 & R300B & 3200-5300 & ... & 8.2 & WHT (ISIS) \\
55914.70 & 201.70 & R158R & 5300-10000 & ... & 15.4 & WHT (ISIS) \\
55918.69 & 205.69 & HK & 14000-25000 & 333 & ... & TNG (NICS) \\
55941.72 & 228.72 & R150V & 4000-9500 & ... & 12.9 & INT (IDS) \\
55942.73 & 229.73 & R150V & 4000-9500 & ... & 12.9 & INT (IDS) \\
55944.75 & 231.75 & R150V & 4000-9500 & ... & 12.9 & INT (IDS) \\
55951.64 & 238.64 & Grism 4 & 3200-9100 & 355 & 16.2 & NOT (ALFOSC) \\
55998.68 & 285.68 & r200 & 6300-10500 & ... & 12.0 & CA-2.2m (CAFOS) \\
56002.58 & 289.58 & Grism 4 & 3200-9100 & 355 & 16.2 & NOT (ALFOSC) \\
56005.62 & 292.62 & Grism 4 & 3200-9100 & 355 & 16.2 & NOT (ALFOSC) \\
56006.60 & 293.60 & Grism 4 & 3200-9100 & 355 & 16.2 & NOT (ALFOSC) \\
56013.14 & 300.14 & R600B & 3680-5300 & ... & 5.7 & WHT (ISIS) \\
56013.14 & 300.14 & R316R & 5756-8850 & ... & 3.0 & WHT (ISIS) \\
56071.56 & 358.56 & R500B & 3440-7600 & 322 & 15.0 & GTC (OSIRIS) \\
56072.61 & 359.61 & R500R & 4800-10000 & 352 & 20.8 & GTC (OSIRIS) \\
56128.47 & 415.47 & R300B & 3600-7000 & 270 & 16.7 & GTC (OSIRIS) \\ [0.5ex]
\hline
\end{tabular}}
\end{center}
\label{t_speclog}
\end{table*}

\begin{table*}[p]
\caption{Pseudo-bolometric UV-to-MIR lightcurve between days 3 and 400 for SN 2011dh calculated from spectroscopic and photometric data with a 1-day sampling between days 3 and 50 and a 5-day sampling between days 50 and 400. Random errors are given in the first parentheses and systematic lower and upper errors (arising from the distance and extinction) in the second parentheses.}
\begin{center}
\scalebox{0.90}{
\begin{tabular}{llllll}
\hline\hline \\ [-1.5ex]
JD (+2400000) & Phase & L & JD (+2400000) & Phase & L \\ [0.5ex]
(d) & (d) & (log erg s$^{-1}$) & (d) & (d) & (log erg s$^{-1}$) \\ [0.5ex]
\hline \\ [-1.5ex]
55717.00 & 4.00 & 41.464 (0.001)  (0.098,0.186) & 55823.00 & 110.00 & 41.244 (0.002)  (0.094,0.164) \\
55718.00 & 5.00 & 41.552 (0.001)  (0.097,0.181) & 55828.00 & 115.00 & 41.204 (0.002)  (0.094,0.164) \\
55719.00 & 6.00 & 41.653 (0.001)  (0.097,0.179) & 55833.00 & 120.00 & 41.163 (0.002)  (0.094,0.164) \\
55720.00 & 7.00 & 41.747 (0.001)  (0.097,0.178) & 55838.00 & 125.00 & 41.121 (0.002)  (0.094,0.164) \\
55721.00 & 8.00 & 41.835 (0.001)  (0.097,0.178) & 55843.00 & 130.00 & 41.078 (0.002)  (0.094,0.164) \\
55722.00 & 9.00 & 41.909 (0.001)  (0.097,0.178) & 55848.00 & 135.00 & 41.033 (0.002)  (0.094,0.165) \\
55723.00 & 10.00 & 41.969 (0.001)  (0.097,0.177) & 55853.00 & 140.00 & 40.990 (0.002)  (0.094,0.165) \\
55724.00 & 11.00 & 42.018 (0.001)  (0.097,0.176) & 55858.00 & 145.00 & 40.948 (0.002)  (0.094,0.165) \\
55725.00 & 12.00 & 42.057 (0.001)  (0.097,0.176) & 55863.00 & 150.00 & 40.906 (0.002)  (0.094,0.165) \\
55726.00 & 13.00 & 42.089 (0.001)  (0.096,0.175) & 55868.00 & 155.00 & 40.863 (0.002)  (0.094,0.165) \\
55727.00 & 14.00 & 42.117 (0.001)  (0.096,0.174) & 55873.00 & 160.00 & 40.818 (0.002)  (0.094,0.165) \\
55728.00 & 15.00 & 42.142 (0.001)  (0.096,0.174) & 55878.00 & 165.00 & 40.773 (0.002)  (0.094,0.164) \\
55729.00 & 16.00 & 42.163 (0.001)  (0.096,0.173) & 55883.00 & 170.00 & 40.728 (0.002)  (0.094,0.164) \\
55730.00 & 17.00 & 42.182 (0.001)  (0.096,0.173) & 55888.00 & 175.00 & 40.683 (0.002)  (0.094,0.164) \\
55731.00 & 18.00 & 42.197 (0.001)  (0.096,0.173) & 55893.00 & 180.00 & 40.639 (0.001)  (0.094,0.164) \\
55732.00 & 19.00 & 42.208 (0.001)  (0.096,0.172) & 55898.00 & 185.00 & 40.597 (0.001)  (0.094,0.164) \\
55733.00 & 20.00 & 42.214 (0.001)  (0.096,0.172) & 55903.00 & 190.00 & 40.555 (0.001)  (0.094,0.164) \\
55734.00 & 21.00 & 42.215 (0.001)  (0.096,0.171) & 55908.00 & 195.00 & 40.516 (0.001)  (0.094,0.164) \\
55735.00 & 22.00 & 42.211 (0.001)  (0.095,0.171) & 55913.00 & 200.00 & 40.477 (0.001)  (0.094,0.164) \\
55736.00 & 23.00 & 42.200 (0.001)  (0.095,0.170) & 55918.00 & 205.00 & 40.438 (0.001)  (0.094,0.164) \\
55737.00 & 24.00 & 42.184 (0.001)  (0.095,0.168) & 55923.00 & 210.00 & 40.403 (0.001)  (0.094,0.164) \\
55738.00 & 25.00 & 42.164 (0.001)  (0.095,0.167) & 55928.00 & 215.00 & 40.368 (0.001)  (0.094,0.164) \\
55739.00 & 26.00 & 42.141 (0.001)  (0.094,0.166) & 55933.00 & 220.00 & 40.334 (0.001)  (0.094,0.164) \\
55740.00 & 27.00 & 42.116 (0.001)  (0.094,0.165) & 55938.00 & 225.00 & 40.300 (0.001)  (0.094,0.164) \\
55741.00 & 28.00 & 42.090 (0.001)  (0.094,0.164) & 55943.00 & 230.00 & 40.267 (0.001)  (0.094,0.164) \\
55742.00 & 29.00 & 42.064 (0.001)  (0.094,0.163) & 55948.00 & 235.00 & 40.235 (0.001)  (0.094,0.163) \\
55743.00 & 30.00 & 42.039 (0.001)  (0.094,0.162) & 55953.00 & 240.00 & 40.204 (0.001)  (0.094,0.163) \\
55744.00 & 31.00 & 42.016 (0.001)  (0.093,0.162) & 55958.00 & 245.00 & 40.172 (0.001)  (0.094,0.163) \\
55745.00 & 32.00 & 41.995 (0.001)  (0.093,0.161) & 55963.00 & 250.00 & 40.139 (0.001)  (0.094,0.163) \\
55746.00 & 33.00 & 41.976 (0.001)  (0.093,0.161) & 55968.00 & 255.00 & 40.106 (0.001)  (0.094,0.163) \\
55747.00 & 34.00 & 41.959 (0.001)  (0.093,0.160) & 55973.00 & 260.00 & 40.072 (0.001)  (0.094,0.163) \\
55748.00 & 35.00 & 41.942 (0.001)  (0.093,0.160) & 55978.00 & 265.00 & 40.039 (0.002)  (0.094,0.163) \\
55749.00 & 36.00 & 41.927 (0.001)  (0.093,0.160) & 55983.00 & 270.00 & 40.005 (0.002)  (0.094,0.163) \\
55750.00 & 37.00 & 41.913 (0.001)  (0.093,0.159) & 55988.00 & 275.00 & 39.972 (0.002)  (0.094,0.163) \\
55751.00 & 38.00 & 41.899 (0.001)  (0.093,0.159) & 55993.00 & 280.00 & 39.939 (0.002)  (0.094,0.163) \\
55752.00 & 39.00 & 41.885 (0.001)  (0.093,0.159) & 55998.00 & 285.00 & 39.906 (0.002)  (0.094,0.163) \\
55753.00 & 40.00 & 41.873 (0.001)  (0.093,0.159) & 56003.00 & 290.00 & 39.872 (0.002)  (0.094,0.163) \\
55754.00 & 41.00 & 41.859 (0.001)  (0.093,0.159) & 56008.00 & 295.00 & 39.838 (0.002)  (0.094,0.163) \\
55755.00 & 42.00 & 41.846 (0.001)  (0.093,0.159) & 56013.00 & 300.00 & 39.804 (0.002)  (0.094,0.163) \\
55756.00 & 43.00 & 41.834 (0.001)  (0.093,0.159) & 56018.00 & 305.00 & 39.770 (0.002)  (0.094,0.163) \\
55757.00 & 44.00 & 41.822 (0.002)  (0.093,0.159) & 56023.00 & 310.00 & 39.737 (0.002)  (0.094,0.163) \\
55758.00 & 45.00 & 41.811 (0.002)  (0.093,0.159) & 56028.00 & 315.00 & 39.703 (0.002)  (0.094,0.163) \\
55759.00 & 46.00 & 41.800 (0.001)  (0.093,0.159) & 56033.00 & 320.00 & 39.670 (0.002)  (0.094,0.163) \\
55760.00 & 47.00 & 41.790 (0.001)  (0.093,0.159) & 56038.00 & 325.00 & 39.638 (0.002)  (0.094,0.163) \\
55761.00 & 48.00 & 41.781 (0.001)  (0.093,0.159) & 56043.00 & 330.00 & 39.605 (0.002)  (0.094,0.163) \\
55762.00 & 49.00 & 41.771 (0.002)  (0.093,0.159) & 56048.00 & 335.00 & 39.572 (0.002)  (0.094,0.163) \\
55763.00 & 50.00 & 41.762 (0.002)  (0.093,0.159) & 56053.00 & 340.00 & 39.540 (0.002)  (0.094,0.163) \\
55768.00 & 55.00 & 41.715 (0.002)  (0.093,0.159) & 56058.00 & 345.00 & 39.508 (0.002)  (0.094,0.163) \\
55773.00 & 60.00 & 41.669 (0.002)  (0.093,0.159) & 56063.00 & 350.00 & 39.476 (0.002)  (0.094,0.163) \\
55778.00 & 65.00 & 41.627 (0.002)  (0.093,0.160) & 56068.00 & 355.00 & 39.445 (0.002)  (0.094,0.163) \\
55783.00 & 70.00 & 41.585 (0.002)  (0.093,0.160) & 56073.00 & 360.00 & 39.413 (0.002)  (0.094,0.163) \\
55788.00 & 75.00 & 41.544 (0.002)  (0.093,0.161) & 56078.00 & 365.00 & 39.382 (0.002)  (0.094,0.163) \\
55793.00 & 80.00 & 41.501 (0.002)  (0.093,0.162) & 56083.00 & 370.00 & 39.351 (0.002)  (0.094,0.163) \\
55798.00 & 85.00 & 41.458 (0.002)  (0.094,0.162) & 56088.00 & 375.00 & 39.320 (0.002)  (0.094,0.163) \\
55803.00 & 90.00 & 41.415 (0.002)  (0.094,0.162) & 56093.00 & 380.00 & 39.289 (0.002)  (0.094,0.163) \\
55808.00 & 95.00 & 41.372 (0.002)  (0.094,0.163) & 56098.00 & 385.00 & 39.255 (0.002)  (0.094,0.163) \\
55813.00 & 100.00 & 41.330 (0.002)  (0.094,0.163) & 56103.00 & 390.00 & 39.220 (0.002)  (0.094,0.163) \\
55818.00 & 105.00 & 41.287 (0.002)  (0.094,0.163) & 56108.00 & 395.00 & 39.186 (0.002)  (0.094,0.164) \\ [0.5ex]
\hline
\end{tabular}}

\end{center}
\label{t_UV_MIR_bol}
\end{table*}

\appendix

\section{Data reductions and calibration}
\label{a_data_reductions}

\paragraph{Template subtraction} The optical and NIR images obtained between days 100 and 500 have been template subtracted after day 300, before which comparison of photometry on original and template subtracted images shows that the background contamination is negligible. The optical templates were constructed by point spread function (PSF) subtraction of the SN from observations acquired after day 600, and the NIR templates by PSF subtraction of the SN from the day 339 WHT observation, which is of excellent quality. For the last day 380 WHT observation we used PSF photometry.

\begin{figure}[tb]
\includegraphics[width=0.48\textwidth,angle=0]{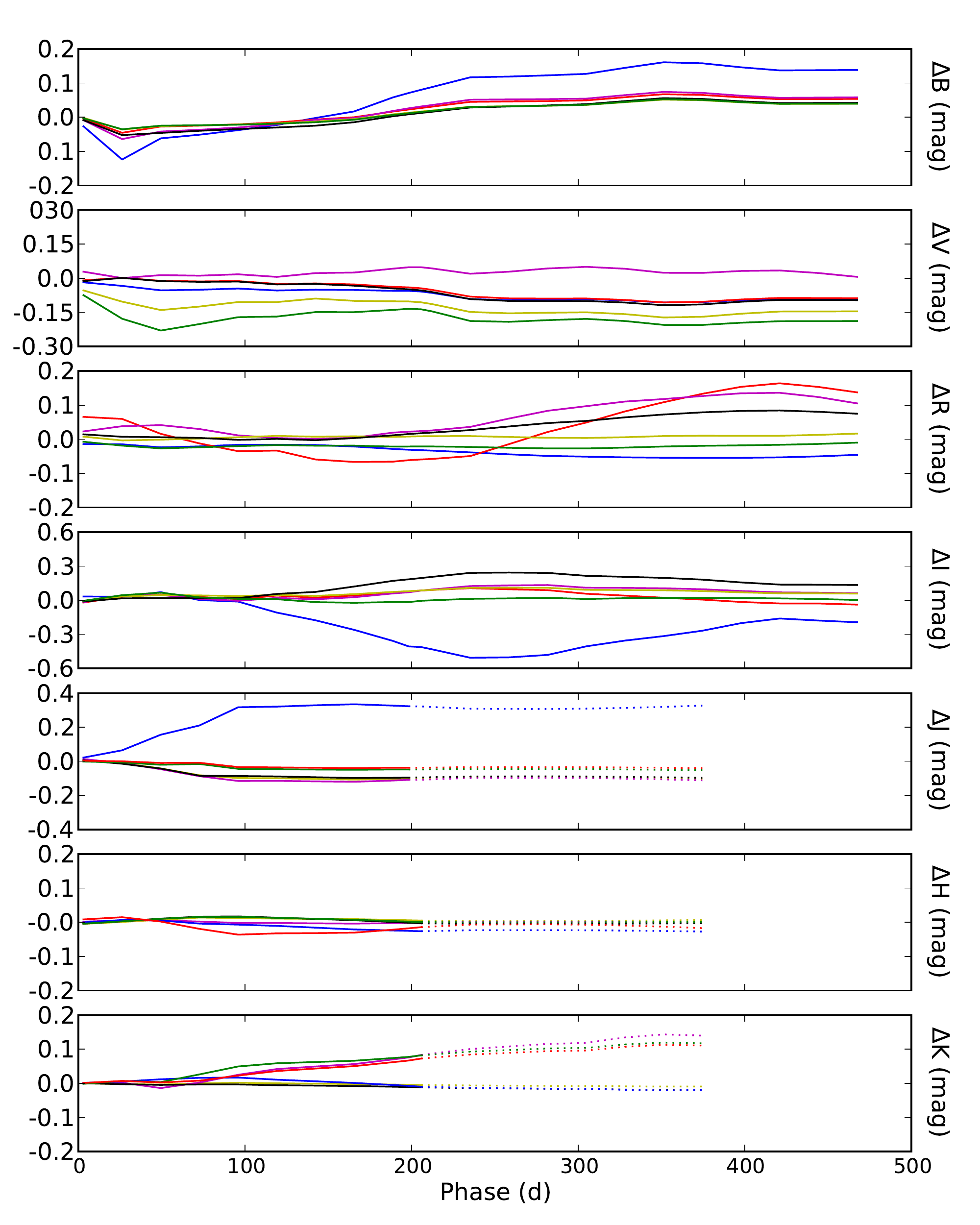}
\caption{Difference between JC $BVRI$ colour and S-corrections for NOT/ALFOSC (black), LT/RATCam (red), CA-2.2m/CAFOS (blue), TNG/LRS (green), AS-1.82m/AFOSC (yellow), and AS-Schmidt (magenta), and difference between 2MASS $JHK$ colour and S-corrections for NOT/NOTCAM (black), TCS/CAIN (red), CA-3.5m/O2000 (blue), TNG/NICS (green), WHT/LIRIS (yellow), and UKIRT/WFCAM (magenta). Results based on extrapolated 2MASS S-corrections are shown as dotted lines.}
\label{f_scorr}
\end{figure}

\paragraph{S-corrections} 
The optical and NIR photometry between days 100 and 500 have been S-corrected, and the accuracy of the photometry depends critically on the accuracy of these corrections. Figure~\ref{f_scorr} shows the difference between colour and S-corrections for the Johnson-Cousins (JC) and 2 Micron All Sky Survey (2MASS) systems for most telescope/instrument combinations used, and S-corrections are clearly necessary for accurate photometry. In some cases, e.g. for the CA-2.2m/CAFOS and NOT/ALFOSC $I$ band and the CA-3.5m/O2000 $J$ band, the differences become as large as 0.3$-$0.5 mag. In particular, the difference between the NOT/ALFOSC and CA-2.2m/CAFOS $I$-band observations are $\sim$0.8 mag at day $\sim$250, mainly because of the strong [\ion{Ca}{ii}] 7291,7323 \AA~and \ion{Ca}{ii} 8498,8542,8662 \AA~lines. As the spectral NIR coverage ends at day $\sim$200, we have assumed that the 2MASS S-corrections do not change after this epoch. This adds uncertainty to the 2MASS photometry after day $\sim$200, but as the 2MASS S-corrections, except for the CA-3.5m/O2000 $J$ band, are generally small and evolve slowly, the errors arising from this approximation are probably modest. The accuracy of the S-corrections can be estimated by comparing S-corrected photometry obtained with different telescope/instrument combinations. The late-time JC and Sloan Digital Sky Survey (SDSS) photometry were mainly obtained with the NOT, but comparisons between S-corrected NOT/ALFOSC, LT/RATCam and CA-2.2m/CAFOS photometry at $\sim$300 days, show differences at the 5 per cent level, suggesting that the precision from the period before day 100 is maintained \citepalias{Erg14a}. The late-time 2MASS photometry was obtained with a number of different telescopes, and although the sampling is sparse, the shape of the lightcurves suggests that the errors in the S-corrections are modest. Additional filter response functions for AT and UKIRT have been constructed as outlined in \citetalias{Erg14a}.

\paragraph{Observations after day 600} 
The results for the observations obtained after day 600 have been adopted from the pre-explosion difference imaging presented in \citetalias{Erg14a}, assuming that the remaining flux at the position of the progenitor originates solely from the SN. The observations were S-corrected using the day 678 spectrum of SN 2011dh \citep{Shi13}. Comparing to results from PSF photometry, where we iteratively fitted the PSF subtracted background, we find differences of $\lesssim$0.1 mag. Given that the pre-explosion magnitudes (which were measured with PSF photometry) are correctly measured, the difference-based magnitudes are likely to have less uncertainty. On the other hand, the PSF photometry does not depend on the the pre-explosion magnitudes. The good agreement found using these two, partly independent methods gives confidence in the results. Further confidence is gained by comparing the S-corrected NOT and HST \citep{Dyk13b} $V$-band observations, for which we find a difference of $\lesssim$0.2 mag. The assumption that the flux at the position of the progenitor originates solely from the SN, is supported by the the depth of absorption features and the line dominated nature of the day 678 spectrum.

\paragraph{Spitzer Telescope observations} For the photometry after day 100 we used a small aperture with a 3 pixel radius, and a correction to the standard aperture given in the IRAC Instrument Handbook determined from the images. All images were template subtracted using archival images as described in \citepalias{Erg14a}. After day 100, background contamination becomes important, and template subtraction is necessary to obtain precision in the photometry. The photometry before day 700 previously published by \citet{Hel13} agrees very well with our photometry, the differences being mostly $\lesssim$5 per cent.

\section{Line measurements}
\label{a_line_measurements}

\paragraph{Line emitting regions} To estimate the sizes of the line emitting regions, we fit the line profile of a spherically symmetric region of constant line emissivity, optically thin in the line (no line scattering contribution) and with a constant absorptive continuum opacity, to the observed, continuum subtracted line profile (see below). The fitting is done by an automated least-square based algorithm, and the method gives a rough estimate of the size of the region responsible for the bulk of the line emission. The absorptive continuum opacity is included to mimic blue-shifts caused by obscuration of receding-side emission. Some lines arise as a blend of more than one line, which has to be taken into account. The [\ion{O}{i}] 6300 \AA~flux was calculated by iterative subtraction of the [\ion{O}{i}] 6364 \AA~flux, from the left to the right, using $\mathrm{F_{6300}}(\lambda)=\mathrm{F_{6300,6364}}(\lambda)-\mathrm{F_{6300}}(\lambda-\Delta\lambda)/\mathrm{R}$, where $\Delta\lambda$ is the wavelength separation between the [\ion{O}{i}] 6300 \AA~and 6364 \AA~lines and $\mathrm{R}$ the [\ion{O}{i}] 6300,6364 \AA~line ratio. This ratio was assumed to be 3, as is supported by the preferred \citetalias{Jer14} steady-state NLTE model and estimates based on small scale fluctuations in the line profiles (Sect.~\ref{s_spec_small_scale}). For all other blended lines, we make a simultaneous fit with the line ratios as free parameters, assuming a common size of the line emitting regions. When fitting the [\ion{Ca}{ii}] 7291,7323 \AA~line we exclude the region >3000 km~s$^{-1}$ redwards 7291 \AA, which could be contaminated by the [\ion{Ni}{ii}] 7378,7411 \AA~line \citepalias{Jer14}.

\paragraph{Line asymmetries} To estimate the asymmetry of a line we calculate the first wavelength moment of the flux (center of flux) for the continuum subtracted line profile (see below). The rest wavelength is assumed to be 6316 \AA~and 7304 \AA~for the [\ion{O}{i}] 6300,6364 \AA~and [\ion{Ca}{ii}] 7291,7323 \AA~lines, respectively. This is appropriate for optically thin emission if we assume the upper levels of the [\ion{Ca}{ii}] 7291,7323 \AA~line to be populated as in local thermal equilibrium (LTE). Optically thin emission for these lines is supported by the preferred \citetalias{Jer14} steady-state NLTE model, the absence of absorption features in the observed spectra and the [\ion{O}{i}] 6300,6364 \AA~line ratio (Sect.~\ref{s_spec_small_scale}). The rest wavelength is assumed to be 5896 \AA~and 8662 \AA~for the \ion{Na}{i} 5890,5896 \AA~and \ion{Ca}{ii} 8498,8542,8662 \AA~lines, respectively. This is appropriate for optically thick emission, where the line emission will eventually scatter in the reddest line. Optically thick emission for the \ion{Na}{i} 5896 \AA~line is supported by the preferred \citetalias{Jer14} steady-state NLTE model and the observed P-Cygni profile. For the \ion{Ca}{ii} 8662 \AA~line this assumption is less justified.

\paragraph{Continuum subtraction} Before fitting the line emitting region or calculating the center of flux, the continuum is subtracted. The flux is determined by a linear interpolation between the minimum flux on the blue and red sides of the smoothed line profile. The search region is set to $\pm{6000}$ km~s$^{-1}$ for most of the lines, $\pm{10000}$ km~s$^{-1}$ for the \ion{Ca}{ii} 8662 \AA~line and $\pm{3000}$ km~s$^{-1}$ for the [\ion{Fe}{ii}] 7155 \AA~line.

\paragraph{Small scale fluctuations} To remove the large scale structure from a line profile we iteratively subtract a 1000 km~s$^{-1}$ box average (i.e. the result of each subtraction is fed into the next one). This procedure is repeated three times, and in tests on the product of synthetic large and small scale structures, the small scale structure is recovered with reasonable accuracy. Using a Monte Carlo (MC) analogue of the \citet{Chu94} model the spatial (relative) RMS of the recovered small scale structure is found to agree well with the actual (relative) RMS of the fluctuations \citep[$\delta_{\mathrm{F}}$ in ][Eq.~11]{Chu94}).

\section{Steady-state NLTE modelling.}
\label{a_nlte_modelling}

\begin{figure}[tb]
\includegraphics[width=0.48\textwidth,angle=0]{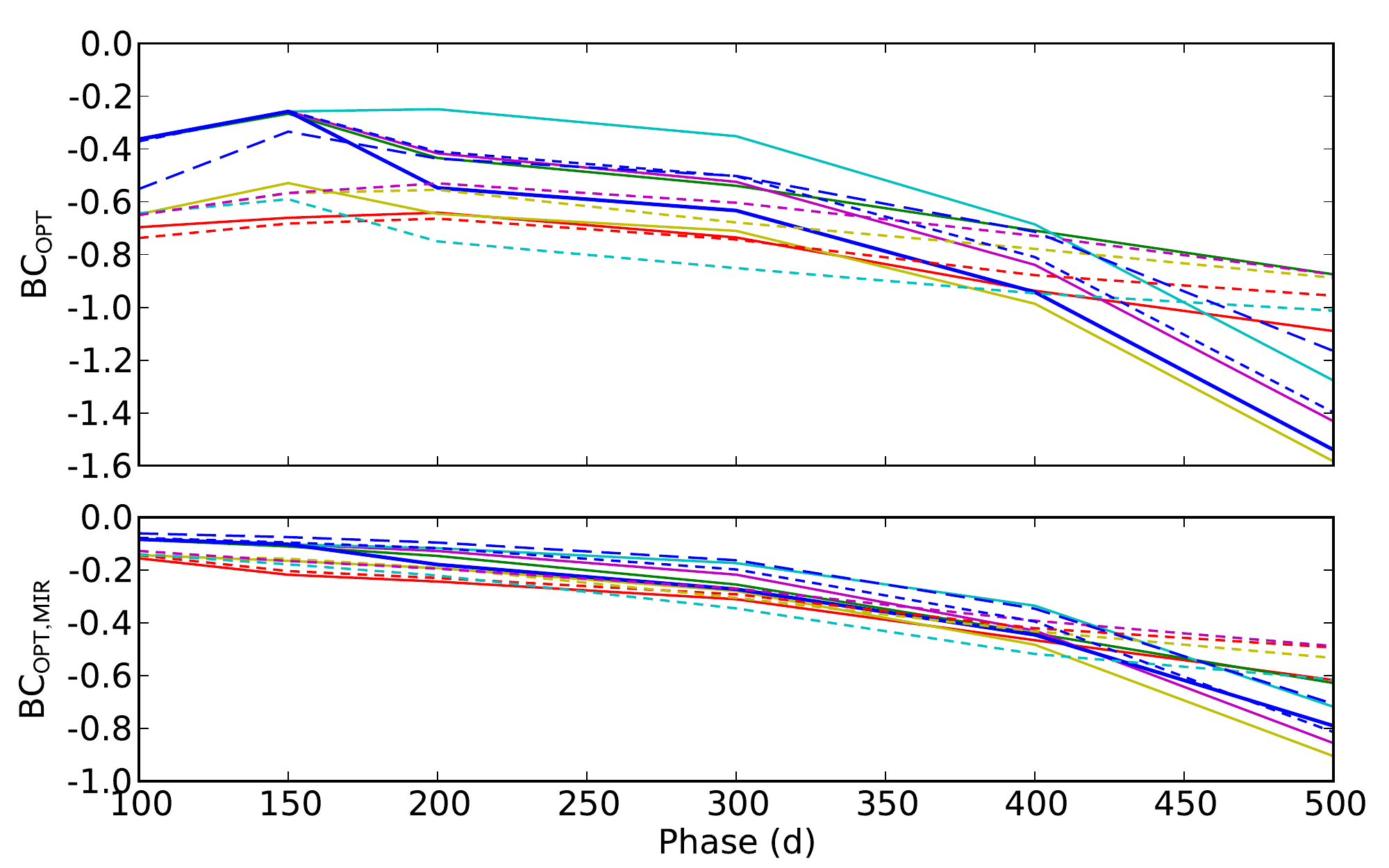}
\caption{Optical (upper panel) and optical-to-MIR (lower panel) BCs between days 100 and 500 for the \citetalias{Jer14} models. The models are shown as in Fig.~\ref{f_opt_nir_mir_mcomp}.}
\label{f_bol_bc_opt_nir_mir_mcomp}
\end{figure}

\begin{figure}[tb]
\includegraphics[width=0.48\textwidth,angle=0]{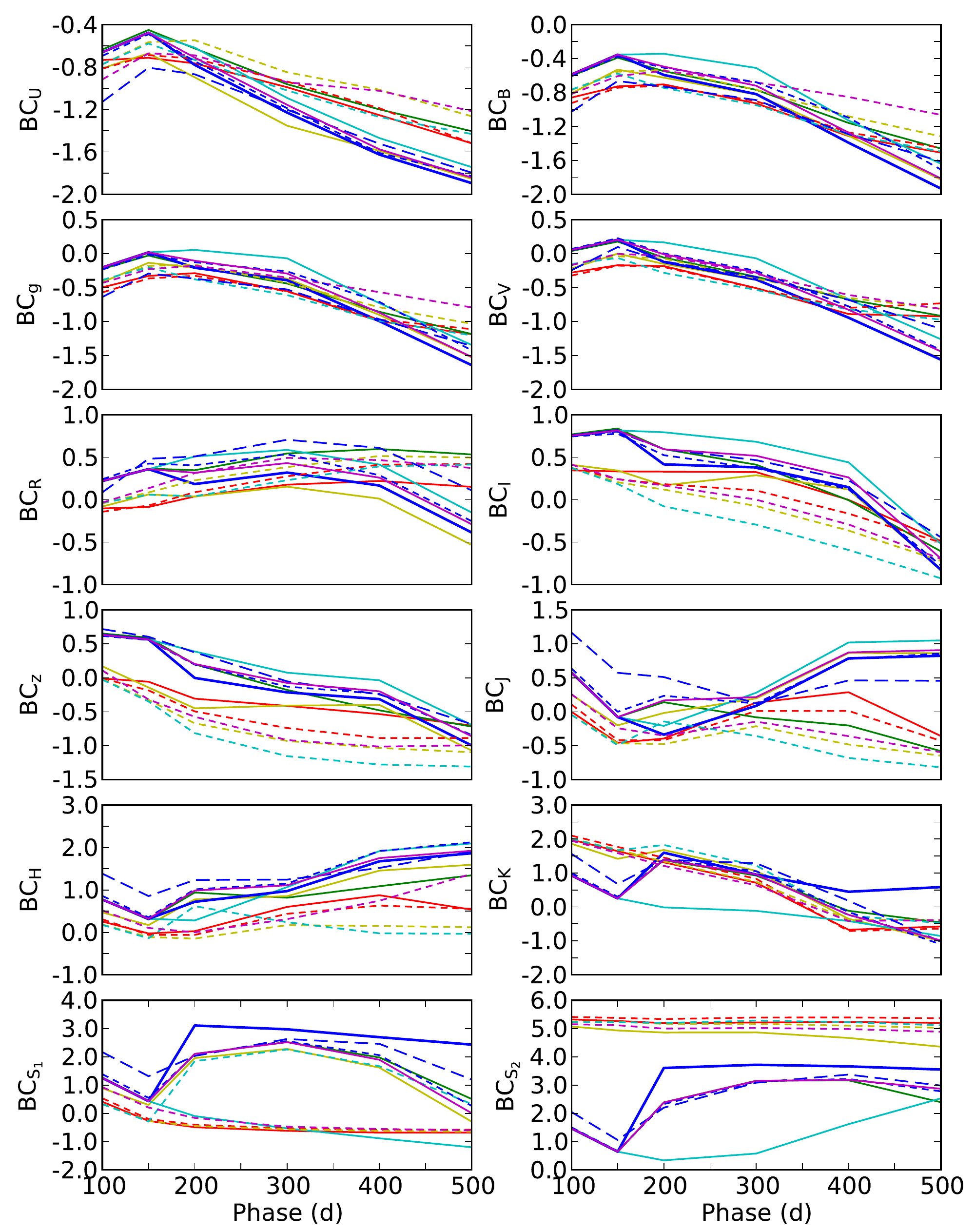}
\caption{Broad-band BCs between days 100 and 500 for the \citetalias{Jer14} models. The models are shown as in Fig.~\ref{f_opt_nir_mir_mcomp}.}
\label{f_bc_opt_nir_mir_mcomp}
\end{figure}

\paragraph{Effects of the model parameters} To analyse the effects of the model parameters on the model lightcurves, a split into a bolometric lightcurve and a bolometric correction (BC) is convenient. In terms of these quantities, the broad-band and pseudo-bolometric\footnote{Pseudo-bolometric luminosity expressed in bolometric magnitudes.} magnitudes are given by $\mathrm{M=M_{Bol}-BC}$. This split is most useful as the bolometric lightcurve depends only on the energy deposition\footnote{Ignoring the <0.05 mag lost due to scattering in the ejecta.}, whereas the BCs depend on how this energy is processed. The energy deposition is independent of molecular cooling and dust and, within the parameter space covered by the \citetalias{Jer14} models, only weakly dependent on the density contrast and the positron trapping. Therefore, the bolometric lightcurves depend significantly only on the initial mass and the macroscopic mixing, whereas the other parameters only significantly affect the BCs. Figures~\ref{f_bol_bc_opt_nir_mir_mcomp} and \ref{f_bc_opt_nir_mir_mcomp} show the pseudo-bolometric and broad-band BCs for the \citetalias{Jer14} models. The optical-to-MIR BCs show small differences whereas the optical and, in particular, the broad-band BCs show considerably larger differences, which makes them well suited to constrain the model parameters. We have investigated the dependence of each BC on each parameter, and in Sects.~\ref{s_modelling_nlte_parameters} and \ref{s_modelling_molecules_dust} we make use of these results. The bolometric lightcurves (not shown) could as well be calculated with HYDE, and their dependence on the initial mass and macroscopic mixing is better investigated with the hydrodynamical model grid.

\paragraph{The optical-to-MIR BC} The optical-to-MIR BCs for the \citetalias{Jer14} models show small differences (<$\pm$0.1 mag) between days 100 and 400, which subsequently increase towards $\pm$0.25 mag at day 500. At day 100 the optical-to-MIR BC is >$-$0.15 mag, which is likely to hold before day 100 as well. In Sect.~\ref{s_modelling_day_0_300}, we take advantage of these facts and use the optical-to-MIR BC for the preferred \citetalias{Jer14} model for all hydrodynamical models. However, as the \citetalias{Jer14} models cover a restricted volume of parameter space as compared to the hydrodynamical model grid, we need to justify this choice further. It is reasonable to assume that the BC depends mainly on the energy deposition per unit mass (determining the heating rate) and the density (determining the cooling rate). Furthermore, we know beforehand that hydrodynamical models giving a bad fit for the period before day 100 will not give a good fit for the period before day 400. Inspecting the hydrodynamical models (for the period before day 100) with a normalized standard deviation in the fit less than 3, we find that these do not span a wide range in (mass averaged) density or energy deposition per mass, and inspecting the \citetalias{Jer14} models we find that these cover about half of this region. Although these quantities evolve quite strongly with time, they scale in a similar way for all models, and this conclusion holds from day 100 to 400. The effect of the steady-state NLTE parameters that do not map onto the hydrodynamical parameter space (dust, molecular cooling, positron trapping, and density contrast) is harder to constrain. The small spread in the BC for the \citetalias{Jer14} models between days 100 and 400, however, make this caveat less worrying. 

\paragraph{Treatment of molecular cooling} Molecular cooling is included in the modelling in a simplified way, and is represented as the fraction of the (radiative and radioactive) heating emitted as molecule (CO and SiO) emission in the O/C and O/Si/S zones. The CO and SiO fundamental and first-overtone band emission is represented as box line profiles between 2.25$-$2.45 (CO first-overtone), 4.4$-$4.9 (CO fundamental), and 4.0$-$4.5 (SiO first-overtone) $\mu$m. The CO first-overtone band overlaps partly with the $K$ band, and the CO fundamental and SiO first-overtone bands overlap with the $S_2$ band, whereas the SiO fundamental band lies outside the $U$ to $S_2$ wavelength range. The fundamental to first-overtone band flux ratios are assumed to be the same as observed for CO in SN 1987A \citep{Bou93}. We have used two configurations, one where the fraction of the heating emitted as molecule emission has been set to one, and one where this fraction has been set to zero. We note that CO fundamental band emission dominates the contribution to the $S_2$ band in all models at all times. This is because the O/Si/S to O/C zone mass ratio and the CO and SiO fundamental to first-overtone band ratios are all $\lesssim$1, decreasing with increasing initial mass and with time, respectively.

\paragraph{Treatment of dust} Dust is included in the modelling in a simplified way, and is represented as a grey absorptive opacity in the core. The absorbed luminosity is re-emitted as blackbody emission with a temperature determined from fits to the photometry. The dust emission is treated separately from the modelling, and is added by post-processing of the model spectra. All the models presented in \citetalias{Jer14} have the same optical depth and dust temperature, determined from the pseudo-bolometric optical lightcurve and fits to the $H$, $K$, and $S_1$ photometry, as described below. In addition we have constructed two new models: 12E, which differs from 12C only in the absence of dust, and 12F, which differs from 12C only in the optical depth and the method used to determine the temperature.

The \citetalias{Jer14} models have an optical depth of 0.25, turned on at day 200, which approximately matches the behaviour of the optical pseudo-bolometric lightcurve. The temperature is constrained to scale as for a homologously expanding surface, representing a large number of optically thick dust clouds \citepalias{Jer14}. Minimizing the sum of squares of the relative flux differences of model and observed $H$, $K$, and $S_1$ photometry at days 200, 300, and 400 (including $H$ only at day 200), we find temperatures of 2000, 1097, and 668 K at days 200, 300, and 400, respectively, given the constraint $\mathrm{T}_\mathrm{dust}<2000$ K. The $S_2$ band was excluded as this band might have a contribution from molecule emission, whereas we know that the contribution from CO first-overtone emission to the $K$ band at day 206 is negligible. However, the fractional area of the emitting surface (x$_\mathrm{dust}$), turns out to be $\sim$20 times smaller than the fractional area needed to reproduce the optical depth, so the derived temperature is not consistent with the assumptions made.
 
Model 12F has an optical depth of 0.44, turned on at day 200, which better matches the behaviour of the optical pseudo-bolometric lightcurve. The constraint on the temperature evolution used for the \citetalias{Jer14} models has been abandoned, and fitting the $H$, $K$, and $S_1$ photometry as described above, we find temperatures of 1229, 931, and 833 K at days 200, 300, and 400, respectively. This shows that, in addition to the inconsistency between the absorbing and emitting area, the scaling of temperature is not well reproduced by the \citetalias{Jer14} models. Because of the better reproduction of the optical pseudo-bolometric lightcurve and the problems with the temperature scaling used for the \citetalias{Jer14} models, we use model 12F instead of model 12C as our preferred model in this paper. Abandoning all attempts to physically explain the temperature leaves us with black box model, parametrized with the optical depth and the temperature. This is a caveat, but also fair, as the original model is not self-consistent. We note that the good spectroscopic agreement found for model 12C in \citetalias{Jer14} does not necessarily apply to model 12F. However, for lines originating from the core, which are expected to be most affected by the higher optical depth of the dust, the flux would be $\sim$17 per cent lower and the blue-shifts $\sim$100 km~s$^{-1}$ higher (Sect.~\ref{s_spec_line_asymmetries}), so the differences are likely to be small.

\section{HYDE and the model grid}
\label{a_hyde}

\paragraph{HYDE} Is a one-dimensional Lagrangian hydrodynamical code based on the flux limited diffusion approximation, following the method described by \citet{Fal77}, and adopting the flux limiter given by \citet{Ber11}. The opacity is calculated from the OPAL opacity tables \citep{Igl96}, complemented with the low temperature opacities given by \citet{Ale94}. In addition we use an opacity floor set to $0.01$ cm$^2$ gram$^{-1}$ in the hydrogen envelope and $0.025$ cm$^2$ gram$^{-1}$ in the helium core, following \citet[][private communication]{Ber12}, who calibrated these values by comparison to the STELLA hydrodynamical code \citep{Bli98}. The electron density, needed in the equation of state, is calculated by solving the Saha equation using the same atomic data as in \citet{Jer11,Jer12}. The transfer of the gamma-rays and positrons emitted in the decay chain of \element[ ][56]{Ni} is calculated with a MC method, using the same grey opacities, luminosities, and decay times as in  \citet{Jer11,Jer12}.

\paragraph{The model grid} Is based on non-rotating solar metallicity helium cores, evolved to the verge of core-collapse with MESA STAR \citep{Pax11}. The configuration used was the default one, and a central density limit of 10$^{9.5}$ gram cm$^{-3}$ was used as termination condition. The evolved models span M$_{\mathrm{He}}$=4.0$-$5.0 M$_\odot$ in 0.25 M$_\odot$ steps and M$_{\mathrm{He}}$=5.0$-$7.0 M$_\odot$ in 0.5 M$_\odot$ steps. Below 4.0 M$_\odot$, stellar models were constructed by a scaling of the 4.0 M$_\odot$ density profile. The SN explosion was parametrized with the injected explosion energy (E), the mass of the \element[ ][56]{Ni} (M$_{\mathrm{Ni}}$), and its distribution. The mass fraction of \element[ ][56]{Ni} ($\mathrm{X}_{\mathrm{Ni}}$) was assumed to be a linearly declining function of the ejecta mass ($\mathrm{m}_{\mathrm{ej}}$) becoming zero at some fraction of the total ejecta mass (Mix$_{\mathrm{Ni}}$), expressed as $\mathrm{X}_{\mathrm{Ni}} \propto 1-\mathrm{m}_{\mathrm{ej}}/(\mathrm{Mix}_{\mathrm{Ni}} \mathrm{M}_{\mathrm{ej}}),~\mathrm{X}_{\mathrm{Ni}} \geq 0$. We note that this expression allows Mix$_{\mathrm{Ni}}$>1, although the interpretation then becomes less clear. The total volume of parameter space spanned is M$_{\mathrm{He}}$=2.5$-$7.0 M$_\odot$, E=0.2$-$2.2$\times$10$^{51}$ erg, M$_{\mathrm{Ni}}$=0.015$-$0.250 M$_\odot$, and Mix$_{\mathrm{Ni}}$=0.6$-$1.6 using a 12$\times$16$\times$13$\times$9 grid. A mass cut with zero velocity was set at 1.5 M$_\odot$, and the material below is assumed to form a compact remnant, although fallback of further material onto this boundary is not prohibited. To calculate bolometric lightcurves between days 100 and 400 we run HYDE in homologous mode, ignore the radiative transfer, and take the bolometric luminosity as the deposited radioactive decay energy (Appendix \ref{a_nlte_modelling}).

\bibliographystyle{aa}
\bibliography{sn2011dh}

\label{lastpage}

\end{document}